\documentstyle[preprint,aps]{revtex}
\newcommand{\be}{\begin{equation}}
\newcommand{\ee}{\end{equation}}
\newcommand{\beq}{\begin{eqnarray}}
\newcommand{\eeq}{\end{eqnarray}}
\newcommand{\otwo}{{\rm O}(2)}
\newcommand{\sotwo}{{\rm SO}(2)}
\newcommand{\sg}{\sigma}
\newcommand{\lop}{{\cal{L}}}
\newcommand{\nop}{{\cal{N}}}
\newcommand{\ord}[1]{{\cal{O}}(#1)}
\newcommand{\pf}{\omega_p}
\newcommand{\gm}[2]{\Gamma_{{#1},{#2}}}
\newcommand{\Gm}[1]{\Gamma_{{#1}}}
\newcommand{\Gmcc}[1]{\Gamma^\ast_{{#1}}}
\newcommand{\gmcc}[2]{\Gamma^\ast_{{#1},{#2}}}
\newcommand{\k}{\kappa}
\newcommand{\glim}{\gamma\rightarrow0^+}
\newcommand{\ind}{\mbox{\rm Ind }}
\newcommand{\bintlim}[2]{{\cal{B}}_{{#1},{#2}}}
\newcommand{\cintlim}[2]{{\cal{C}}_{{#1},{#2}}}
\newcommand{\sintlim}{{\cal{S}}}
\newtheorem{theorem}{Theorem}[section]
\newtheorem{lemma}{Lemma}[section]
\begin{document}
\title{Amplitude Equations for Electrostatic Waves:\\
universal singular behavior in the limit of weak instability}
\author{John David Crawford}
\address{Department of Physics and Astronomy\\
University of Pittsburgh\\
Pittsburgh, Pennsylvania  15260}
\date{September 30, 1994}
\maketitle
\begin{abstract}
An amplitude equation for an unstable mode in a collisionless plasma is derived
from the dynamics on the unstable manifold of the equilibrium $F_0(v)$.\\ The
mode eigenvalue arises from a simple zero of the dielectric
$\epsilon_{{k}}(z)$;  as the linear growth rate $\gamma$ vanishes, the
eigenvalue merges with the continuous spectrum on the imaginary axis and
disappears. The evolution of the
mode amplitude $\rho(t)$ is studied using an expansion in $\rho$. As $\glim$,
the expansion coefficients diverge, but these singularities are absorbed by
rescaling the
amplitude: $\rho(t)\equiv\gamma^2\,r(\gamma t)$. This renders the theory finite
and also indicates that the electric field
exhibits trapping scaling $E\sim\gamma^2$. These singularities and scalings are
independent of the specific $F_0(v)$ considered. The asymptotic dynamics of
$r(\tau)$ can depend on $F_0$ only through $\exp{i\xi}$ where $d\epsilon_{{k}}
/dz=|{\epsilon'_{{k}}}|\exp{-i\xi/2}$. Similar results also hold for the
electric field and distribution function.
\end{abstract}


\section{Introduction}

The evolution of an unstable electrostatic mode is a fundamental
problem in collisionless plasma theory. Although quite idealized, this
evolution
involves many features that are essential to more complicated and realistic
problems,
in particular there is a singular interaction between the wave and resonant
particles.
This resonance drives the initial growth of the unstable linear mode,
and then trapping of resonant particles by the finite amplitude wave
 marks the onset of strong nonlinear effects that saturate the instability.
These
nonlinear effects are difficult to treat analytically while maintaining the
self-consistent relationship between electric field and particles, and
calculations on this
problem have emphasized special regimes which allow simplifying approximations,
e.g.
a ``bump on tail'' distribution or instability driven by a small cold
beam.\cite{frieman,bald,dru,oni,owm,dewar,sim1,den,sim2,janssen,burnap}

In this paper, I describe a new approach which simplifies the problem by
restricting attention
to the dynamics occuring on the unstable manifold of the equilibrium.
Physically this
restriction means I consider initial conditions
in which {\em only} the unstable modes are excited, rather than allowing
arbitrary
initial conditions comprised of all linear modes. Mathematically the unstable
manifold is finite-dimensional and this reduction in dimension provides a
considerable simplification.

The restriction on initial conditions is compensated by the freedom from
inessential
assumptions about the equilibrium $F_0(v,\mu)$. Here $\mu$ denotes any
parameters such as density or temperature
that determine the properties of $F_0$; it is not necessary to make a specific
choice for $\mu$, rather we let
$F_0(v,\mu_c)$ denote the critical equilibrium. For $\mu<\mu_c$, $F_0(v,\mu)$
is linearly stable and
for $\mu>\mu_c$ there is an unstable mode (or modes) with linear growth rate
$\gamma$. The limit
$\mu\rightarrow\mu_c$ from the unstable regime will usually be denoted $\glim$.
Thus I am able to give a unified
treatment of instabilities in beam-plasma systems with warm or cold beams as
well as two-stream instabilities
for equilibria with counterstreaming components of equal density.

An additional motivation for this approach is the possibility that the dynamics
on the manifold
will exhibit universal features of the instability.  This hope arises from
experience with simpler
bifurcations in dissipative dynamical systems. Consider, for example, a Hopf
bifurcation\cite{guc,jdc2}
in which an equilibrium loses stability as an isolated conjugate pair
of eigenvalues $(\lambda,\lambda^\ast)$ cross the imaginary axis
while all other modes remain stable.  In this situation, there is a
two-dimensional unstable manifold
associated with the unstable modes, and the isolation of critical eigenvalues
permits the dynamics near the equilibrium to be rigorously reduced to the
two-dimensional
dynamics on the unstable manifold. In polar variables $(\rho,\theta)$, this
reduced two-dimensional system has the form
\beq
\dot{\rho}&=&\gamma\rho+a_1\rho^3+a_2\rho^5+\ord{\rho^7}\label{eq:hopf1}\\
\dot{\theta}&=&\omega+{a'_1}\rho^2+{a'_2}\rho^4+\ord{\rho^6}\label{eq:hopf2}
\eeq
where $A(t)=\rho(t)\,e^{-i\theta(t)}$
is the amplitude of the unstable linear mode and
$\lambda=\gamma-i\omega$ is the critical eigenvalue; the evolution of $\rho(t)$
decouples from
the phase $\theta(t)$ so that the dynamics can be easily analyzed.  Provided
the cubic coefficient $a_1$ is non-zero at the onset of instability, these
equations have a universal structure in the limit of weak instability $\glim$.
More specifically, by setting
$\rho(t)=\sqrt{\gamma}\;r(\gamma t)$ we obtain from (\ref{eq:hopf1})
\be
\frac{dr}{d\tau}=r+a_1r^3+\gamma a_2 r^5+\ord{\gamma^2}\label{eq:hopf3}
\ee
where $\tau=\gamma t$. As $\glim$, the terms of higher order in $r$ vanish
leaving $\dot{r}=r+a_1 r^3$, an asymptotic
equation of universal form reflecting the specific problem under consideration
only through the
coefficient $a_1$.  In this way, the unstable manifold dynamics for Hopf
bifurcation
reveals the slow time scale $\tau$ and,
more interestingly, a universal scaling behavior $\rho\sim\sqrt{\gamma}$ for
the mode amplitude.

There are many key differences between such a simple dissipative bifurcation
and the bifurcation
arising from the appearance of
an unstable mode in the linear spectrum of a Vlasov equilibrium. The Vlasov
equation is a Hamiltonian
dynamical system and the spectrum for a stable equilibrium is pure
imaginary.\cite{morrison,marwein,holm} The eigenvalues describing
the unstable modes are not isolated at the onset of instability, in fact they
appear
in the spectrum for the first time at onset and are embedded in the continuous
spectrum on the
imaginary axis.\cite{cra1} Furthermore, without dissipation, one does not
expect the unstable  manifold
associated with the instability to be attracting.

Despite these important differences, it is feasible to adapt the derivation of
(\ref{eq:hopf1}) - (\ref{eq:hopf2}) to the bifurcation of an electrostatic mode
and to obtain
the corresponding equations for the dynamics on the unstable manifold. The
$\glim$ limit of these
equations is remarkable and provides a striking contrast with the familiar
limiting behavior
in Hopf bifurcation.\cite{jdc} For an electrostatic mode, the coefficients
$(a_j, {a'_j})$ are singular as $\glim$
with asymptotic form $\gamma^{-(4j-1)}$. These divergences are not unphysical,
rather they signal the presence of a different scaling from that
characterizing a Hopf bifurcation;  by setting $\rho(t)=\gamma^2\,r(\gamma t)$
one can absorb the
singular behavior at every order leaving rescaled equations for $r$ that are
well behaved as $\glim$.
Unlike the rescaled equation for Hopf bifurcation (\ref{eq:hopf3}), in the
Vlasov case the terms that
are higher order in $r$ are {\em not} higher order in $\gamma$, rather at each
order
$b_j\,r^{2j}$ the rescaled
coefficient $b_j=\gamma^{(4j-1)}\;a_j$ is order unity as $\glim$.
Hence the equation for $dr/d\tau$ does not truncate, and the
$\glim$ equation retains an infinite set of terms. The first two coefficients
have been calculated and shown to be {\em independent} of $F_0(v,\mu_c)$ as
$\glim$ with values $b_1=-1/4$ and $b_2=13/64$. The coefficients at higher
order are not explicitly determined, however one can prove that at each order
there is a universal function $Q_j$,
independent of $F_0(v,\mu_c)$, such that as $\glim$
\be
b_j=\mbox{\rm Re}\,Q_j(e^{i\xi(\mu_c)})
\ee
where the phase $e^{i\xi}\equiv{\epsilon'_k}^\ast/\epsilon'_k$ is defined in
terms of
the $\glim$ limit of the derivative of the dielectric function $\epsilon_k$.
The identification of $e^{i\xi}$
as a scaling variable for the $\glim$ regime which captures any remaining
dependence  on the underlying equilibrium is a novel result of this work.

The scaling $\rho(t)\sim\gamma^2$ for the amplitude of the unstable mode
implies that the wave electric field follows the so-called trapping
scaling $E\sim\gamma^2$ in the limit $\glim$. The terminology arises from the
equivalent scaling
$\omega_b\sim\gamma$ between the bounce frequency of trapped particles and the
growth rate ($\omega_b^2\equiv ekE_k/m$). Trapping scaling has been a
characteristic feature of previous numerical simulations\cite{den,sim2} as well
as recent experiments\cite{tsu}; however the theoretical results on the scaling
of the saturated electric field are divided between theories predicting ``Hopf
scaling'' $E\sim\sqrt{\gamma}$\cite{frieman,sim1,janssen,burnap}, and theories
predicting trapping scaling.\cite{bald,dru,oni,owm,dewar,endnote} Perhaps the
best known analysis leading to a prediction of Hopf scaling is a controversal
paper by Simon and Rosenbluth.\cite{sim1} Their work treats a one mode bump on
tail instability by perturbatively expanding the Vlasov equation and seeking a
time-periodic nonlinear solution. The perturbation theory leads to singular
results and the final expressions are rendered finite by prescribing a
regularization procedure. Subsequent perturbation theories have encountered
comparable difficulties and proposed similar
prescriptions.\cite{janssen,burnap}

The approach I develop in this work differs crucially from these investigations
in the interpretation and treatment of the singular behavior of the expansion
at $\gamma=0$. The derivation of the amplitude
dynamics on the unstable manifold gives nonlinear coefficients as integrals
over velocity, and as $\glim$ these integrals diverge due to pinching
singularities that develop at the phase velocity of the mode. As already
mentioned, these
divergences can be absorbed by simply rescaling the wave amplitude and this
rescaling reveals an electric field that follows the trapping scaling. If,
on the other hand, one were to regularize the integrals by somehow discarding
the divergent part then the resulting equations for the mode amplitude would
indeed scale as in dissipative Hopf bifurcation. This, in essence, is the step
taken in the theories that predict Hopf scaling for the electric field.

The theory of invariant manifolds for equilibria of infinite-dimensional
dynamical systems, such as partial differential equations, has been developed
extensively in recent years with rigorous results establishing the existence
and properties of these structures for various classes of evolution equations.
Examples utilizing a variety of techniques are found in
\cite{marsmac,henry,carr,chow,mielke,renardy}, and there is an introductory
review by Vanderbauwhede and Iooss.\cite{vi} However, this theory does not yet
treat equations such as the Vlasov-Poisson system and it seems to be an open
problem to rigorously construct invariant manifolds for Vlasov equilibria.
In this paper, unstable manifolds serve a heuristic role by motivating certain
procedures for constructing the expansions in the mode amplitude. Although
these same expansions could certainly be set up without mentioning the
manifolds,
the dynamical systems viewpoint does seem to bring additional insight. Future
development of a rigorous invariant manifold theory for the Vlasov equation may
provide mathematical justification for these expansions. The present analysis
is solely concerned with understanding the properties of the instability as
represented by the amplitude expansions.

The remainder of the Introduction is devoted to defining our notation and
summarizing relevant well known facts concerning the spectrum and
eigenfunctions of the Vlasov-Poisson equation. In Section II the description of
the unstable manifold is briefly reviewed and the equations necessary to obtain
the dynamics on the manifold are derived. These equations are solved using
power series in the amplitude of the unstable mode in Section III and a
detailed analysis of the lowest order term in this expansion reveals the
divergence mentioned above. This divergence is shown to imply the trapping
scaling for the mode amplitude. In Sections IV-VI, the structure of the
expansions is examined to all orders. The increasing strength of the
divergences is calculated and a detailed analysis is made of how the dynamics
on the unstable manifold depends on the critical equilibrium $F_0(v,\mu_c)$ in
the $\glim$ limit. The mode amplitude dynamics, the electric field, and to a
large extent the distribution function depend on $F_0(v,\mu_c)$ only through
the derivative of the dielectric function $\epsilon'_k$. This conclusion
indicates a degree of universality to the dynamics of a weakly unstable
electrostatic mode that has not been previously appreciated.

\subsection{Notation}

For a neutral collisionless plasma with a fixed ion density $n_0$,
the electron distribution function $F(x,v,t)$ and the electrostatic potential
$\Phi(x,t)$
satisfy the dimensionless Vlasov-Poisson equations (in one dimension)
\be
\frac{\partial F}{\partial t}+v\frac{\partial F}{\partial x}+\frac{\partial
\Phi}{\partial x}\frac{\partial F}{\partial v}=0\label{eq:vlasov}
\ee
\be
\frac{\partial^2\Phi}{\partial
x^2}=\int^\infty_{-\infty}\,dv\,F(x,v,t)-1.\label{eq:poisson}
\ee
The plasma length is $L$ and periodic boundary conditions are assumed
with normalization
\be
\int^{L/2}_{-L/2}\,dx\,\int^\infty_{-\infty}\,dv\,F(x,v,t)=L.\label{eq:Fnorm}
\ee
Here $x$, $t$ and $v$ are measured in units of $u/\pf$, $\pf^{-1}$ and $u$,
respectively, where $u$ is a chosen velocity scale and $\pf^2=4\pi e^2n_0/m$.
The electron charge and mass are $-e$ and $m$ and the ions are singly charged.
The dimensionless distribution function and potential are measured in units of
$u^{-1}$ and $mu^2/e$ respectively.

An inner product between two functions $G_1(x,v)$ and $G_2(x,v)$ is defined by
\be
\left(G_1,G_2\right)\equiv\int^{L/2}_{-L/2}\,dx\,<G_1,G_2>\label{eq:innerxv}
\ee
where $<G_1,G_2>$ denotes the integration over $v$ alone:
\be
<G_1,G_2>\equiv\,\int^\infty_{-\infty}\,dv\, G_1(x,v)^\ast
G_2(x,v).\label{eq:innerv}
\ee
With periodic boundary conditions the allowed wavevectors are multiples of
$k_c=2\pi/L$, and the Fourier expansion of $G(x,v)$ will be written as
\be
G(x,v)=\sum^{\infty}_{k=-\infty}\,e^{ikk_cx}\,G_k(v).\label{eq:fexp}
\ee

Let $F_0(v,\mu)$ denote a parametrized family of equilibria,
normalized to unit density
\be
\int^\infty_{-\infty}\,dv\,F_0(v,\mu)=1,\label{eq:Feqnorm}
\ee
and re-express the distribution function relative to this family
$F(x,v,t)=F_0(v,\mu)+f(x,v,t)$, then $f$ satisfies
\be
\frac{\partial f}{\partial t}+v\frac{\partial f}{\partial x}+\frac{\partial
\Phi}{\partial x}\frac{\partial F_0}{\partial v}+\frac{\partial \Phi}{\partial
x}\frac{\partial f}{\partial v}=0\label{eq:vp1}
\ee
\be
\frac{\partial^2\Phi}{\partial
x^2}=\int^\infty_{-\infty}\,dv\,f(x,v,t)\label{eq:vp2}
\ee
and
\be
\int^{L/2}_{-L/2}\,dx\,\int^\infty_{-\infty}\,dv\,f(x,v,t)=0.\label{eq:fnorm}
\ee
With the Fourier series for $f$, equations (\ref{eq:vp1}) - (\ref{eq:vp2}) can
be combined
\be
\frac{\partial f}{\partial t}=\lop\,f+\nop(f)\label{eq:dynsys}
\ee
where the linear operator is defined by
\be
\lop\,f=\sum^{\infty}_{k=-\infty}\,e^{ikk_cx}\,(L_k f_k)(v)
\ee
\be
(L_k f_k)(v)=\left\{\begin{array}{cc}0&k=0\\
-ikk_c\left[vf_k(v)+k^{-2}\,\eta(v,\mu)\int^\infty_{-\infty}\,dv'\,f_k(v')
\right]&k\neq0\end{array}\right.
\ee
with
\be
\eta(v,\mu)=-k_c^{-2}\frac{\partial F_0}{\partial v }(v,\mu),
\ee
and
\be
\nop(f)=\frac{i}{k_c}\sum^{\infty}_{k=-\infty}\,e^{ikk_cx}\,
{\sum^{\infty}_{l=-\infty}}'\,
\frac{1}{l}\frac{\partial f_{k-l}}{\partial v}
\int^\infty_{-\infty}\,dv'\,f_l(v').\label{eq:nop}
\ee
Here and below a primed summation omits the $l=0$ term.

Symmetries of the model (\ref{eq:vlasov}) - (\ref{eq:poisson}) and the
equilibrium $F_0(v,\mu)$ are important qualitative features of the problem.
Spatial translation,
${\cal T}_a:(x,v)\rightarrow (x+a,v),$
and reflection,
${\kappa:(x,v)\rightarrow (-x,-v)},$
act as operators on the distribution function in the usual way: if $\alpha$
denotes an arbitrary transformation then $(\alpha\cdot f)(x,v)\equiv
f(\alpha^{-1}\cdot(x,v))$. The operators $\lop$ and $\nop$ commute with ${\cal
T}_a$ due to the spatial homogeneity of $F_0$,
and if $F_0(v,\mu)=F_0(-v,\mu)$,
then $\lop$ and $\nop$ also commute with the reflection operator $\k$.
Together ${\cal T}_a$ and $\k$ generate the symmetry group of the circle
$\otwo$ and without $\kappa$ the symmetry drops to $\sotwo$.

The spectral theory for $\lop$ is well established, and the needed results are
simply recalled to establish the notation.\cite{cra1,vkamp,case} The
eigenvalues $\lambda=-ikk_c z$ of $\lop$ are determined by the roots
$\Lambda_{k}(z,\mu)=0$ of the ``spectral function'',
\be
\Lambda_{k}(z,\mu)\equiv 1+
\frac{1}{k^2}\int^\infty_{-\infty}\,dv\,\frac{\eta(v,\mu)}{v-z}.
\label{eq:specfcn}
\ee
Unless it is necessary to manipulate the parameter dependence of
$\Lambda_{k}(z,\mu)$, the argument $\mu$ will be suppressed. If only
translation symmetry is present, these eigenvalues are generically complex;
when the equilibrium is also reflection-symmetric then real eigenvalues can
arise, for example in a two-stream instability.\cite{cowley}

The spectral function is analytic in the upper and
lower half planes with a branch cut on the real axis
along the support of $\eta$. For $z=r\pm i\epsilon$,
the discontinuity across the cut is given by the Plemelj formula\cite{musk}
\be
\lim_{\epsilon\rightarrow0^+}\,\Lambda_{k}(r\pm i\epsilon)=
1+ \frac{1}{k^2}\left[{\mbox{\rm P.V.}}
\int^\infty_{-\infty}\,dv\,\frac{\eta(v,\mu)}{v-r}\right]
\pm i\pi\eta(r,\mu).\label{eq:plemej}
\ee
The analytic continuation of $\Lambda_{k}(z)$ from $\mbox{\rm Im}\;z>0$ to
$\mbox{\rm Im}\;z<0$ yields the dielectric function
$\epsilon_{{k}}(z)$\cite{endnote1}, defined in the usual way via the Landau
contour.\cite{text}   Since our analysis focuses on the regime $\mbox{\rm
Im}\;z\geq0$, the notations $\Lambda_{k}(z)$ and $\epsilon_{{k}}(z)$ are
interchangeable in the subsequent discussion.

The branch cut of $\Lambda_{k}(z)$ corresponds to a continuous spectrum for
$\lop$ on the imaginary axis. As $\mu$ varies the roots of $\Lambda_{k}(z)$
typically vary; in particular, roots
can appear or disappear through the branch cut.
The appearance of a root at the cut
corresponds to the birth of eigenvalues embedded in the continuous spectrum and
this occurs for the critical parameter values $\mu_c$ marking the threshold of
linear instability. From (\ref{eq:plemej}) such a real root $r$ must satisfy
\beq
\eta(r,\mu_c)&=&0\label{eq:nmode1}\\
1+ \frac{1}{k^2}\left[{\mbox{\rm P.V.}}
\int^\infty_{-\infty}\,dv\,\frac{\eta(v,\mu_c)}{v-r}\right]
&=&0.
\eeq

Corresponding to the eigenvalue $\lambda=-ikk_c z$ is an eigenfunction
\be
\Psi(x,v)=e^{ikk_cx}\,\psi(v)\label{eq:efcn1}
\ee
with
\be
\psi(v)=\left(-\frac{1}{k^2}\right)\frac{\eta(v,\mu)}{v-z}.\label{eq:efcn2}
\ee
There is also an adjoint eigenfunction $\tilde{\Psi}(x,v)$
satisfying $(\tilde{\Psi},\Psi)=1$ given by
\be
\tilde{\Psi}(x,v)=\frac{1}{L}e^{ikk_cx}\tilde{\psi}(v)\label{eq:aefcn1}
\ee
where
\be
\tilde{\psi}(v)\equiv-\left(\frac{1}{\Lambda'_{k}(z)(v-z)}\right)^\ast.
\label{eq:aefcn2}
\ee
The normalization in (\ref{eq:aefcn2}) assumes that the root of
$\Lambda_{k}(z)$ is simple and is chosen so that $<\tilde{\psi},\psi>=1$.

\section{Amplitude equation on the unstable manifold}

Since Landau damping is weakest at long wavelengths, one expects that
instability will occur first at $k_c=2\pi/L$ as $\mu$ is varied through
$\mu_c$. This point has recently been treated pedagogically by Shadwick and
Morrison.\cite{shad}
The critical eigenvalue is then $\lambda=-ik_cz_0$ corresponding to
$\Lambda_{1}(z_0)=0$ for $k=1$. The root $z_0$ determines the phase
velocity $v_p$ and the growth rate $\gamma$ of the linear mode
\be
z_0=\frac{i\gamma}{k_c} + v_p;\label{eq:z0def}
\ee
both $v_p$ and $\gamma$ depend on $\mu$. However it is more convenient to take
$\gamma$ as the independent parameter and regard $\mu(\gamma)$, $z_0(\gamma)$,
and $v_p(\gamma)$ as functions of the growth rate. Thus $\Lambda_{1}(z_0)=0$ is
understood to mean
\be
\Lambda_{1}(z_0(\gamma),\mu(\gamma))= 1+\int_{-\infty}^{\infty}\,
\frac{dv\,\eta(v,\mu(\gamma))}{v-v_p(\gamma)-i\gamma/k_c}=0.\label{eq:paramrt}
\ee
In Appendix A, this equation is solved for $v_p(\gamma)$ and $\mu(\gamma)$ to
first order in $\gamma$.
{}From (\ref{eq:z0def}), $\lambda=-ik_cz_0(\gamma)$ becomes
$\lambda=\gamma-i\omega(\gamma)$ where $\omega(\gamma)=k_cv_p(\gamma)$.
The notation $\glim$ for the weak growth rate regime always refers to the joint
limit
\be
(\gamma, \omega(\gamma), \mu(\gamma))\rightarrow(0, \omega(0), \mu(0))\equiv(0,
\omega_c, \mu_c).\label{eq:glimdef}
\ee
Subsequently, the $\gamma$ argument for $v_p$, $\omega$, $z_0$, and $\mu$  will
generally be suppressed when it is not explicitly required.

I assume that $z_0$ is a simple root:
\be
\Lambda'_{1}(z_0)\equiv\frac{d\,\Lambda_{1}}{dz}(z_0)\neq0;\label{eq:simple}
\ee
in addition, at $z=z_0$ all derivatives of $\Lambda_{k}(z)$ are assumed to have
finite limits:
$\lim_{\gamma\rightarrow0^+}|\Lambda^{(j)}_{k}(z_0)|<\infty$
where
\be
\Lambda^{(j)}_{k}(z)\equiv\frac{d^j\,\Lambda_{k}}{dz^j}(z)=
\frac{j!}{k^2} \int^\infty_{-\infty}\,dv\,
\frac{\eta(v,\mu)}{(v-z)^{j+1}}.\label{eq:specfcnder}
\ee
{}From (\ref{eq:specfcn}) and (\ref{eq:paramrt}) $\Lambda_{k}(z_0)$ can be
evaluated for arbitrary $k$
\be
\Lambda_{k}(z_0)=\frac{k^2-1}{k^2};\label{eq:specfcnid}
\ee
this identity will be needed below. Since $\Lambda_{k}(z)$ and
$\epsilon_{{k}}(z)$ are identical for $\mbox{\rm Im}\;z \geq0$, the relations
(\ref{eq:simple}) - (\ref{eq:specfcnid}) are unchanged if $\Lambda_{k}(z_0)$ is
replaced by $\epsilon_{{k}}(z_0)$.

The unstable mode corresponding to $z_0$ is
\be
\Psi_1(x,v)=e^{ik_cx}\,\psi_c(v)\equiv e^{ik_cx}\,
\left(\frac{-\eta(v,\mu)}{v-z_0}\right).\label{eq:lefcn}
\ee
When $F_0(v,\mu)$ lacks reflection symmetry, then this wave typically has a
non-zero phase velocity and $\lambda$ is complex. In this case the identities
$\Lambda_{k}(z)=\Lambda_{-k}(z)$ and $\Lambda_{k}(z)^\ast=\Lambda_{k}(z^\ast)$
imply three additional modes: $\Psi_1^\ast$, $\Psi_2$, and $\Psi_2^\ast$ where
\be
\Psi_2(x,v)=e^{ik_cx}\,
\left(\frac{-\eta(v,\mu)}{v-z_0^\ast}\right).
\ee
These eigenfunctions correspond to eigenvalues $\lambda^\ast$, $-\lambda^\ast$,
and $-\lambda$, respectively, and fill out the eigenvalue quartet
characteristic
of Hamiltonian systems.

In the event that $F_0(v,\mu)$ is reflection-symmetric in $v$, then both real
and complex eigenvalues may occur. If $\lambda$ is complex, then since $\kappa$
and $\lop$ commute, $\Psi_1$ and
\be
(\kappa\cdot\Psi_1)(x,v)=e^{-ik_cx}\,\psi_c(-v)
\ee
are linearly independent eigenvectors for the same eigenvalue
$\lambda=-ik_cz_0$. Thus $\lambda$ generically has multiplicity two. The same
considerations hold for $\Psi_1^\ast$, $\Psi_2$, and $\Psi_2^\ast$ so the
entire quartet has double mutiplicity. When $\lambda$
is real as in the symmetric two-stream instability, then one can show that
$\kappa\cdot\Psi_1=\Psi_1^\ast$ and $\lambda$ is again multiplicity two. The
eigenvectors $\Psi_2$ and $\kappa\cdot\Psi_2=\Psi_2^\ast$ correspond to
$-\lambda$.

\subsection{Prototypical example}

A convenient and explicit family of equilibria, satisfying (\ref{eq:Feqnorm}),
is
\be
F_0(v,\mu)=\frac{1}{\pi}\left[\frac{n}{(v-u_p)^2+1}+
\frac{\Delta(1-n)}{(v-u_b)^2+\Delta^2}\right]\label{eq:family}
\ee
with parameter set $\mu=(n,u_p,u_b,\Delta)$. In this example, one component,
the plasma, has density $n\,n_0$ and the second component, the beam, has
density $(1-n)n_0$; each component has its own drift velocity and the beam has
thermal width $\Delta$. The thermal width of the plasma has been taken as the
velocity unit. If $n=0.5$, $\Delta=1$ and $u_p=-u_b$, then the family has
reflection symmetry.

In the four-dimensional parameter space
the threshold of linear instability corresponds to a three-dimensional surface
which is denoted by $\mu_c$. The equilibria $F_0(v,\mu_c)$ on this surface have
eigenvalues embedded in the continuous spectrum.

An illustrative realization of a one mode beam-plasma instability is shown in
Fig. 1 for a system with $L=2\pi$, $n=0.8$, $u_p=0.0$, $\Delta=0.3$, and $u_b$
varied to produce the instability of the $k=1$ mode. The spectrum of $\lop$ for
the stable (a), critical (b) and unstable (c) regimes is illustrated in Fig. 2.

\subsection{Critical linear modes}

In this paper only the simplest instabilities having two unstable
eigenvectors are considered. This setting nevertheless encompasses both the
case of a reflection-symmetric instability with a real eigenvalue (two-stream)
and the case of a complex conjugate eigenvalue pair without reflection symmetry
(beam-plasma). In either case, the components of the distribution function
along the critical eigenvectors $\Psi_1$ and $\Psi_1^\ast$ are separated out by
writing
\be
f(x,v,t)=\left[A(t)\Psi(x,v) + cc\right] + S(x,v,t)\label{eq:linmodes}
\ee
where $A(t)=(\tilde{\Psi},f)$ is the mode amplitude for $\Psi$ and
$(\tilde{\Psi},S)=0$.
In (\ref{eq:linmodes}) the subscript
on $\Psi_1$ has been dropped, and $\tilde{\Psi}=\exp(ik_cx)\,\tilde{\psi_c}/L$
is the adjoint function for $z_0$ given in (\ref{eq:aefcn1}). The action of the
translations ${\cal T}_a$ and reflection $\kappa$ on the distribution function
implies an action by these operators on the mode amplitudes: from
(\ref{eq:linmodes}) we have
\beq
{\cal T}_a\cdot A&=&e^{-ik_c a}A\label{eq:Atrans}\\
\kappa\cdot A&=&A^\ast.\label{eq:Aref}
\eeq
When these transformations are symmetries, these relations are useful for
organizing the amplitude expansions below.

The Vlasov equation (\ref{eq:dynsys}) determines
the dynamics for $A$ and $S$:
\beq
\dot{A}&=&\lambda\,A+(\tilde{\Psi},\nop(f))\label{eq:adot}\\
\frac{\partial S}{\partial t}&=&\lop
S+\nop(f)-\left[(\tilde{\Psi},\nop(f))\,\Psi + cc\right]\label{eq:Sdot}
\eeq
where
\be
(\tilde{\Psi},\nop(f))=-\frac{i}{k_c}\,{\sum^{\infty}_{l=-\infty}}'
\frac{1}{l}<{\partial_v}\,\tilde{\psi_c}, f_{1-l}>
\int^\infty_{-\infty}\,dv'\,f_l(v').\label{eq:projnl}
\ee
In writing (\ref{eq:adot}) I have used the adjoint relationship
$(\tilde{\Psi},\lop S)=
(\lop^\dagger\tilde{\Psi},S)=\lambda^\ast(\tilde{\Psi},S)=0$ and in
(\ref{eq:projnl}) an integration by parts $<\tilde{\psi_c},{\partial_v}\,
f_{k-l}>= -<{\partial_v}\,\tilde{\psi_c},f_{k-l}>$ moves the velocity
derivative in (\ref{eq:nop}) onto $\tilde{\psi_c}$.

\subsection{Amplitude equation on the unstable manifold}

In (\ref{eq:adot}) and (\ref{eq:Sdot}) the critical modes are linearly
decoupled from the other degrees of freedom but remain coupled to $S$ through
the nonlinear terms. For $\gamma>0$, by restricting to the dynamics on the
unstable manifold, one can decouple the nonlinear terms as well and obtain from
(\ref{eq:adot}) an autonomous description of the dynamics of $A$ as a
two-dimensional flow. This reduction to a two-dimensional submanifold is
analogous to the familiar procedure of center manifold reduction in dissipative
bifurcation theory; here the unstable manifold partially compensates for the
absence of a low-dimensional center manifold at criticality.\cite{guc,jdc2}

The essential properties of an unstable manifold $W^u$ are briefly described,
more detail can be found in the extensive dynamical systems
literature.\cite{guc,jdc2} The unstable modes $\Psi$ and $\Psi^\ast$ span a
two-dimensional unstable subspace $E^u$ and the remaining spectrum of $\lop$
determines the center subspace $E^c$ (for spectrum on the imaginary axis) and
a two-dimensional stable subspace $E^s$, spanned by the two stable modes; see
Fig. 2(c). These
subspaces are invariant under the linear flow\cite{endnote2}, $\partial_t
f=\lop f$, but this invariance is lost when the nonlinear terms $\nop (f)$
couple $E^u$ to $E^c\oplus E^s$. However there are nonlinear manifolds present
for the full nonlinear flow that are analogous to the subspaces of linear
theory. Specifically, the unstable manifold $W^u$ is invariant under the
nonlinear evolution, $\partial_t f= \lop f+\nop (f)$, and tangent to $E^u$ at
the equilibrium $F_0(v,\mu)$; hence $W^u$ is also two-dimensional. Solutions on
this manifold $f^u(x,v,t)$ asymptotically approach
$F_0$ as $t\rightarrow-\infty$.

The invariance of the unstable manifold means that restricting the Vlasov
equation to $W^u$ yields an autonomous two-dimensional dynamical system
describing the evolution of initial conditions on the manifold, $f^u(x,v,0)\in
W^u$.  Near the equilibrium this restriction is tractable since the
tangency between $W^u$ and the unstable subspace allows the manifold to be
described by a function,
\beq
H:E^u&\rightarrow& E^c\oplus E^s\\
(A,A^\ast)&\rightarrow&H(x,v,A,A^\ast)\label{eq:graph}
\eeq
which measures the ``distance'' from $E^u$ to $W^u$; see Fig. 3. Thus near
$F_0$, for trajectories $f^u(x,v,t)$ on $W^u$, the evolution of the
non-critical modes $S(x,v,t)$ in (\ref{eq:linmodes}) is controlled by the
critical modes:
\be
S(x,v,t)=H(x,v,A(t),A^\ast(t)),\label{eq:Su}
\ee
and these trajectories can be described entirely in terms of $H$ and the
evolution of $A(t)$:
\be
f^u(x,v,t)=\left[A(t)\Psi(x,v) + cc\right] +
H(x,v,A(t),A^\ast(t)).\label{eq:fu}
\ee

{\em If} $H$ is known, then using (\ref{eq:fu}), the general equations
(\ref{eq:adot}) and (\ref{eq:Sdot}) can be restricted to the unstable manifold:
\beq
\dot{A}&=&\lambda\,A+(\tilde{\Psi},\nop(f^u))\label{eq:aeqn}\\
\left.\frac{\partial S}{\partial t}\right|_{f^u} &=&\lop
H+\nop(f^u)-\left[(\tilde{\Psi},\nop(f^u))\,\Psi + cc\right].\label{eq:Sdotwu}
\eeq
Now (\ref{eq:aeqn}) defines an {\em autonomous} two-dimensional flow describing
the self-consistent nonlinear evolution of the unstable mode; this is the
amplitude equation I wish to study.

Translation symmetry forces the right hand side of (\ref{eq:aeqn}) to have the
form
\be
\lambda\,A+(\tilde{\Psi},\nop(f^u))=A\,p(\sigma,\mu)\label{eqn:pdef}
\ee
where $\sigma\equiv|A|^2$ and the function $p(\bullet,\mu)$ is not constrained
by the translation symmetry.\cite{endnote3} Typically $p(\sigma,\mu)$ is
complex-valued, however when
$F_0$ is reflection-symmetric then $p(\sigma,\mu)$ is forced to be real.

It is convenient to  view (\ref{eqn:pdef}) as expressing $p$ in terms of $H$,
and make this
connection more explicit by evaluating $(\tilde{\Psi},\nop(f^u))$. This
calculation requires the Fourier series for $H$
\be
H(x,v,A,A^\ast)=\sum^{\infty}_{k=-\infty}\,e^{ikk_cx}\,H_k(v,A,A^\ast),
\label{eq:Hexpand}
\ee
and the Fourier components of $f^u$
\be
f^u_k(v)=\left[A\psi_c(v)\,\delta_{k,1} +
A^\ast\psi_c(v)^\ast\,\delta_{k,-1}\right] + H_k(v,A,A^\ast).\label{eq:fufc}
\ee
The evaluation of $(\tilde{\Psi},\nop(f^u))$ from (\ref{eq:projnl}) is
simplified by first noting that the components of $H$ are forced by translation
symmetry to have the form
\beq
H_0(v,A,A^\ast)&=&\sg\,h_0(v,\sg)\nonumber\\
H_1(v,A,A^\ast)&=&A\sg\,h_1(v,\sg)\label{eq:hdef}\\
H_k(v,A,A^\ast)&=&A^k\,h_k(v,\sg)\;\;\;\;{\mbox{for}}\;\;k\geq2,\nonumber
\eeq
where $H_{-k}=H_k^\ast$ and the functions $h_k$ are not constrained by the
translations.\cite{cra5} However, if reflection symmetry also
holds, then
\be
h_k(-v,\sg)=h_k(v,\sg)^\ast.
\ee
Combining (\ref{eq:fufc}) - (\ref{eq:hdef}) with $(\tilde{\Psi},\nop(f^u))$ in
(\ref{eq:projnl}) yields
\beq
(\tilde{\Psi},\nop(f^u))&=&\frac{-iA\sigma}{k_c}\left\{\rule{0in}{0.3in}
<\partial_v\tilde{\psi_c},(h_0-h_2)>+\frac{\Gm
2}{2}<\partial_v\tilde{\psi_c},\psi_c^\ast>\right.\label{eq:peval}\\
&&+\sigma\left[\rule{0in}{0.2in}\Gm 1<\partial_v\tilde{\psi_c},h_0>-\Gmcc
1<\partial_v\tilde{\psi_c},h_2>+\frac{\Gm
2}{2}<\partial_v\tilde{\psi_c},h_1^\ast>-\frac{\Gmcc
2}{2}<\partial_v\tilde{\psi_c},h_3>\right]\nonumber\\
&&+\sum_{l=3}^{\infty}\left.\;\frac{\sigma^{l-2}}{l}[\Gm
l<\partial_v\tilde{\psi_c},h_{l-1}^\ast>-\sigma\Gmcc
l<\partial_v\tilde{\psi_c},h_{l+1}>]\nonumber
\rule{0in}{0.3in}\right\}
\eeq
where, on the
right hand side, the velocity integral of $h_{k}$ is denoted by
\be
\Gm k(\sg)\equiv\int^\infty_{-\infty}\,dv'\,h_{k}(v',\sg).\label{eq:Gdef}
\ee
Now comparing (\ref{eqn:pdef}) and (\ref{eq:peval}) provides the desired
expression for $p$
\beq
p(\sigma,\mu)&=&\lambda-\frac{i\sigma}{k_c}\left\{\rule{0in}{0.3in}
<\partial_v\tilde{\psi_c},(h_0-h_2)>+\frac{\Gm
2}{2}<\partial_v\tilde{\psi_c},\psi_c^\ast>\right.\label{eq:ph}\\
&&+\sigma\left[\Gm 1<\partial_v\tilde{\psi_c},h_0>-\Gmcc
1<\partial_v\tilde{\psi_c},h_2>+\frac{\Gm
2}{2}<\partial_v\tilde{\psi_c},h_1^\ast>-\frac{\Gmcc
2}{2}<\partial_v\tilde{\psi_c},h_3>\right]\nonumber\\
&&+\sum_{l=3}^{\infty}\left.\;\frac{\sigma^{l-2}}{l}[\Gm
l<\partial_v\tilde{\psi_c},h_{l-1}^\ast>-\sigma\Gmcc
l<\partial_v\tilde{\psi_c},h_{l+1}>]\nonumber
\rule{0in}{0.3in}\right\}.
\eeq
In order to exploit this expression for $p$, it is necessary to determine $H$
or equivalently to determine the functions $h_k$.

\subsection{Representation of the unstable manifold}

An equation for $H$ follows by requiring consistency between
(\ref{eq:Su}) and (\ref{eq:Sdotwu}); setting the time derivative of
(\ref{eq:Su}) equal to the right hand side of (\ref{eq:Sdotwu}) gives
\be
\frac{\partial H}{\partial A}\,\dot{A}+\frac{\partial H}{\partial
A^\ast}\,\dot{A}^\ast = \lop\,H+ \nop(f^u)-\left[(\tilde{\Psi},\nop(f^u))\,\Psi
+ cc\right]
\label{eq:Heqn}
\ee
which is to be solved for $H$ subject to $H(x,v,0,0)=0$
and
\be
\frac{\partial H}{\partial A}(x,v,0,0)=\frac{\partial H}{\partial
A^\ast}(x,v,0,0)=0.
\ee
These latter conditions are implied by the tangency between $W^u$ and $E^u$ at
the equilibrium $(A,A^\ast)=(0,0)$, and are automatically satisfied in this
case by virtue of (\ref{eq:hdef}).

With the previous expression for $(\tilde{\Psi},\nop(f^u))$ in (\ref{eq:peval})
and the notation in (\ref{eq:hdef}) for the Fourier components of $H$, the
components of (\ref{eq:Heqn}) take the form
\beq
\lefteqn{\frac{\partial H_0}{\partial A}\,\dot{A}+\frac{\partial H_0}{\partial
A^\ast}\,\dot{A}^\ast=}\hspace{0.5in}\label{eq:hfc0}\\
&&\frac{i\sg}{k_c}\,\frac{\partial}{\partial v}
\left\{\left[\psi_c^\ast+\sg( h_1^\ast -\psi_c \Gmcc 1)+\sg^2\, h_1^\ast \Gm 1
+\sum^{\infty}_{l=2}\, \frac{\sg^{l-1}}{l}
 h_l^\ast \Gm l\right]-cc\right\}\nonumber
\eeq
\beq
\lefteqn{\frac{\partial H_{1}}{\partial A}\,\dot{A}+\frac{\partial
H_1}{\partial A^\ast}\,\dot{A}^\ast - L_1\,H_1
=}\hspace{1.0in}\label{eq:hfc1}\\
&&\frac{iA\sg}{k_c}\,{\cal P}_\perp\frac{\partial}{\partial v}
\left\{\rule{0in}{0.3in} h_0-h_2+\frac{1}{2}\psi_c^\ast\, \Gm 2
+\sg\left[h_0 \Gm 1- h_2\Gmcc 1+\frac{1}{2}h_1^\ast\Gm 2
-\frac{1}{2}h_3\Gmcc 2 \right]\right.\nonumber\\
&&\hspace{1.0in}+\sum^{\infty}_{l=3}\left.\frac{\sg^{l-2}}{l}
\left[h_{l-1}^\ast \Gm l-\sg h_{l+1} \Gmcc l\right]
\rule{0in}{0.3in}\right\}\nonumber
\eeq
\beq
\lefteqn{\frac{\partial H_{2}}{\partial A}\,\dot{A}+\frac{\partial
H_2}{\partial A^\ast}\,\dot{A}^\ast - L_2\,H_2=}\hspace{1.0in}\label{eq:hfc2}\\
&&\frac{iA^2}{k_c}\,\frac{\partial}{\partial v}
\left\{\rule{0in}{0.3in}\psi_c+
\sg\left[h_1+\psi_c\Gm 1 -h_3
+\frac{1}{2} h_0 \Gm 2  +\frac{1}{3} \psi_c^\ast \Gm 3 \right]\right.
\nonumber\\
&&\hspace{1.25in}
\sg^2\left[h_1\Gm 1 - h_3 \Gmcc 1
-\frac{1}{2} h_4 \Gmcc 2 +\frac{1}{3} h_1^\ast \Gm 3 \right]\nonumber\\
&&\hspace{1.0in}\left.
-\frac{\sg^3}{3} h_5 \Gmcc 3 +
\sum^{\infty}_{l=4}\frac{\sg^{l-2}}{l}\left[h_{l-2}^\ast \Gm l-\sg^2 h_{l+2}
\Gmcc l\right]\rule{0in}{0.3in}\right\}\nonumber
\eeq
and
\beq
\lefteqn{\frac{\partial H_{k}}{\partial A}\,\dot{A}+\frac{\partial
H_k}{\partial A^\ast}\,\dot{A}^\ast - L_k\,H_k=}\hspace{1.0in}\label{eq:hfck}\\
&&\frac{iA^k }{k_c}\,\frac{\partial}{\partial v}
\left\{\rule{0in}{0.3in}h_{k-1}+\frac{\psi_c}{k-1}\Gm {k-1}
+\sum^{k-2}_{l=2}\, \frac{h_{k-l}}{l}\Gm l\right.
\nonumber\\
&&\hspace{0.75in}+\sg\left[h_{k-1}\Gm 1-h_{k+1}
+\frac{h_1}{k-1}\Gm {k-1}
+\frac{h_0}{k}\Gm k+\frac{\psi_c^\ast}{k+1}\Gm {k+1}\right]
\nonumber\\
&&\hspace{0.75in}+\sg^2\left[-h_{k+1}\Gmcc 1
+\frac{h_1^\ast}{k+1}\Gm {k+1}\right]
+\sum^{\infty}_{l=k+2}\frac{\sg^{l-k}}{l}h_{l-k}^\ast
\Gm l\nonumber\\
&&\hspace{1.5in}\left.-\sum^{\infty}_{l=2}\frac{\sg^{l}}{l}h_{k+l}
\Gmcc l\rule{0in}{0.3in}\right\}\nonumber
\eeq
for $k=0, 1, 2$, and $k>2$, respectively. In the $k=1$ component
(\ref{eq:hfc1}),  ${\cal P}$ is the projection operator onto the $\psi_c(v)$
component of a function $g(v)$,
\be
({\cal P}g)(v)\equiv<\tilde{\psi}_c,g>\psi_c(v),\label{eq:Pdef}
\ee
and the orthogonal projection is denoted by ${\cal P}_\perp\equiv I-{\cal P}$.

The expressions for $\dot{A}\partial_AH_k+\dot{A}^\ast\partial_{A^\ast}H_k$ can
also be evaluated in terms of the functions $h_k(v,\sigma)$ and
$p(\sigma,\mu)$:
\beq
\frac{\partial H_0}{\partial A}\,\dot{A}+\frac{\partial H_0}{\partial
A^\ast}\,\dot{A}^\ast&=&\sigma(p+p^\ast)\left[h_0+\sigma\frac{\partial
h_0}{\partial {\sigma}}\right]\label{eq:hfc0sim}
\eeq
\beq
\frac{\partial H_{1}}{\partial A}\,\dot{A}+\frac{\partial H_1}{\partial
A^\ast}\,\dot{A}^\ast - L_1\,H_1&=& A\sigma\left[
\{(2p+p^\ast)-L_1\}h_1+(p+p^\ast)\sigma\frac{\partial h_1}{\partial
{\sigma}}\right]\label{eq:hfc1sim}
\eeq
and
\beq
\frac{\partial H_{k}}{\partial A}\,\dot{A}+\frac{\partial H_k}{\partial
A^\ast}\,\dot{A}^\ast - L_k\,H_k&=&
A^k\left[
\{kp-L_k\}h_k+(p+p^\ast)\sigma\frac{\partial h_k}{\partial {\sigma}}\right]
\label{eq:hfcksim}
\eeq
for $k=0, 1$, and $k\geq2$, respectively. By combining (\ref{eq:hfc0}) -
(\ref{eq:hfck}) and (\ref{eq:hfc0sim}) - (\ref{eq:hfcksim}), the component
equations of (\ref{eq:Heqn}) reduce to a simpler form:
\beq
\lefteqn{(p+p^\ast)\left[h_0+\sigma\frac{\partial h_0}{\partial
{\sigma}}\right]=}\hspace{0.5in}\label{eq:hfc0f}\\
&&\frac{i}{k_c}\,\frac{\partial}{\partial v}
\left\{\left[\psi_c^\ast+\sg( h_1^\ast -\psi_c \Gmcc 1)+\sg^2\, h_1^\ast \Gm 1
+\sum^{\infty}_{l=2}\, \frac{\sg^{l-1}}{l}
 h_l^\ast \Gm l\right]-cc\right\}\nonumber
\eeq
\beq
\lefteqn{\left[
\{(2p+p^\ast)-L_1\}h_1+(p+p^\ast)\sigma\frac{\partial h_1}{\partial
{\sigma}}\right]=}\hspace{1.0in}\label{eq:hfc1f}\\
&&\frac{i}{k_c}\,{\cal P}_\perp\frac{\partial}{\partial v}
\left\{\rule{0in}{0.3in}h_0-h_2+\frac{1}{2}\psi_c^\ast\, \Gm 2
+\sg\left[h_0 \Gm 1- h_2\Gmcc 1+\frac{1}{2}h_1^\ast\Gm 2
-\frac{1}{2}h_3\Gmcc 2 \right]\right.\nonumber\\
&&\hspace{1.0in}+\sum^{\infty}_{l=3}\left.\frac{\sg^{l-2}}{l}
\left[h_{l-1}^\ast \Gm l-\sg h_{l+1} \Gmcc l\right]
\rule{0in}{0.3in}\right\}\nonumber
\eeq
\beq
\lefteqn{\left[
\{2p-L_2\}h_2+(p+p^\ast)\sigma\frac{\partial h_2}{\partial
{\sigma}}\right]=}\hspace{1.0in}\label{eq:hfc2f}\\
&&\frac{i}{k_c}\,\frac{\partial}{\partial v}
\left\{\rule{0in}{0.3in}\psi_c+
\sg\left[h_1+\psi_c\Gm 1 -h_3
+\frac{1}{2} h_0 \Gm 2  +\frac{1}{3} \psi_c^\ast \Gm 3 \right]\right.
\nonumber\\
&&\hspace{1.25in}
\sg^2\left[h_1\Gm 1 - h_3 \Gmcc 1
-\frac{1}{2} h_4 \Gmcc 2 +\frac{1}{3} h_1^\ast \Gm 3 \right]\nonumber\\
&&\hspace{1.0in}\left.
-\frac{\sg^3}{3} h_5 \Gmcc 3 +
\sum^{\infty}_{l=4}\frac{\sg^{l-2}}{l}\left[h_{l-2}^\ast \Gm l-\sg^2 h_{l+2}
\Gmcc l\right]\rule{0in}{0.3in}\right\}\nonumber
\eeq
and
\beq
\lefteqn{\left[
\{kp-L_k\}h_k+(p+p^\ast)\sigma\frac{\partial h_k}{\partial
{\sigma}}\right]=}\hspace{1.0in}\label{eq:hfckf}\\
&&\frac{i}{k_c}\,\frac{\partial}{\partial v}
\left\{\rule{0in}{0.3in}h_{k-1}+\frac{\psi_c}{k-1}\Gm {k-1}
+\sum^{k-2}_{l=2}\, \frac{h_{k-l}}{l}\Gm l\right.
\nonumber\\
&&\hspace{0.75in}+\sg\left[h_{k-1}\Gm 1-h_{k+1}
+\frac{h_1}{k-1}\Gm {k-1}
+\frac{h_0}{k}\Gm k+\frac{\psi_c^\ast}{k+1}\Gm {k+1}\right]
\nonumber\\
&&\hspace{0.75in}+\sg^2\left[-h_{k+1}\Gmcc 1
+\frac{h_1^\ast}{k+1}\Gm {k+1}\right]
+\sum^{\infty}_{l=k+2}\frac{\sg^{l-k}}{l}h_{l-k}^\ast
\Gm l\nonumber\\
&&\hspace{1.5in}\left.-\sum^{\infty}_{l=2}\frac{\sg^{l}}{l}h_{k+l}
\Gmcc l\rule{0in}{0.3in}\right\}\nonumber
\eeq
for $k=0, 1, 2$, and $k>2$, respectively.

Together with (\ref{eq:ph}) these component equations determine the functions
$p(\sigma,\mu)$ and $\{h_k(\sigma,\mu)\}_{k=0}^{\infty}$. From a practical
point of view, all that has been achieved to this point is a reduction of the
problem to the analysis of functions of a {\em single} real variable, i.e.
$\sigma$.  However, in the study of the amplitude equation (\ref{eq:aeqn}),
this reduction does provide a useful simplification which is exploited in the
discussion below.

\section{Expansions, recursion relations, and pinching singularities}

The amplitude equation on the unstable manifold,
\be
\dot{A}=A\;p(\sigma,\mu),\label{eq:modeeqn}
\ee
is analyzed by expressing $p$ as a power series in the mode amplitude,
\be
p(\sg,\mu)=\sum^{\infty}_{j=0}\,p_j(\mu)\,\sg^j,\label{eq:pseries}
\ee
whose coefficients $p_j(\mu)$ are calculated from (\ref{eq:ph}) using an
analogous series for $h_k$
\beq
h_k(v,\sg)=\sum^{\infty}_{j=0}\,h_{k,j}(v)\,\sg^j.\label{eq:hseries}
\eeq
For notation, denote the integral over $h_{k,j}(v)$ by
\be
\gm kj\equiv\int^\infty_{-\infty}\,dv\,h_{k,j}(v)\label{eq:G}
\ee
so that
\be
\Gm k(\sg)=\sum^{\infty}_{j=0}\,\gm kj\,\sg^j\label{eq:Gseries}
\ee
from (\ref{eq:Gdef}), then expanding the expression for $p$ in (\ref{eq:ph})
gives the coefficients $p_j(\mu)$ as
\beq
p_0&=&\lambda\label{eq:p0}\\
p_j&=&\frac{i}{k_c}[{\cal A}_{j}+{\cal B}_{j}]\;\;\;
\mbox{for}\;\;j\geq 1\label{eq:pj}
\eeq
where
\beq
{\cal A}_j&=&-<\partial_v\tilde{\psi_c},(h_{0,j-1}-h_{2,j-1})>
-\frac{1}{2}<\partial_v\tilde{\psi_c}, \psi_c^\ast>\gm 2{j-1}\label{eq:ap}\\
&&\hspace{0.2in}
-\sum^{j-2}_{l=0}\left(<\partial_v\tilde{\psi_c}, h_{0,j-l-2}>\gm 1l
-<\partial_v\tilde{\psi_c}, h_{2,j-l-2}>\gmcc {1}{l}\right)\nonumber\\
{\cal B}_j&=&-\sum^{j-2}_{l=0}
\left\{\frac{1}{2}<\partial_v\tilde{\psi_c}, h_{1,j-l-2}^\ast> \gm 2l
+ \sum^{l}_{m=0}\left[\frac{\gm {j-l+1}{m}}{j-l+1}
<\partial_v\tilde{\psi_c}, h_{j-l,l-m}^\ast> \right.\right.\label{eq:bp}\\
&&\left.\left.\hspace{3.0in}
-\frac{\gmcc {j-l}{m}}{j-l}
<\partial_v\tilde{\psi_c},h_{j-l+1,l-m}>\right]\right\}.\nonumber
\eeq
Here and below, a summation is understood to be omitted if the lower limit
exceeds the upper limit. The organization of terms between (\ref{eq:ap}) and
(\ref{eq:bp}) will turn out to distinguish different singular behaviors in the
$\gamma\rightarrow0^+$ limit:
${\cal A}_j\sim\gamma^{-(4j-1)}$ and ${\cal B}_j\sim\gamma^{-(4j-2)};$ thus the
${\cal B}_j$ terms are sub-dominant in the weak growth rate regime.

The leading term in (\ref{eq:pseries}) is the eigenvalue $p_0=\lambda$,
and higher order terms
are determined by calculating $h_{k,j}(v)$ from (\ref{eq:hfc0f}) -
(\ref{eq:hfckf}); the results are summarized below. For $k\neq0$ the
coefficients $h_{k,j}(v)$ are expressed in terms of the resolvent operator
$R_k(w)\equiv(w-L_k)^{-1}$. For an arbitrary complex number $w$, the resolvent
acts on a function $g(v)$ by\cite{cra2}
\be
(R_k(w)\,g)(v)=\frac{1}{ikk_c(v-iw/kk_c)}
\left[g(v)-\frac{\eta(v,\mu)}{k^2\,\Lambda_{k}(iw/kk_c)}
\int^{\infty}_{-\infty} \,dv'\,\frac{g(v')}{v'-iw/kk_c}\right].\label{eq:resol}
\ee

\subsection{Series coefficients for $h_k(v,\sg)$}

For $k=0$ the coefficients $h_{0,l}$ are found from (\ref{eq:hfc0f}); inserting
the expansions for $h_k(v,\sigma)$  and $\Gm k(\sg)$ into (\ref{eq:hfc0f}) and
setting the coefficient of $\sigma^l$ to zero yields
\be
h_{0,l}(v)=\frac{I_{0,l}(v)}{(1+l)(\lambda+\lambda^\ast)}.\label{eq:h0lfcn}
\ee
The functions $I_{0,l}(v)$ are given by
\beq
I_{0,0}(v)&=& \frac{i}{k_c}\,\frac{\partial }{\partial v}(\psi_c^\ast-\psi_c)
\label{eq:I00}
\eeq
and for $l\geq1$
\beq
I_{0,l}(v)&=& -\sum^{l-1}_{j=0}\,(1+j)(p_{l-j}+p_{l-j}^\ast)\,h_{0,j}(v)\\
&&+\frac{i}{k_c}\,\frac{\partial}{\partial v}
\left\{\left[h_{1,l-1}^\ast -
\psi_c \gmcc {1}{l-1} +\sum^{l-2}_{j=0}\,h_{1,j}^\ast\gm {1}{l-j-2}
\right.\right.\nonumber\\
&&\hspace{1.0in}\left.\left.
+\sum^{l-1}_{j=0}\,\sum^{j}_{j'=0}\,\frac{h_{l-j+1,j'}^\ast}{l-j+1}
\gm {l-j+1}{j-j'}\right]-cc\right\}.\nonumber
\eeq

For $k\geq1$, following the same procedure in (\ref{eq:hfc1f}) -
(\ref{eq:hfckf}) determines the corresponding expressions for $h_{k,l}$; these
coefficients have the form
\be
h_{k,l}(v)=R_k(w_{k,l})\,I_{k,l}\label{eq:coeffhkl}
\ee
where $R_k(w)$ is given in (\ref{eq:resol}) and
\be
w_{k,l}\equiv(l+\delta_{k,1})(\lambda+\lambda^\ast)+k\lambda=
2(l+\delta_{k,1})\,\gamma+k\lambda.\label{eq:wkl}
\ee
For $k=1$, the functions $I_{1,l}(v)$ are given by
\beq
\lefteqn{I_{1,l}(v)=
-\sum^{l-1}_{j=0}\left[(2+j)p_{l-j}+(1+j)p_{l-j}^\ast\right]h_{1,j}}
\hspace{0.5in}\label{eq:I1l}
\\
&&+ \frac{i}{k_c}\,{\cal P}_\perp\frac{\partial}{\partial v}
\left\{\rule{0in}{0.3in}
h_{0,l}-h_{2,l}+ \frac{1}{2}\psi_c^\ast\gm 2l\right.\nonumber\\
&&\hspace{1.0in}
+\sum^{l-1}_{j=0}\left[
h_{0,j}\gm {1}{l-j-1}-
h_{2,j}\gmcc {1}{l-j-1}+\frac{1}{2}h_{1,j}^\ast\gm {2}{l-j-1}\right.
\nonumber\\
&&\hspace{1.25in}\left.-\frac{1}{2}h_{3,j}\gmcc {2}{l-j-1}
+\sum^{j}_{m=0}
\left(\frac{h_{l-j+1,m}^\ast}{l-j+2} \gm {l-j+2}{j-m}\right)\right]\nonumber\\
&&\hspace{1.0in}\left.
-\sum^{l-2}_{j=0}\sum^{j}_{m=0}
\left[\frac{h_{l-j+2,m}}{l-j+1} \gmcc {l-j+1}{j-m}\right]
\rule{0in}{0.3in}\right\}.\nonumber
\eeq
For $k=2$, from (\ref{eq:hfc2f}), the functions $I_{2,l}(v)$ are given by
\beq
I_{2,0}(v)&=& \frac{i}{k_c}\,\frac{\partial}{\partial v} \psi_c\label{eq:I20}
\eeq
for $l=0$, and for $l\geq1$
\beq
\lefteqn{I_{2,l}(v)=-\sum^{l-1}_{j=0}
\left[(2+j)p_{l-j}+jp_{l-j}^\ast\right]\,h_{2,j}}\hspace{0.5in}
\label{eq:I2l}\\
&&+\frac{i}{k_c}\,\frac{\partial}{\partial v}\left\{
h_{1,l-1}+\psi_c\gm {1}{l-1} -h_{3,l-1}
+\frac{1}{3} \psi_c^\ast \gm {3}{l-1}+ \frac{1}{2} \sum^{l-1}_{j=0} h_{0,j}
\gm {2}{l-j-1} \right.
\nonumber\\
&&\hspace{0.5in}+\sum^{l-2}_{j=0}\left[h_{1,j}\gm {1}{l-j-2} - h_{3,j}\gmcc
{1}{l-j-2}
-\frac{1}{2} h_{4,j} \gmcc {2}{l-j-2} +\frac{1}{3} h_{1,j}^\ast \gm
{3}{l-j-2}\right.
\nonumber\\
&&\hspace{2.5in}
+\sum^{j}_{m=0}\left.\frac{h_{l-j,m}^\ast}{l-j+2}
\gm {l-j+2}{j-m}\right]
\nonumber\\
&&\hspace{0.5in}\left.
-\sum^{l-3}_{j=0}
\frac{h_{5,j} }{3} \gmcc {3}{l-j-3}
-\sum^{l-4}_{j=0}\sum^{j}_{m=0}\frac{h_{l-j+2,m}}{l-j}
\gmcc {l-j}{j-m}\right\}.\nonumber
\eeq
For $k>2$, the functions $I_{k,l}(v)$ are given by
\beq
I_{k,l}(v)&=&-\sum^{l-1}_{j=0}\,[(k+j)p_{l-j}+j\,p_{l-j}^\ast]\,h_{k,j}(v)
\label{eq:Ikl}\\
&&\hspace{0.25in}+\frac{i}{k_c}\,\frac{\partial}{\partial v}
\left\{
h_{k-1,l}+\frac{\psi_c}{k-1}\gm {k-1}l
+\sum^{k-2}_{l'=2}\,\sum^{l}_{j=0}\, \frac{h_{k-l',j}}{l'}\gm {l'}{l-j}
\right.\nonumber\\
&&\hspace{1.0in}-h_{k+1,l-1}+\frac{\psi_c^\ast}{k+1}\gm {k+1}{l-1}.\nonumber\\
&&\hspace{1.0in}
+\sum^{l-1}_{j=0}\,\left[{h_{k-1,j}}\gm {1}{l-j-1}+\frac{h_{1,j}}{k-1}\gm
{k-1}{l-j-1}+\frac{h_{0,j}}{k}\gm {k}{l-j-1}\right]
\nonumber\\
&&\hspace{1.0in}+
\sum^{l-2}_{j=0}\,\left[-h_{k+1,j}\gmcc {1}{l-j-2}+\frac{h^\ast_{1,j}}{k+1}\gm
{k+1}{l-j-2}\right.\nonumber\\
&&\hspace{1.0in}\left.\left.
+\sum^{j}_{m=0}\,\left(\frac{h^\ast_{l-j,m}}{l+k-j}\gm {l+k-j}{j-m}
-\frac{h_{k+l-j,m}}{l-j}\gmcc {l-j}{j-m}\right)\right]\right\};
\nonumber
\eeq
in this last expression, if a subscript is negative the term is understood to
be omitted, e.g. for $l=0$, $h_{k+1,l-1}$ is omitted.

The arguments $w_{k,l}$ to the resolvent (\ref{eq:coeffhkl})
determine poles of $h_{k,l}(v)$ located at $v=z_{k,l}$ where
$z_{k,l}\equiv{iw_{k,l}}/{kk_c}$. For $k\geq1$, these poles always fall in the
upper half plane above the phase velocity $v_p$:
\be
z_{k,l}= z_0+\frac{i\gamma d_{k,l}}{k_c}=v_p
+\frac{i\gamma(1+d_{k,l})}{k_c} \label{eq:respole}
\ee
where $d_{k,l}\equiv 2(l+\delta_{k,1})/k$.

The relations (\ref{eq:pj}) and (\ref{eq:h0lfcn}) - (\ref{eq:Ikl}) can be
solved systematically to calculate $p_j$ and $h_{k,l}$ to any order. The
leading coefficient $p_0$ is determined by linear
theory  (\ref{eq:p0}) and from the linear eigenfunction $\psi_c$ one can also
calculate $h_{0,0}$ and $h_{2,0}$, c.f. (\ref{eq:I00}) and (\ref{eq:I20}),
respectively. These two coefficients then suffice to calculate $p_1$ from
(\ref{eq:pj}).
{}From $\{p_1, h_{0,0}, h_{2,0}\}$, the coefficients $h_{1,0}$ and $h_{3,0}$
can be determined and then $h_{0,1}$ and $h_{2,1}$. This provides the input
to calculate $p_2$, and from this point on the structure of the calculation to
all orders falls into a recognizable pattern that is summarized in Table I.

The expressions (\ref{eq:h0lfcn}) and (\ref{eq:coeffhkl}) for $h_{k,l}$ allow a
useful evaluation of $\gm kl$ in terms of $I_{k,l}$. For $k=0$ the coefficient
vanishes identically $\gm 0l=0,$
and for $k>0$, from (\ref{eq:G}) and (\ref{eq:coeffhkl}), one has
\beq
\gm kl&=&\int^\infty_{-\infty}\,dv\,R_k(w_{k,l})\,I_{k,l}\nonumber\\
&=&\frac{1}{ikk_c}\left[1-\frac{1}{k^2\Lambda_{k}(z_{k,l})}
\int^\infty_{-\infty}\,dv\,\frac{\eta(v,\mu)}{v-z_{k,l}}\right]
\int^\infty_{-\infty}\,dv'\,\frac{I_{k,l}(v')}{v'-z_{k,l}}\nonumber\\
&=&\frac{1}{ikk_c}\left[\frac{k^2\Lambda_{k}(z_{k,l})-
(\Lambda_{1}(z_{k,l})-1)}{k^2\Lambda_{k}(z_{k,l})}\right]
\int^\infty_{-\infty}\,dv'\,\frac{I_{k,l}(v')}{v'-z_{k,l}}\nonumber\\
&=&-\left[\frac{ik/k_c}{k^2-1+\Lambda_{1}(z_{k,l})}\right]
\int^\infty_{-\infty}\,dv'\,\frac{I_{k,l}(v')}{v'-z_{k,l}}\label{eq:gmkj}
\eeq
where in the last step the identity
$k^2\Lambda_{k}(z_{k,l})=k^2-1+\Lambda_{1}(z_{k,l})$ has been used. The final
expression in (\ref{eq:gmkj}) will prove helpful in analyzing the form of $\gm
kl$ as $\glim$.

\subsection{Analysis of the cubic coefficient}

It is instructive at this point to evaluate and examine the cubic coefficient
$p_1(\mu)$ in detail. From (\ref{eq:pj}) we have
\beq
p_1&=&-\frac{i}{k_c}\left[<\partial_v\tilde{\psi_c},(h_{0,0}-h_{2,0})>
+\frac{1}{2}<\partial_v\tilde{\psi_c}, \psi_c^\ast>\gm
2{0}\right],\label{eq:p1coeff}
\eeq
so that $h_{0,0}$ and $h_{2,0}$ must be found from (\ref{eq:h0lfcn}) and
(\ref{eq:coeffhkl}). A straightforward calculation yields
\beq
h_{0,0}(v)&=&\frac{1}{k_c^2}\frac{\partial}{\partial v}
\left[\frac{-\eta(v,\mu)}{(v-z_0)\,(v-z_0^\ast)}\right]\label{eq:h00}\\
h_{2,0}(v)&=&\frac{1}{2k_c^2}\left[\frac{\partial_v\psi_c}{(v-z_0)} +
\frac{\Lambda^{(2)}_{1}(z_0)\,\eta(v,\mu)}{6(v-z_0)}\right]\label{eq:h20}
\eeq
where $\Lambda^{(2)}_{1}(z_0)$ is the second derivative defined in
(\ref{eq:specfcnder}).
Integrating these functions over velocity yields $\gm00=0$ and
\be
\gm20=\frac{-\Lambda^{(2)}_{1}(z_0)}{3\,k_c^2}.
\ee
The remaining integrals in (\ref{eq:p1coeff}) are expressible in terms of the
derivatives of $\Lambda_{1}(z_0)$:
\beq
<\partial_v\tilde{\psi_c},
\psi_c^\ast>&=&\frac{ik_c}{2\gamma}\label{eq:cubint1}\\
<\partial_v\tilde{\psi_c}, h_{0,0}>&=&-\frac{ik_c}{4\gamma^3}
\left[1-\frac{i\gamma\Lambda^{(2)}_{1}(z_0)}{k_c\Lambda'_{1}(z_0)}
-\frac{2\gamma^2\Lambda^{(3)}_{1}(z_0)}{3k_c^2\Lambda'_{1}(z_0)}\right]
\label{eq:cubint2}\\
<\partial_v\tilde{\psi_c}, h_{2,0}>&=&
\frac{2(\Lambda^{(2)}_{1}(z_0))^2-3\Lambda^{(4)}_{1}(z_0)}
{48k_c^2\Lambda'_{1}(z_0)}.\label{eq:cubint3}
\eeq
Thus we find from (\ref{eq:p1coeff})
\beq
p_1&=&-\frac{1}{4\gamma^3}\left[\rule{0in}{0.3in}1-
\frac{i\gamma\Lambda^{(2)}_{1}(z_0)}{k_c\Lambda'_{1}(z_0)}-
\frac{\gamma^2}{3k_c^2}
\left(\frac{2\Lambda^{(3)}_{1}(z_0)-
\Lambda'_{1}(z_0)\Lambda^{(2)}_{1}(z_0)}{\Lambda'_{1}(z_0)}\right)
\right.\nonumber\\
&&\hspace{0.75in}\left.+\frac{i\gamma^3}{12k_c^3\,\Lambda'_{1}(z_0)}
\left(3\Lambda^{(4)}_{1}(z_0)-{2}(\Lambda^{(2)}_{1}(z_0))^2\right)
\rule{0in}{0.3in}\right]
\eeq
which simplifies to
\be
p_1=\frac{1}{\gamma^3}\left[-\frac{1}{4}+\ord\gamma\right]\label{eq:p1sing}
\ee
as $\gamma\rightarrow0^+$.

The cubic coefficient is obviously singular in the weak growth rate regime and
the strength of the singularity is set by the contribution from $h_{0,0}(v)$
which is the nonlinear correction to the equilibrium $F_0$ at this
order.\cite{endnote4} The following summary of the calculation of
${<\partial_v\tilde{\psi_c}, h_{0,0}>}$ in (\ref{eq:cubint2}) pinpoints the
origin of the singularity. From (\ref{eq:h00}) and the definition of
$\tilde{\psi_c}$, one has
\beq
<\partial_v\tilde{\psi_c}, h_{0,0}>&=&
\int_{-\infty}^{\infty}\,dv\,\frac{\partial\tilde{\psi_c}}{\partial
v}^\ast(v)h_{0,0}(v) \nonumber\\
&=&\frac{1}{k_c^2\Lambda'_{1}(z_0)}
\int_{-\infty}^{\infty}\,dv\,\frac{1}{(v-z_0)^2}\frac{\partial}{\partial v}
\left[\frac{-\eta(v,\mu)}{(v-z_0)\,(v-z_0^\ast)}\right]\nonumber\\
&=&-\frac{2}{k_c^2\Lambda'_{1}(z_0)}
\int_{-\infty}^{\infty}\,dv\,\frac{\eta(v,\mu)}{(v-z_0)^4\,(v-z_0^\ast)}.
\label{eq:calc1}
\eeq
Since $z_0=v_p+{i\gamma}/{k_c}$ in (\ref{eq:calc1}), there is clearly a
pinching singularity at $v_p$ when $\glim$. By expanding the integrand in
partial fractions,
\beq
\frac{1}{(v-z_0)^4\,(v-z_0^\ast)}&=&\frac{(k_c/2i\gamma)}{(v-z_0)^4}
-\frac{(k_c/2i\gamma)^2}{(v-z_0)^3}+\frac{(k_c/2i\gamma)^3}{(v-z_0)^2}\\
&&\hspace{0.35in}-\left(\frac{k_c}{2i\gamma}\right)^4
\left[\frac{1}{v-z_0}-\frac{1}{v-z_0^\ast}\right],\nonumber
\eeq
(\ref{eq:calc1}) becomes
\beq
<\partial_v\tilde{\psi_c}, h_{0,0}>&=&-\frac{2}{k_c^2\Lambda'_{1}(z_0)}
\left\{-\left(\frac{k_c}{2i\gamma}\right)^4
\left[\rule{0in}{0.2in}(\Lambda_{1}(z_0)-1)-(\Lambda_{1}(z_0^\ast)-1)\right]
\right.\label{eq:exph00}\\
&&\hspace{1.0in}
+\left(\frac{k_c}{2i\gamma}\right)^3\Lambda'_{1}(z_0)
-\left(\frac{k_c}{2i\gamma}\right)^2\frac{\Lambda^{(2)}_{1}(z_0)}{2}\nonumber\\
&& \hspace{1.5in}
\left.+\left(\frac{k_c}{2i\gamma}\right)\frac{\Lambda^{(3)}_{1}(z_0)}{ 6}
\right\}.\nonumber
\eeq
Since $\Lambda_{1}(z_0)=0$ and $\Lambda_{1}(z_0^\ast)=0$, the $\gamma^{-4}$
terms vanish {\em identically} and do not contribute
to the $\glim$ limit; thus (\ref{eq:exph00}) reduces to (\ref{eq:cubint2}) and
a $\gamma^{-3}$ singularity.

The interpretation of the $\gamma^{-3}$ singularity is suggested by examining
the form of the mode equation (\ref{eq:modeeqn}), truncated at cubic order, in
the $\glim$ limit:
\be
\dot{A}=A\left[\lambda-\frac{1}{4\gamma^3}
\left(1-\frac{i\gamma\Lambda^{(2)}_{1}(z_0)}{k_c\Lambda'_{1}(z_0)}+
\ord{\gamma^2}\right)|A|^2+\cdots\right]
\label{eq:cubictrun}
\ee
where $\lambda=\gamma-i\omega$; in terms of amplitude and phase variables
$A=\rho e^{-i\theta}$ this reads
\beq
\dot{\rho}&=&\rho\left[\gamma-\frac{1}{4\gamma^3}\rho^2+\ord{\rho^4}\right]\\
\dot{\theta}&=&\omega-\mbox{\rm Re}
\left(\frac{\Lambda^{(2)}_{1}(z_0)}{\Lambda'_{1}(z_0)}\right)
\frac{\rho^2}{4k_c\gamma^2}+\ord{\rho^4}.
\eeq
The manifest singularities at $\gamma=0$ can be removed by introducing a
rescaled amplitude variable:
\be
\rho(t)\equiv\gamma^2\;r(\gamma t)\label{eq:trapping}
\ee
which evolves on the time scale $\tau\equiv\gamma t$. In terms of $r(\tau)$,
the mode equation becomes
\beq
\frac{dr}{d\tau}&=&
r\left[1-\frac{1}{4}r^2+\ord{\gamma^8r^4}\right]\label{eq:rtrun}\\
\frac{d\theta}{d t}{}&=&\omega-\frac{\gamma^2}{4k_c}\mbox{\rm Re}
\left(\frac{\Lambda^{(2)}_{1}(z_0)}{\Lambda'_{1}(z_0)}\right) {r^2}
+\ord{\gamma^8r^4}.\label{eq:thetatrun}
\eeq
For the terms shown explicitly in (\ref{eq:rtrun}) - (\ref{eq:thetatrun}), the
$\glim$ limit is now well behaved; this  suggests that the effect of the
singularities in the coefficients is to produce the
scaling $\rho\sim\gamma^2$ in the dynamics of the mode amplitude. Our ansatz in
(\ref{eq:trapping}) bears some resemblence to the simpler case of Hopf
bifurcation discussed in the Introduction but the scaling is quantitatively
different indicating a much stronger nonlinear effect and a smaller nonlinearly
evolving mode.

It is also important to note that to this order the result (\ref{eq:trapping})
has a remarkable degree of universality. The $\gamma^{-3}$ singularity in
(\ref{eq:p1sing}) sets the scaling in (\ref{eq:trapping}) and the coefficient,
$-1/4$, of this singularity is completely independent of the underlying
equilibrium $F_0(v,\mu_c)$. Thus, for example, the singularity is the same for
a beam-plasma instablility (complex $\lambda$) as for a two-stream instability
(real $\lambda$). Indeed to this order, the rescaled amplitude equation for
$dr/d\tau$ is also completely independent of $F_0$ as $\glim$.

These remarks tacitly assume that the neglected higher order terms in
(\ref{eq:cubictrun}) do not alter conclusions reached for the truncated
equations. The remainder of the paper is devoted to a systematic analysis of
the higher order terms in the series for $p$ and $h_{k,l}$. It will turn out
that the coefficients
$p_j$ for $j\geq2$ are also singular and that the naive estimate
$\ord{\gamma^8}$ in (\ref{eq:rtrun}) and (\ref{eq:thetatrun}) is not correct.
In fact the terms that are higher order in $r$ are {\em not} higher order in
$\gamma$, rather they appear to be as important for the dynamics of $A(t)$ as
the cubic term $A|A|^2$. Nevertheless the singularities to {\em all} orders are
absorbed by the scaling in (\ref{eq:trapping}). The dependence of the higher
order terms on $F_0(v,\mu_c)$ as $\glim$ will be discussed later.

\section{Singularity structure of the expansion}

The detailed calculation of $p_j$ rapidly becomes prohibitively
laborious, but the recursion relations determining the higher order
coefficients in terms of lower order quantities (cf. Table I) can be analyzed
to obtain useful information. Most basic is the question: for $j>1$, how
singular is $p_j$ as $\glim$ ? This issue requires an accurate estimate of the
build up of pinching singularities in the integrals found in $p_j$. For this
purpose I introduce an ``index'' which allows the divergence of a given
integral to be assessed by a simple counting procedure.

\subsection{Definition of the index}

For $n>0$ define
\be
D_n(\alpha,v)\equiv\frac{1}{(v-\alpha_1)(v-\alpha_2)\cdots(v-\alpha_n)}
\label{eq:Ddef}
\ee
where  $\alpha\equiv(\alpha_1,\ldots,\alpha_n)$ and define
$D_0(\alpha,v)\equiv1$. Evaluating $p_j$ for $j\geq1$ involves integrands of
the form
\be
{\cal G}(v,\mu)=D_m(\beta,v)^\ast\,D_n(\alpha,v)\,\frac{\partial^l\eta(v,\mu)}
{\partial v^l}\label{eq:indatom}
\ee
with $m+n\geq1$. The poles in (\ref{eq:indatom}) are given by
\beq
\alpha_j&=&z_0+i\gamma\nu_j/k_c\hspace{0.5in}j=1,\ldots,n\label{eq:poles1}\\
\beta^\ast_j&=&
z_0^\ast-i\gamma\zeta_j/k_c\hspace{0.5in}j=1,\ldots,m;\label{eq:poles2}
\eeq
hence they lie along the vertical line Re $v=v_p$ at locations determined by
the numbers $\nu_j\geq0$ and $\zeta_j\geq0$ which are assumed to be independent
of $F_0$ for all $j$; in particular $\nu_j$ and $\zeta_j$ are independent of
$\gamma$. The function ${\cal G}(v,\mu)$ depends on $\mu$ through $z_0$,
$\gamma$, and $\eta(v,\mu)$.

The {\em index} of ${\cal G}(v,\mu)$ in (\ref{eq:indatom}) is defined to be
\be
\ind [{\cal G}]\equiv m+n+l-2.\label{eq:inddef}
\ee
Since we assume $m+n\geq1$, $\ind [{\cal G}]\geq-1$. If $mn\neq0$, then
as $\glim$, the integral of ${\cal G}$ diverges due to a pinching singularity
at the phase velocity $v_p$. When $\ind [{\cal G}]\geq0$, the index of ${\cal
G}$ gives the maximum possible strength of this divergence.

\begin{lemma} For ${\cal G}(v,\mu)$ in \mbox{\rm (\ref{eq:indatom})} with
$m+n\geq2$ and $J=\ind [{\cal G}]$, the integral of ${\cal G}$ satisfies
\be
\lim_{\glim}\; \left|\gamma^{J}\; \int^\infty_{-\infty}\,dv\,{\cal
G}(v,\mu)\right|<\infty.\label{eq:sing0}
\ee
If $J$ is replaced by $J-1$, then the limit \mbox{\rm (\ref{eq:sing0})}
diverges in general unless $mn=0$ in which case the limit is zero for any
$J>0$.
\end{lemma}
\noindent {\em {\bf Proof}.}
\begin{quote} Since
\beq
\lefteqn{\int^\infty_{-\infty}\,dv\,D_m(\beta,v)^\ast\,D_n(\alpha,v)\,
\frac{\partial^l\eta(v,\mu)} {\partial v^l}=}\\
&&\int^\infty_{-\infty}\,dv\, \frac{\partial^{l-1}\eta(v,\mu)} {\partial
v^{l-1}}\,\left\{\sum^{m}_{j=1}D_{m+1}(\beta,\beta_j,v)^\ast\,D_n(\alpha,v)+
\sum^{n}_{j=1}D_{m}(\beta,v)^\ast\,D_{n+1}(\alpha,\alpha_j,v)\right\},
\nonumber
\eeq
we can reduce to the case with $l=0$. If $mn=0$ then there is no pinching
singularity and the integral in (\ref{eq:sing0}) has a finite limit, cf.
(\ref{eq:specfcnder}); hence the limit is zero if $J>0$. Assume that $mn>0$,
then for integrals
$\int^\infty_{-\infty}\,dv\,D_m(\beta,v)^\ast\,D_n(\alpha,v)\,\eta(v,\mu)$ with
$m+n=2$ the integration can be easily done and (\ref{eq:sing0}) explicitly
verified. For integrals with  $m+n>2$, by expanding the integrand in partial
fractions, they
can be re-expressed in terms of integrals with $m+n-1$. A simple induction
argument then establishes (\ref{eq:sing0}).  In Appendix B the limit in
(\ref{eq:sing0}) for $l=0$ is shown to be non-zero in general so, barring an
accidental cancellation, if $J$ is replaced by $J-1$ then the modified limit
will diverge as $\gamma^{-1}$.
{\bf $\Box$}\end{quote}

The definition in (\ref{eq:inddef}) can be generalized in some obvious ways
without sacrificing its usefulness. Suppose $q(\mu)$ is a function of $\mu$
with the asymptotic behavior
\be
\lim_{\glim}\;q(\mu)\sim\gamma^{-\delta},
\ee
then we define the index of $q(\mu)\,{\cal G}(v,\mu)$ to be
\be
\ind [q\,{\cal G}]\equiv \ind [{\cal G}]+\delta.\label{eq:prod}
\ee
Estimates of the form (\ref{eq:sing0}) still hold for $q{\cal G}$ with
$J=\mbox{\rm max}\,\{\delta,\ind [q\,{\cal G}]\}$; the anomalous case
$J=\delta$ arises when $\ind [{\cal G}]=-1$ since the integral of ${\cal G}$ is
then nonsingular and the asymptotic behavior of $q\int dv\,{\cal G}$ is
determined by $q(\mu)$. Finally if ${\cal G}_1(v,\mu)$ and ${\cal G}_2(v,\mu)$
have indices satisfying $\ind [{\cal G}_1]\geq\ind [{\cal G}_2]$ then
we define the index of the sum to be the larger index:
\be
\ind [{\cal G}_1+{\cal G}_2]\equiv\ind [{\cal G}_1].\label{eq:sum}
\ee

\subsection{Examples}

As a simple example of the index, note that the functions $\psi_c$, $h_{0,0}$,
and $h_{2,0}$ have indices given by
\beq
\ind [\psi_c]&=&-1\label{eq:ex1}\\
\ind [h_{0,0}]&=&1\label{eq:ex2}\\
\ind [h_{2,0}]&=&1;\label{eq:ex3}
\eeq
similarly the integrands in (\ref{eq:cubint1}) - (\ref{eq:cubint3}) give
\beq
\ind [\partial_v\tilde{\psi_c}^\ast\psi_c^\ast]&=&1\label{eq:ex4}\\
\ind [\partial_v\tilde{\psi_c}^\ast h_{0,0}]&=&3\\\label{eq:ex5}
\ind [\partial_v\tilde{\psi_c}^\ast h_{2,0}]&=&3.\label{eq:ex6}
\eeq

{}From Lemma IV.1, this information immediately tells us that the singularity
of $p_1$ cannot be worse than $\gamma^{-3}$.

A further useful observation is the effect of various operators on the index.
\begin{lemma} For ${\cal G}(v,\mu)$ as in \mbox{\rm (\ref{eq:indatom})}, the
operators $\partial_v {\cal G}$, ${\cal P}_\perp {\cal G}$ and
$R_k(w_{k,l}){\cal G}$
either leave the index unchanged or else increase it by one:
\beq
\ind [\partial_v {\cal G}]&=&\ind [{\cal G}]+1\\
\ind [{\cal P}_\perp {\cal G}]&=&\ind [{\cal G}]\\
\ind [R_k(w_{k,l}){\cal G}]&=&\ind [{\cal G}]+1.\label{eq:resind}
\eeq
\end{lemma}
\noindent {\em {\bf Proof}.}\begin{quote}The first identity is obvious, and the
second follows immediately from the definition of ${\cal P}_\perp$ below
(\ref{eq:Pdef}). For the third identity, the resolvent (\ref{eq:resol}) gives
\be
R_k(w_{k,l})\,{\cal G}=\frac{1}{ikk_c(v-z_{k,l})}
\left[{\cal G}(v,\mu)-\frac{\eta(v,\mu)}{k^2\,\Lambda_{k}(z_{k,l})}
\int^{\infty}_{-\infty} \,dv'\,\frac{{\cal
G}(v',\mu)}{v'-z_{k,l}}\right].\label{eq:reseqn}
\ee
For the first term the index is increased by one,
$\ind [{\cal G}/(v-z_{k,l})]=\ind [{\cal G}]+1.$
In the second term,
$\ind [\eta(v,\mu)/(v-z_{k,l})]=-1,$
so the function multiplying the integral
\be
q(\mu)\equiv\frac{1}{k^2\Lambda_{k}(z_{k,l})} \int^{\infty}_{-\infty}
\,dv'\,\frac{{\cal G}(v',\mu)}{v'-z_{k,l}}
\ee
will determine the index and the singular behavior of the corresponding
velocity integral (recall the discussion below (\ref{eq:prod})). Since
\be
\int^{\infty}_{-\infty} \,dv'\,\frac{{\cal G}(v',\mu)}{v'-z_{k,l}}
\sim{\frac{1}{\gamma^{(1+\ind [{\cal G}])}}}
\ee
and
\be
k^2\Lambda_{k}(z_{k,l})=k^2-1+\Lambda_{1}(z_{k,l})=k^2-1+
\Lambda'_{1}(z_0)\left(\frac{i\gamma d_{k,l}}{k_c}\right)+
\ord{\gamma^2},
\ee
$q(\mu)$ has the asymptotic form
\be
\lim_{\glim}\left|\gamma^{J}\;q(\mu)\right|<\infty
\ee
with $J=\ind [{\cal G}]+1+\delta_{k,1}$.
Hence the index of the second term in (\ref{eq:reseqn}) is $\ind [{\cal G}]+1$
and $\ind [{\cal G}]$ for $k=1$ and $k\neq1$, respectively. In either case, the
overall index is given by (\ref{eq:resind}).

{\bf $\Box$}\end{quote}

\subsection{Estimate of singularities}

The main result on the singularity structure of the amplitude expansions for
$p(\sigma,\mu)$ and $h_{k,l}(v,\sigma)$ can now be proved.

\begin{theorem} For $k\geq0$ and $l\geq0$, the indices of $I_{k,l}$ and
$h_{k,l}$ are given by
\beq
\ind [I_{k,l}]&=&\ind [h_{k,l}]-1\label{eq:Iindex}\\
\ind [h_{k,l}]&=&2k+4l-3+4(\delta_{k,0}+\delta_{k,1}).\label{eq:hindex}
\eeq
The asymptotic behavior of the coefficients $\gm kl$ and $p_j$ satisfies
\beq
\lim_{\glim}\; \left|\gamma^{J}\;\gm kl\right|&<&\infty
\hspace{0.25in}\mbox{\em where}\;\;J=\ind [h_{k,l}]+\delta_{k,1}
\label{eq:gmklasy}\\
\lim_{\glim}\;
\left|\gamma^{4j-1}\;p_j(\mu)\right|&<&\infty\hspace{0.25in}\mbox{\em for}
\;\;j\geq1.\label{eq:pjasy}
\eeq
\end{theorem}
\noindent {\em {\bf Proof}.}\begin{quote}
\begin{enumerate}
\item Since $h_{0,l}=I_{0,l}/2\gamma(1+l)$ and $h_{k,l}=R_k(w_{k,l})I_{k,l}$,
it  follows immediately from (\ref{eq:prod}) and (\ref{eq:resind}) that the
indices of $I_{k,l}$ and $h_{k,l}$ differ by one as stated in (\ref{eq:Iindex})
- {\em provided} the index of $I_{k,l}$ is well defined. The induction argument
outlined below shows that each function $I_{k,l}$ is obtained via the recursion
relations as a sum of terms each having a well defined index. Thus $I_{k,l}$
will always have a well defined index.

\item  The estimate in (\ref{eq:gmklasy}) follows from the formula for $\gm kl$
in (\ref{eq:gmkj}):
\be
\gm kl=-\frac{(ik/k_c)}{k^2-1+\Lambda_{1}(z_{k,l})}
\int^\infty_{-\infty}\,dv\,\frac{I_{k,l}(v)}{v-z_{k,l}};
\ee
as $\glim$, the integral cannot diverge more strongly than
\be
\int^\infty_{-\infty}\,dv\,\frac{I_{k,l}(v)}{v-z_{k,l}}\sim
\left(\frac{1}{\gamma}\right)^{\ind [h_{k,l}]},
\ee
and the function multiplying the integral is nonsingular unless $k=1$ in which
case it is $\ord{\gamma^{-1}}$.

\item The index formula for $h_{k,l}$ (\ref{eq:hindex}) and the estimate for
$p_j(\mu)$ in (\ref{eq:pjasy}) are proved by induction using the recursion
relations. The induction argument is organized by the pattern shown in Table I.
At the top of Table I, the results in (\ref{eq:hindex}) and (\ref{eq:pjasy})
have been explicitly verified for $\{h_{0,0},h_{2,0}\}$ and $p_1$ in Section
III.B.

\item Moving downward in Table I, the next quantities to consider are $h_{1,0}$
and $h_{3,0}$ from (\ref{eq:I1l}) and (\ref{eq:Ikl}), respectively:
\beq
h_{1,0}&=&\frac{i}{k_c}R_1(w_{1,0})\,{\cal P}_\perp\frac{\partial}{\partial v}
\left\{h_{0,0}-h_{2,0}+ \frac{1}{2}\psi_c^\ast\gm 20\right\}\\
h_{3,0}&=&\frac{i}{k_c}R_3(w_{3,0})\,\frac{\partial}{\partial v}
\left\{ h_{2,0}+\frac{\psi_c}{2}\gm {2}0\right\}.
\eeq
Reading right to left in these expressions, the index for each of the
quantities in braces $\{\cdots\}$ is 1, and this index is increased to 2 by
$\partial_v$ and further increased to 3 by the resolvent; hence $\ind
[h_{1,0}]=3$ and $\ind [h_{3,0}]=3$
in agreement with (\ref{eq:hindex}). Next, according to Table I, are $h_{0,1}$
and $h_{2,1}$; from (\ref{eq:h0lfcn}) and (\ref{eq:I2l}) they are:
\beq
h_{0,1}&=&\frac{1}{4\gamma}\left[
-(p_{1}+p_{1}^\ast)\,h_{0,0}(v)\rule{0in}{0.25in}\right.\\
&&\hspace{0.25in}\left.
+\frac{i}{k_c}\,\frac{\partial}{\partial v}
\left\{\left[h_{1,0}^\ast -\psi_c \gmcc {1}{0} +\frac{h_{0,0}^\ast}{2}\gm
{2}{0}\right]-cc\right\}\right]\nonumber\\
&&\nonumber\\
h_{2,1}&=&R_2(w_{2,1})\left[-2p_{1}\,h_{2,0}\rule{0in}{0.25in}\right.\\
&&\hspace{0.25in}\left.
+\frac{i}{k_c}\,\frac{\partial}{\partial v}
\left\{h_{1,0}+\psi_c\gm {1}{0} -h_{3,0}
+\frac{1}{3} \psi_c^\ast \gm {3}{0}+ \frac{1}{2} h_{0,0} \gm {2}{0}\right\}
\right].\nonumber
\eeq
For $h_{0,1}$ the expression in braces $\{\cdots\}$ has index 3 and the larger
expression in brackets $[\cdots]$ has index 4, so $\ind [h_{0,1}]=5$ in
agreement with (\ref{eq:hindex}). Similarly the index of $h_{2,1}$ is seen to
have the correct value $\ind [h_{2,1}]=5$.

\item This provides enough information to
verify (\ref{eq:pjasy}) for $p_2$. From (\ref{eq:pj}) this coefficient is
$p_2={i}[{\cal A}_{2}+{\cal B}_{2}]/{k_c}$
where
\beq
{\cal A}_{2}&=&-\left[<\partial_v\tilde{\psi_c},(h_{0,1}-h_{2,1})>
+\frac{1}{2}<\partial_v\tilde{\psi_c}, \psi_c^\ast>\gm 2{1}\right.
\label{eq:a2}\\
&&\left.\hspace{0.25in}+<\partial_v\tilde{\psi_c}, h_{0,0}>\gm 10
-<\partial_v\tilde{\psi_c}, h_{2,0}>\right]\nonumber\\
{\cal B}_{2}&=&-\left[
\frac{1}{2}<\partial_v\tilde{\psi_c}, h_{1,0}^\ast> \gm 20
+\frac{\gm {3}{0}}{3}<\partial_v\tilde{\psi_c}, h_{2,0}^\ast>\right.\\
&&\left.\hspace{0.5in}-\frac{\gmcc {2}{0}}{2}
<\partial_v\tilde{\psi_c},h_{3,0}>\right].\nonumber
\eeq
Since multiplying by $\partial_v\tilde{\psi_c}^\ast$ increases the index
of an integrand by two, cf. (\ref{eq:ex1}) - (\ref{eq:ex6}), it is
straightforward to verify that $p_2$ satisfies the estimate in
(\ref{eq:pjasy}). For example, consider the first term in (\ref{eq:a2}). Since
$\ind [h_{0,1}]=\ind [h_{2,1}]=5$, the integrand has index
$\ind[\partial_v\tilde{\psi_c}^\ast(h_{0,1}-h_{2,1})]=7$ so the worst possible
singularity for the integral is
\be
<\partial_v\tilde{\psi_c},(h_{0,1}-h_{2,1})>\sim\frac{1}{\gamma^7}
\ee
which is consistent with (\ref{eq:pjasy}). A similar verification for the
remaining terms in $p_2$ is straightforward.

\item According to Table I, for fixed $N\geq1$, given the functions
\be
\{h_{k,l}(v)\;|\;k=0,\ldots, N+1\;\;\mbox{\rm and}\;\;l=0,\ldots,l_{max}(N,k)\}
\label{eq:assumpt1}
\ee
where $l_{max}(N,k)\equiv N-k+1-2(\delta_{k,0}+\delta_{k,1})$ and the lower
order coefficients
\be
\{p_j\;|\;j=1,\ldots, N\},\label{eq:assumpt2}
\ee
then one can calculate the additional functions
\be
\{h_{k,l}(v)\;|\;k=0,\ldots, N+2\;\;\mbox{\rm for}\;\;l=l_{max}(N+1,k)\}
\label{eq:iterate1}
\ee
and the next coefficient $p_{N+1}$. For $N=2$, it has been shown that the
functions (\ref{eq:assumpt1}) and the coefficients (\ref{eq:assumpt2}) satisfy
(\ref{eq:hindex}) and (\ref{eq:pjasy}), respectively.

\item The assumption for the induction step is now clear.
For arbitrary $N\geq2$, assume that the functions (\ref{eq:assumpt1})
satisfy (\ref{eq:hindex}) and also
that the coefficients (\ref{eq:assumpt2}) satisfy the estimate in
(\ref{eq:pjasy}). Then from this hypothesis, the recursion relations
determining the functions in (\ref{eq:iterate1}) and for $p_{N+1}$ imply that
these new quantities will also satisfy (\ref{eq:hindex}) and (\ref{eq:pjasy}),
respectively. Verifying this conclusion involves only a routine evaluation of
the indices arising from the recursion relations as was done above for $N=2$.
This completes the proof of (\ref{eq:hindex}) and (\ref{eq:pjasy}).
\end{enumerate}
{\bf $\Box$}\end{quote}

\subsection{Implications of Theorem IV.1}

The index formula for $h_{k,l}(v)$ and the asymptotic estimate on $\gm kl$
imply that in the the general expression for $p_j$ (\ref{eq:pj}) the dominant
terms are in ${\cal A}_j$. More precisely, one finds ${\cal
A}_j\sim\gamma^{-(4j-1)}$ and ${\cal B}_j\sim\gamma^{-(4j-2)}$ as $\glim$.
The estimates in (\ref{eq:gmklasy}) and (\ref{eq:pjasy}) motivate the following
asymptotic definitions for $\gm kl$ and $p_j$:
\beq
c_{k,l}(\mu_c)&\equiv&\lim_{\glim}\left[\gamma^{(2k+4l-3+5\delta_{k,1})}\,\gm
kl\right]
\hspace{0.25in}\mbox{\rm for} \;\;k\geq 1\;\;
\mbox{\rm and} \;\;l\geq0\label{eq:ckldef}\\
b_j(\mu_c)&\equiv&\lim_{\glim}\left[\gamma^{4j-1}\,p_j(\mu)\right]
=\lim_{\glim}\left[\gamma^{4j-1}\,{\cal A}_j(\mu)\right]
\hspace{0.25in}\mbox{\rm for} \;\;j\geq1.\label{eq:bjdef}
\eeq
In this notation, the asymptotic form of $\Gm k$ for $k\geq1$ is
\be
\Gm k=\left(\frac{1}{\gamma}\right)^{2k-3+5\delta_{k,1}}
\left[\Gamma^c_k+\ord\gamma\right]
\ee
where
\be
\Gamma^c_k(\mu_c)\equiv\sum_{l=0}^{\infty}\,c_{k,l}(\mu_c)r^{2l}.
\label{eq:gmkcrit}
\ee

\subsubsection{Mode amplitude equation}

The full mode amplitude equation,
\be
\frac{dA}{dt}=\lambda A+A\sum_{j=1}^{\infty}p_j(\mu)\sigma^j
\ee
when written in polar variables $A=\rho\,e^{-i\theta}$ gives
\beq
\frac{d\rho}{dt}&=&\rho\gamma+\rho\sum^\infty_{j=1}\,
[\mbox{\rm Re}\;p_j(\mu)]\,\rho^{2j}\label{eq:rhodot}\\
\frac{d\theta}{dt}&=&
\omega-\sum^\infty_{j=1}[\mbox{\rm Im}\;p_j(\mu)]\,\,\rho^{2j}.
\label{eq:thetadot}
\eeq
The $\glim$ limit can be taken using the asymptotic behavior for $p_j$
(\ref{eq:bjdef}) given by Theorem IV.1 and the rescaling
$\rho(t)=\gamma^2\;r(\gamma t)$  suggested by the low order analysis
(\ref{eq:trapping}):
\beq
\frac{dr}{d\tau}&=&r\left\{1+\sum^\infty_{j=1}\,
\left[\mbox{\rm Re}\;b_j(\mu_c)+{\cal{O}}(\gamma)\right]\,r^{2j}\right\}
\label{rescaled1}\\
\frac{d\theta}{dt}&=&\omega-
\gamma\,\sum^\infty_{j=1}\,\left[\mbox{\rm
Im}\;b_j(\mu_c)+{\cal{O}}(\gamma)\right]\,r^{2j}
\label{rescaled2}
\eeq
where $\omega=k_cv_p$ is the linear mode frequency. These equations are
non-singular at every order as $\glim$, and the final asymptotic equations are
\beq
\frac{dr}{d\tau}&=& rR(r,\mu_c)\label{eq:g=0}
\label{asym1}\\
\frac{d\theta}{dt}&=&\omega,
\label{asym2a}
\eeq
where
\be
R(r,\mu_c)\equiv \left\{1+\sum^\infty_{j=1}\,
[\mbox{\rm Re}\;b_j(\mu_c)]\,r^{2j}\right\}.\label{eq:g=0b}
\ee
If the equilibrium is reflection-symmetric, then $\omega=0$ and
$\left[\mbox{\rm
Im}\;b_j(\mu_c)+{\cal{O}}(\gamma)\right]=0$ so the phase equation
(\ref{rescaled2}) becomes
${d\theta}/{dt}=0.$

Note that, in contrast to the rescaled amplitude equation for Hopf bifurcation
(\ref{eq:hopf3}), in this case the higher order terms in $r$ are not higher
order in $\gamma$; thus there is no apparent justification for a truncation
of the expansion after the leading nonlinear terms. The dependence of
$R(r,\mu_c)$  on $\mu_c$ is analyzed in Section V.

\subsubsection{Electric field}

The electric field,
\be
E(x,t)=\sum_{k=1}^{\infty}\;(e^{ikk_cx}\;E_k(t)+\;cc),
\ee
is obtained from the potential $E=-\partial_x\Phi$, and from Poisson's equation
(\ref{eq:vp2}) the Fourier components are
\be
ikk_cE_k(t)=-\int_{-\infty}^{\infty}\,dv\;f^u_k(v,t).
\ee
With (\ref{eq:fufc}) for $f^u_k(v,t)$ this becomes
\be
E_k(t)=\left\{\begin{array}{cc}\frac{i}{k_c}A(1+\sigma\Gm 1)&k=1\\
&\\
\frac{i}{kk_c}A^k\;\Gm k&k>1;\end{array}\right.
\ee
hence the field is given by
\be
E(x,t)=\left(\frac{i}{k_c}\left[\gamma^2\,r(\tau)\,
(1+\gamma^4\,r^4\Gm 1)e^{i(k_cx-\theta(t))}+
\sum_{k=2}^{\infty}\;\frac{\gamma^{2k}\Gm k\,r^k}{k}e^{ik(k_cx-\theta(t))}
\right]+cc\right).
\ee

In the $\glim$ limit, this gives
\beq
E(x,t)&=&\frac{i\gamma^2}{k_c}\left[\rule{0in}{0.35in}r(\tau)\,
\left(1+\Gamma^c_1r(\tau)^4+\ord\gamma\right)e^{i(k_cx-\theta(t))}\right.
\label{efield}\\
&&\hspace{0.7in}+\left.\gamma
\sum_{k=2}^{\infty}\;\frac{\left(\Gamma^c_k\,r(\tau)^k+\ord\gamma\right)}
{k}e^{ik(k_cx-\theta(t))}\right]+cc\nonumber
\eeq
where $\Gamma^c_k$ is defined in (\ref{eq:gmkcrit}). This result describes an
electric field that obeys trapping scaling $E\sim\gamma^2$ and also indicates
that all higher spatial harmonics are uniformly $\ord\gamma$ relative to the
wavenumber of the unstable mode.

\section{Asymptotic recursion relations}

{}From the definition of $b_j$ (\ref{eq:bjdef}) one expects these limits to
depend on the parameters $\mu_c$ through the underlying equilibrium
$F_0(v,\mu_c)$. A striking feature of the explicitly calculated cubic
coefficient (\ref{eq:p1sing}) was the result
\be
b_1=-\frac{1}{4};\label{eq:cubicb1}
\ee
thus $b_1$ is a pure number and independent of $\mu_c$.
It is natural to ask what happens for the higher order singularities.
The dependence of $b_j$ on $\mu_c$ can be systematically investigated by taking
the $\glim$ limit of the recursion relations directly rather than calculating
individual coefficients $p_j$ and studying their asymptotic behavior.

Let $\hat{h}_{k,l}$ and $\hat{I}_{k,l}$ denote the terms of maximum index in
$h_{k,l}$ and $I_{k,l}$ respectively. From Theorem IV.1, this definition
implies
\beq
h_{k,l}&=&\hat{h}_{k,l}+\;[\mbox{\rm terms of index
$2k+4l-4+4(\delta_{k,0}+\delta_{k,1})$ or less}]\label{eq:hlead}\\
I_{k,l}&=&\hat{I}_{k,l}+\;[\mbox{\rm terms of index
$2k+4l-5+4(\delta_{k,0}+\delta_{k,1})$ or less}]
\eeq
Since only terms of maximum index can contribute to the limit in
(\ref{eq:bjdef}), the expression for $p_j$ in (\ref{eq:pj}) yields an explicit
formula for $b_j$
\beq
ik_c\,b_j&=&\lim_{\gamma\rightarrow0^+}\gamma^{4j-1}
\left[\rule{0in}{0.3in}
<\partial_v\tilde{\psi_c},(\hat{h}_{0,j-1}-\hat{h}_{2,j-1})>\right.
\label{eq:bj}\\
&&\hspace{0.75in}\left.
+\sum^{j-2}_{l=0}\left(<\partial_v\tilde{\psi_c}, \hat{h}_{0,j-l-2}>\gm 1l
-<\partial_v\tilde{\psi_c}, \hat{h}_{2,j-l-2}>\gmcc {1}{l}\right)\right].
\nonumber
\eeq
Thus to study $b_j$, only recursion relations for $\hat{h}_{k,l}$ are required.

\subsection{Truncated recursion relations}

{}From (\ref{eq:h0lfcn}) - (\ref{eq:Ikl}), these relations are obtained by
simply discarding terms of submaximal index. This truncation yields the
following expressions:
\be
\hat{h}_{0,l}=\frac{\hat{I}_{0,l}}{(1+l)(\lambda+\lambda^\ast)}\label{eq:h0hat}
\ee
for $k=0$, with
\beq
\hat{I}_{0,l}(v)&=& \left\{\begin{array}{lc}
\frac{i}{k_c}\,\frac{\partial }{\partial v}(\psi_c^\ast-\psi_c)\hspace{1.0in}&
l=0\hspace{0.0in}\\
&\\
-\sum^{l-1}_{j=0}\,(1+j)(p_{l-j}+p_{l-j}^\ast)\,\hat{h}_{0,j}(v)&
l>0\hspace{0.0in}\\
\hspace{0.1in}+\frac{i}{k_c}\,\frac{\partial}{\partial v}
\left\{\left(\hat{h}_{1,l-1}^\ast-\psi_c \gmcc {1}{l-1}
+\sum^{l-2}_{j=0}\,\hat{h}_{1,j}^\ast\gm {1}{l-j-2}\right)-cc\right\}&
\end{array}\right.\label{eq:I0hat}
\eeq
For $k\geq1$, we obtain $\hat{h}_{k,l}(v)$ by calculating
$R_k(w_{k,l})\,\hat{I}_{k,l}$ and discarding any terms of index less than $\ind
[h_{k,l}]$.  The required forms for $\hat{I}_{k,l}$ are
\beq
\hat{I}_{1,l}(v)&=& \frac{i}{k_c}\,{\cal P}_\perp\frac{\partial}{\partial v}
\left\{\hat{h}_{0,l}-\hat{h}_{2,l}
+\sum^{l-1}_{j=0}\left(\hat{h}_{0,j}\gm {1}{l-j-1}-
\hat{h}_{2,j}\gmcc {1}{l-j-1}\right)\right\}\label{eq:I1hat}\\
&&\hspace{2.0in}-\sum^{l-1}_{j=0}
\left[(2+j)p_{l-j}+(1+j)p_{l-j}^\ast\right]\hat{h}_{1,j}.
\nonumber\\
&&\nonumber\\
\hat{I}_{2,l}(v)&=& \left\{\begin{array}{lc}
\frac{i}{k_c}\,\frac{\partial}{\partial v} \psi_c\hspace{1.0in}&
l=0\hspace{0.0in}\\
&\\
-\sum^{l-1}_{j=0}\left[(2+j)p_{l-j}+jp_{l-j}^\ast\right]\,\hat{h}_{2,j}&
l>0\hspace{0.0in}\\
\hspace{0.5in}+\frac{i}{k_c}\,\frac{\partial}{\partial v}
\left\{\hat{h}_{1,l-1} -\hat{h}_{3,l-1}+\psi_c\gm {1}{l-1}\right.&\\
\hspace{1.25in}+\sum^{l-2}_{j=0}\left.\left[\hat{h}_{1,j}\gm {1}{l-j-2} -
\hat{h}_{3,j}\gmcc {1}{l-j-2}\right]\right\}&
\end{array}\right.\label{eq:I2hat}
\eeq
and for $k\geq3$
\beq
\hat{I}_{k,l}(v)&=&\frac{i}{k_c}\,\frac{\partial }{\partial v}
\left\{\hat{h}_{k-1,l}-\hat{h}_{k+1,l-1}
+\sum^{l-1}_{j=0}\,\hat{h}_{k-1,j}\gm {1}{l-j-1}
-\sum^{l-2}_{j=0}\,\hat{h}_{k+1,j}\gmcc {1}{l-j-2}\right\}\label{eq:Iklhat}\\
&&\hspace{2.0in}
-\sum^{l-1}_{j=0}\,[(k+j)p_{l-j}+j\,p_{l-j}^\ast]\,\hat{h}_{k,j}(v).\nonumber
\eeq

\subsection{Integrated recursion relations}

The expression for $b_j$ (\ref{eq:bj}) depends only on certain integrals of
$\hat{h}_{k,l}$ which can be obtained directly from an appropriate integrated
form of the truncated relations. For non-negative integers $(m,n)$, consider
limits of the form:
\beq
\sintlim(m,n;\beta,\alpha)&\equiv&
\left(\frac{i}{k_c}\right)^{m+n-2}
\lim_{\glim}\left(
\frac{\gamma^{m+n-2}}{\Lambda'_{1}(z_0)}
\int^\infty_{-\infty}\,dv\,D_m(\beta,v)^\ast\,D_n(\alpha,v)\,
\eta(v,\mu)\right)\label{eq:slimdef}\\
\bintlim kl(m,n;\beta,\alpha)&\equiv&\left(\frac{i}{k_c}\right)^{m+n-1}
\lim_{\glim}\left(
\frac{\gamma^J}{\Lambda'_{1}(z_0)}
\int^\infty_{-\infty}\,dv\,D_m(\beta,v)^\ast\,D_n(\alpha,v)\,
\hat{h}_{k,l}(v)\right)\label{eq:bintlim}\\
\cintlim kl(m,n;\beta,\alpha)&\equiv&\left(\frac{i}{k_c}\right)^{m+n-1}
\lim_{\glim}\left(
\frac{\gamma^J}{\Lambda'_{1}(z_0)}
\int^\infty_{-\infty}\,dv\,D_m(\beta,v)^\ast\,D_n(\alpha,v)\,
\hat{h}_{k,l}(v)^\ast\right)
\eeq
where $J\equiv\ind [{h}_{k,l}]+m+n$ and $m+n\geq2$ is required in
(\ref{eq:slimdef}). The resonance denominators in
$D_m(\beta,v)^\ast\,D_n(\alpha,v)$ are defined using our previous notation
(\ref{eq:Ddef}), (\ref{eq:poles1}), and (\ref{eq:poles2}); in particular, the
pole locations are:
\beq
\alpha_j&=&z_0+i\gamma\nu_j/k_c\hspace{0.5in}j=1,\ldots,n\\
\beta_j^\ast&=&z_0^\ast-i\gamma\zeta_j/k_c\hspace{0.5in}j=1,\ldots,m,
\eeq
and the sets of poles are indicated with the notation
$\alpha=(\alpha_1,\ldots,\alpha_n)$, $\beta=(\beta_1,\ldots,\beta_m)$,
$\nu=(\nu_1,\ldots,\nu_n)$, and $\zeta=(\zeta_1,\ldots,\zeta_m)$. In using the
notation above, e.g. for $\sintlim(m,n;\beta,\alpha)$, it is understood that
the first $m$ arguments after the semi-colon correspond to the poles in the
lower half-plane $D_m(\beta,v)^\ast$ and the remaining $n$ arguments denote the
poles in the upper half-plane $D_n(\alpha,v)$.

The latter two limits are related by
\be
\cintlim kl(m,n;\beta,\alpha)=e^{i\xi}\,[\bintlim kl(n,m;\alpha,\beta)]^\ast
(-1)^{m+n-1}\label{eq:cintid}
\ee
where the phase $e^{i\xi}$ is defined by
\be
e^{i\xi(\mu_c)}\equiv
\lim_{\glim}\left(\frac{\Lambda'_{1}(z_0)^\ast}{\Lambda'_{1}(z_0)}\right).
\label{eq:phase}
\ee
Note that since $\Lambda'_{k}(z_0)=\epsilon'_{k}(z_0)$ for $\gamma\geq0$, this
is the phase mentioned in the Introduction.

{}From (\ref{eq:bj}), the coefficients of the leading singularities are given
by
\beq
b_j&=&\bintlim 2{j-1}(0,2;z_0,z_0)-\bintlim 0{j-1}(0,2;z_0,z_0)
\label{eq:bjfinal}\\
&&\hspace{1.0in}+\sum^{j-2}_{l=0}\,[c_{1,l}^\ast\;\bintlim
2{j-l-2}(0,2;z_0,z_0) -c_{1,l}\;\bintlim 0{j-l-2}(0,2;z_0,z_0)]\nonumber
\eeq
where $c_{1,l}$ is defined from (\ref{eq:ckldef}) by the limit
\be
c_{1,l}=\lim_{\glim}\left(\gamma^{4(l+1)}\,\gm 1l\right).\label{eq:cl}
\ee
This expression is evaluated below, c.f. (\ref{eq:clrecur}).

The properties of $\sintlim(m,n;\beta,\alpha)$ in  (\ref{eq:slimdef}) are
analyzed in Appendix B; the main conclusion needed here concerns the dependence
on $\mu_c$.
\begin{lemma} For non-negative integers $(m,n)$ such that $m+n\geq2$,
$\sintlim(m,n;\beta,\alpha)$ has the form
\be
\sintlim(m,n;\beta,\alpha)=
d(m,n;\zeta,\nu)+(-1)^{m+n}d(n,m;\nu,\zeta)\,e^{i\xi(\mu_c)}\label{eq:sint1}
\ee
where the real-valued functions $d(m,n;\zeta,\nu)$ are independent of
$F_0(v,\mu_c)$. Thus $\sintlim(m,n;\alpha,\beta)$ depends on $\mu_c$ only
through the phase $\exp(i\xi)$.
\end{lemma}
\noindent {\em {\bf Proof}.} See Appendix B.
{\bf $\Box$}

The analysis of $\bintlim kl(m,n;\beta,\alpha)$ follows the pattern for the
calculation of ${h}_{k,l}(v)$ in Table I. From the definition
(\ref{eq:bintlim}), $\bintlim 00(m,n;\beta,\alpha)$ and $\bintlim
20(m,n;\beta,\alpha)$
can be evaluated in terms of $\sintlim(m,n;\alpha,\beta)$:
\beq
\bintlim 00(m,n;\beta,\alpha)&=&\sum_{i=1}^{m}\;
\sintlim(m+2,n+1;\beta,\beta_i,z_0,\alpha,z_0)\label{eq:B00}\\
&&\hspace{1.5in}
+\sum_{i=1}^{n}\;\sintlim(m+1,n+2;\beta,z_0,\alpha,\alpha_i,z_0)\nonumber\\
\bintlim 20(m,n;\beta,\alpha)&=&
\frac{1}{2}\left[\sum_{i=1}^{m}\;
\sintlim(m+1,n+2;\beta,\beta_i,\alpha',z_0)\right.\label{eq:B20}\\
&&\hspace{2.0in}\left.
+\sum_{i=1}^{n+1}\;\sintlim(m,n+3;\beta,\alpha',\alpha'_i,z_0) \right]
\nonumber
\eeq
where $\alpha'=(\alpha,z_{2,0})$. This expression for $\bintlim
20(m,n;\beta,\alpha)$ uses the identities:
\be
\lim_{\glim}\Lambda_{k}(z_{k,l})=\frac{k^2-1}{k^2}\label{eq:lamid}
\ee
from (\ref{eq:specfcnid}), and
\be
\lim_{\glim}
\left({\gamma^{m+n}}
\int^\infty_{-\infty}\,dv\,\frac{D_m^\ast\,D_n}{v-z_{k,l}}\eta(v,\mu)\right)=
-\delta_{m,0}\;\delta_{n,0}.\label{eq:intid}
\ee
The latter identity is verified by noting that the integrand has index $m+n-1$
so if $m+n>0$ we get zero, and when $m=n=0$ then the integral reduces to
$\Lambda_{1}(z_{k,l})-1$. From these expressions for $\bintlim 00$ and
$\bintlim 20$ their dependence on $\mu_c$ is easily characterized.

\begin{lemma} For all non-negative integers $(m,n)$, $\bintlim
00(m,n;\beta,\alpha)$ and $\bintlim 20(m,n;\beta,\alpha)$ depend on $\mu_c$
only through the phase $\exp(i\xi)$.
\end{lemma}
\noindent {\em {\bf Proof}.} \begin{quote} This follows immediately from
(\ref{eq:B00}) - (\ref{eq:B20}) and Lemma V.1.
{\bf $\Box$}
\end{quote}

Recursion relations for the remaining limits $\bintlim kl(m,n;\beta,\alpha)$
and also for $c_{k,l}$ in (\ref{eq:ckldef}) follow from the truncated relations
in (\ref{eq:h0hat}) - (\ref{eq:Iklhat}); this derivation is briefly summarized
for $k=0, 1, 2,$ and $k\geq3$. These relations are somewhat complicated, but
provide the basis for Theorem V.1 below.

\subsubsection{$\bintlim 0l(m,n;\beta,\alpha)$ for $l\geq1$}

{}From the definition (\ref{eq:bintlim}) and substituting for
$\hat{h}_{0,l}(v)$ from (\ref{eq:h0hat})
\be
\bintlim 0l(m,n;\beta,\alpha)= \left(\frac{i}{k_c}\right)^{m+n-1}
\lim_{\glim}\left(
\frac{\gamma^{4l+m+n+1}}{2\gamma(l+1)\Lambda'_{1}(z_0)}
\int^\infty_{-\infty}\,dv\,D_m(\beta,v)^\ast\,D_n(\alpha,v)\,
\hat{I}_{0,l}(v)\right)\label{eq:bint0l}
\ee
Using (\ref{eq:I0hat}) the integral becomes
\beq
\lefteqn{\int^\infty_{-\infty}\,dv\,
D_m(\beta,v)^\ast\,D_n(\alpha,v)\,\hat{I}_{0,l}(v)=}\\
&&-\sum^{l-1}_{j=0}\,(1+j)(p_{l-j}+p_{l-j}^\ast)\,
\int^\infty_{-\infty}\,dv\,D_m(\beta,v)^\ast\,D_n(\alpha,v)\,
\hat{h}_{0,j}(v)\nonumber\\
&&-\frac{i}{k_c}\,\int^\infty_{-\infty}\,dv\,
\frac{\partial }{\partial v}(D_m^\ast\,D_n)
\left[\left(\hat{h}_{1,l-1}^\ast-\psi_c \gmcc {1}{l-1}
+\sum^{l-2}_{j=0}\,\hat{h}_{1,j}^\ast\gm {1}{l-j-2}\right)-cc\right]\nonumber
\eeq
With the substitution
\be
-\frac{\partial }{\partial
v}(D_m^\ast\,D_n)=\sum_{i=1}^{m}\frac{D_m^\ast\,D_n}{v-\beta_i^\ast}+
\sum_{i=1}^{n}\frac{D_m^\ast\,D_n}{v-\alpha_i},
\ee
the limit in (\ref{eq:bint0l}) can be rewritten as
\beq
\lefteqn{2(l+1)\bintlim 0l(m,n;\beta,\alpha)=}\\
&&-\sum^{l-1}_{j=0}\,(1+j)
\left[\lim_{\glim}\gamma^{4(l-j)-1}(p_{l-j}+p_{l-j}^\ast)\right]
\,\left(\frac{i}{k_c}\right)^{m+n-1}
\lim_{\glim}\left(
\frac{\gamma^{4j+m+n+1}}{\Lambda'_{1}(z_0)}
\int^\infty_{-\infty}\,dv\,D_m^\ast\,D_n\,
\hat{h}_{0,j}(v)\right)\nonumber\\
&&
\hspace{0.5in}+\left(\frac{i}{k_c}\right)^{m+n}\,
\sum_{i=1}^{m}\lim_{\glim}\left(
\frac{\gamma^{4l+m+n}}{\Lambda'_{1}(z_0)}
\int^\infty_{-\infty}\,dv\,\frac{D_m^\ast\,D_n}{v-\beta_i^\ast}
\left[\hat{h}_{1,l-1}^\ast-\psi_c \gmcc {1}{l-1}
+\cdots\right]\right)\nonumber\\
&&
\hspace{0.5in}+\left(\frac{i}{k_c}\right)^{m+n}\,
\sum_{i=1}^{n}\lim_{\glim}\left(
\frac{\gamma^{4l+m+n}}{\Lambda'_{1}(z_0)}
\int^\infty_{-\infty}\,dv\,\frac{D_m^\ast\,D_n}{v-\alpha_i}
\left[\hat{h}_{1,l-1}^\ast-\psi_c \gmcc {1}{l-1}
+\cdots\right]\right),\nonumber
\eeq
and then evaluated to obtain
\beq
\lefteqn{2(l+1)\bintlim 0l(m,n;\beta,\alpha)=
-\sum^{l-1}_{j=0}\,(1+j)(b_{l-j}+b_{l-j}^\ast)\,\bintlim 0j(m,n;\beta,\alpha)}
\label{eq:B0lrecur}\\
&&
\hspace{0.8in}+\sum_{i=1}^{m}\left\{\rule{0in}{0.3in}
\cintlim 1{l-1}(m+1,n;\beta,\beta_i,\alpha)
-\bintlim 1{l-1}(m+1,n;\beta,\beta_i,\alpha)\right.\nonumber\\
&&
\hspace{1.0in}-c_{1,l-1}^\ast\,\sintlim(m+1,n+1;\beta,\beta_i,z_0,\alpha)
+c_{1,l-1}\,\sintlim(m+2,n;\beta,\beta_i,z_0,\alpha)
\nonumber\\
&&\hspace{1.0in}
\left.+\sum^{l-2}_{j=0}\,\left[c_{1,l-j-2}\,\cintlim
1{j}(m+1,n;\beta,\beta_i,\alpha)-c_{1,l-j-2}^\ast\,\bintlim
1{j}(m+1,n;\beta,\beta_i,\alpha)\right]\right\}\nonumber\\
&&\hspace{0.8in}
+\sum_{i=1}^{n}\left\{\rule{0in}{0.3in}
\cintlim 1{l-1}(m,n+1;\beta,\alpha,\alpha_i)
-\bintlim 1{l-1}(m,n+1;\beta,\alpha,\alpha_i)\right.\nonumber\\
&&\hspace{1.0in}-c_{1,l-1}^\ast\,\sintlim(m,n+2;\beta,z_0,\alpha,\alpha_i)
+c_{1,l-1}\,\sintlim(m+1,n+1;\beta,z_0,\alpha,\alpha_i)
\nonumber\\
&&\hspace{1.0in}
\left.+\sum^{l-2}_{j=0}\,\left[c_{1,l-j-2}\,\cintlim
1{j}(m,n+1;\beta,\alpha,\alpha_i)-c_{1,l-j-2}^\ast\,\bintlim
1{j}(m,n+1;\beta,\alpha,\alpha_i)\right]\right\}.\nonumber
\eeq

\subsubsection{$\bintlim 1l(m,n;\beta,\alpha)$ for $l\geq0$}

{}From the definition (\ref{eq:bintlim})
\be
\bintlim 1l(m,n;\beta,\alpha)=\left(\frac{i}{k_c}\right)^{m+n-1}
 \lim_{\glim}\left(
\frac{\gamma^{4l+m+n+3}}{\Lambda'_{1}(z_0)}
\int^\infty_{-\infty}\,dv\,D_m(\beta,v)^\ast\,D_n(\alpha,v)\,
\hat{h}_{1,l}(v)\right).
\ee
Using $\hat{h}_{1,l}=R_1(w_{1,l})\,\hat{I}_{1,l}$ and (\ref{eq:gmkj})
for $\gm 1l$, this can be rewritten as
\beq
\bintlim 1l(m,n;\beta,\alpha)&=&
-\,c_{1,l}\,\sintlim(m,n+1;\beta,\alpha,z_{1,l})\label{b1lrecur}\\
&&
-\left(\frac{i}{k_c}\right)^{m+n}
\lim_{\glim}\left(
\frac{\gamma^{4l+m+n+3}}{\Lambda'_{1}(z_0)}\int^\infty_{-\infty}\,dv\,
\frac{D_m(\beta,v)^\ast\,D_n(\alpha,v)\,\hat{I}_{1,l}(v)}{v-z_{1,l}}\right)
\nonumber
\eeq
where we have used  (\ref{eq:cl}).
Using the expression for $\hat{I}_{1,l}(v)$ in (\ref{eq:I1hat}) we obtain for
the second term
\beq
\lefteqn{\left(\frac{i}{k_c}\right)^{m+n}\lim_{\glim}\left(
\frac{\gamma^{4l+m+n+3}}{\Lambda'_{1}(z_0)}\int^\infty_{-\infty}\,dv\,
\frac{D_m(\beta,v)^\ast\,D_n(\alpha,v)\,\hat{I}_{1,l}(v)}{v-z_{1,l}}\right)=}
\label{eq:Ihat1}\\
&&
-\sum^{l-1}_{j=0}\left[(2+j)b_{l-j}+(1+j)b_{l-j}^\ast\right]
\bintlim 1j(m,n+1;\beta,\alpha,z_{1,l})\nonumber\\
&&
+\left[\rule{0in}{0.2in}
\bintlim 0l(m,n+2;\beta,\alpha,z_{1,l},z_{1,l})-
\bintlim 2l(m,n+2;\beta,\alpha,z_{1,l},z_{1,l})\right.\nonumber\\
&&\hspace{0.75in}
+\sum^{l-1}_{j=0}\left(\left.c_{1,l-j-1}
\bintlim 0j(m,n+2;\beta,\alpha,z_{1,l},z_{1,l})
-c_{1,l-j-1}^\ast\bintlim 2j(m,n+2;\beta,\alpha,z_{1,l},z_{1,l})\right)
\rule{0in}{0.2in}\right]\nonumber\\
&&
-\sintlim(m,n+2;\beta,\alpha,z_{1,l},z_0)
\left[\rule{0in}{0.2in} \bintlim 0l(0,2;z_0,z_0)-\bintlim
2l(0,2;z_0,z_0)\right.\nonumber\\
&&\hspace{1.75in}
+\sum^{l-1}_{j=0}\left(\left.c_{1,l-j-1}\bintlim 0j(0,2;z_0,z_0)
-c_{1,l-j-1}^\ast\bintlim 2j(0,2;z_0,z_0)\right)
\rule{0in}{0.2in}\right]\nonumber\\
&&
+\sum_{i=1}^{m}\left\{\rule{0in}{0.2in}
\bintlim 0l(m+1,n+1;\beta,\beta_i,\alpha,z_{1,l})-
\bintlim 2l(m+1,n+1;\beta,\beta_i,\alpha,z_{1,l})\right.\nonumber\\
&&\hspace{0.75in}
+\sum^{l-1}_{j=0}\left(\left.c_{1,l-j-1}\bintlim
0j(m+1,n+1;\beta,\beta_i,\alpha,z_{1,l})
-c_{1,l-j-1}^\ast\bintlim 2j(m+1,n+1;\beta,\beta_i,\alpha,z_{1,l})\right)
\rule{0in}{0.2in}\right\}\nonumber\\
&&
+\sum_{i=1}^{n}\left\{\rule{0in}{0.2in}
\bintlim 0l(m,n+2;\beta,\alpha,\alpha_i,z_{1,l})-
\bintlim 2l(m,n+2;\beta,\alpha,\alpha_i,z_{1,l})\right.\nonumber\\
&&\hspace{0.75in}
+\sum^{l-1}_{j=0}\left(\left.c_{1,l-j-1}\bintlim
0j(m,n+2;\beta,\alpha,\alpha_i,z_{1,l})
-c_{1,l-j-1}^\ast\bintlim 2j(m,n+2;\beta,\alpha,\alpha_i,z_{1,l})\right)
\rule{0in}{0.2in}\right\}.\nonumber
\eeq

\subsubsection{$\bintlim 2l(m,n;\beta,\alpha)$ for $l\geq1$}

{}From the definition (\ref{eq:bintlim})
\be
\bintlim 2l(m,n;\beta,\alpha)=\left(\frac{i}{k_c}\right)^{m+n-1}
 \lim_{\glim}\left(
\frac{\gamma^{4l+m+n+1}}{\Lambda'_{1}(z_0)}
\int^\infty_{-\infty}\,dv\,
D_m(\beta,v)^\ast\,D_n(\alpha,v)\,\hat{h}_{2,l}(v)\right).
\ee
Using $\hat{h}_{2,l}=R_2(w_{2,l})\,\hat{I}_{2,l}$, the integral becomes
\beq
\int^\infty_{-\infty}\,dv\,
D_m(\beta,v)^\ast\,D_n(\alpha,v)\,\hat{h}_{2,l}(v)&=&
\frac{-i}{2k_c}\left[
\int^\infty_{-\infty}\,dv\,
\frac{D_m^\ast\,D_n\,\hat{I}_{2,l}(v)}{v-z_{2,l}}\right.\\
&&\left.
-\frac{1}{4\Lambda_{2}(z_{2,l})}
\int^\infty_{-\infty}\,dv\,
\frac{D_m^\ast\,D_n\,\eta(v)}{v-z_{2,l}}
\int^\infty_{-\infty}\,dv'\,
\frac{\hat{I}_{2,l}(v')}{v'-z_{2,l}}
\right],\nonumber
\eeq
and with (\ref{eq:lamid}) - (\ref{eq:intid})
we obtain
\be
\bintlim 2l(m,n;\beta,\alpha)=
-\frac{1}{2}\left(\frac{i}{k_c}\right)^{m+n}
\left(1+\frac{\delta_{m,0}\;\delta_{n,0}}{3}\right)
\lim_{\glim}\left(
\frac{\gamma^{4l+m+n+1}}{\Lambda'_{1}(z_0)}
\int^\infty_{-\infty}\,dv\,\frac{
D_m^\ast\,D_n\,\hat{I}_{2,l}(v)}{v-z_{2,l}}\right).\label{eq:B2leqn}
\ee
Let $\alpha'=(\alpha,z_{2,l})$ and define
$D_{n+1}(\alpha',v)=D_{n}(\alpha,v)/(v-z_{2,l})$, then using (\ref{eq:I2hat})
the limit in (\ref{eq:B2leqn}) yields:
\beq
\lefteqn{\left(\frac{i}{k_c}\right)^{m+n}\lim_{\glim}\left(
\frac{\gamma^{4l+m+n+1}}{\Lambda'_{1}(z_0)}
\int^\infty_{-\infty}\,dv\,\frac{
D_m(\beta,v)^\ast\,D_n(\alpha,v)\,\hat{I}_{2,l}(v)}{v-z_{2,l}}\right)=}
\label{eq:I2leqn}\\
&&
-\sum^{l-1}_{j=0}\left[(2+j)b_{l-j}+jb_{l-j}^\ast\right]\,\bintlim
2j(m,n+1;\beta,\alpha')\nonumber\\
&&+\sum_{i=1}^{m}\left\{\rule{0in}{0.2in}
\bintlim 1{l-1}(m+1,n+1;\beta,\beta_i,\alpha')-
\bintlim 3{l-1}(m+1,n+1;\beta,\beta_i,\alpha')\right.\nonumber\\
&&\hspace{0.75in}-c_{1,l-1}\sintlim(m+1,n+2;\beta,\beta_i,\alpha',z_0)
\nonumber\\
&&\hspace{0.75in}+\sum^{l-2}_{j=0}\left.\left[
c_{1,l-j-2}\bintlim 1{j}(m+1,n+1;\beta,\beta_i,\alpha')-
c_{1,l-j-2}^\ast\bintlim 3{j}(m+1,n+1;\beta,\beta_i,\alpha')\right]
\rule{0in}{0.2in}\right\}\nonumber\\
&&
+\sum_{i=1}^{n+1}\left\{\rule{0in}{0.2in}
\bintlim 1{l-1}(m,n+2;\beta,\alpha',\alpha_i')-
\bintlim 3{l-1}(m,n+2;\beta,\alpha',\alpha_i')\right.\nonumber\\
&&\hspace{0.75in}-c_{1,l-1}\sintlim(m,n+3;\beta,\alpha',\alpha_i',z_0)
\nonumber\\
&&\hspace{0.75in}+\sum^{l-2}_{j=0}\left.\left[
c_{1,l-j-2}\bintlim 1{j}(m,n+2;\beta,\alpha',\alpha_i')-
c_{1,l-j-2}^\ast\bintlim 3{j}(m,n+2;\beta,\alpha',\alpha_i')\right]
\rule{0in}{0.2in}\right\}.\nonumber
\eeq

\subsubsection{$\bintlim kl(m,n;\beta,\alpha)$ for $k\geq3$, $l\geq0$}

{}From the definition (\ref{eq:bintlim})
\be
\bintlim kl(m,n;\beta,\alpha)=\left(\frac{i}{k_c}\right)^{m+n-1}
\lim_{\glim}\left(
\frac{\gamma^{J}}{\Lambda'_{1}(z_0)}
\int^\infty_{-\infty}\,dv\,
D_m(\beta,v)^\ast\,D_n(\alpha,v)\,\hat{h}_{k,l}(v)\right)
\ee
where $J=2k+4l-3+m+n$. Using $\hat{h}_{k,l}=R_k(w_{k,l})\,\hat{I}_{k,l}$
\beq
\int^\infty_{-\infty}\,dv\,
D_m(\beta,v)^\ast\,D_n(\alpha,v)\,\hat{h}_{k,l}(v)&=&
\frac{-i}{kk_c}\left[
\int^\infty_{-\infty}\,dv\,
\frac{D_m^\ast\,D_n\,\hat{I}_{k,l}(v)}{v-z_{k,l}}\right.\\
&&\left.
-\frac{1}{k^2\Lambda_{k}(z_{k,l})}
\int^\infty_{-\infty}\,dv\,
\frac{D_m^\ast\,D_n\,\eta(v)}{v-z_{k,l}}
\int^\infty_{-\infty}\,dv'\,
\frac{\hat{I}_{k,l}(v')}{v'-z_{k,l}}
\right],\nonumber
\eeq
and with (\ref{eq:lamid}) - (\ref{eq:intid})
we obtain
\be
\bintlim kl(m,n;\beta,\alpha)=
-\frac{1}{k}\left(\frac{i}{k_c}\right)^{m+n}
\left(1+\frac{\delta_{m,0}\;\delta_{n,0}}{k^2-1}\right)
\lim_{\glim}\left(
\frac{\gamma^{J}}{\Lambda'_{1}(z_0)}
\int^\infty_{-\infty}\,dv\,\frac{
D_m(\beta,v)^\ast\,D_n(\alpha,v)\,\hat{I}_{k,l}(v)}{v-z_{k,l}}\right).
\label{eq:Bkleqn}
\ee
Let $\alpha'=(\alpha,z_{k,l})$ and define
$D_{n+1}(\alpha',v)=D_{n}(\alpha,v)/(v-z_{k,l})$, then from (\ref{eq:Iklhat})
we find
\beq
\lefteqn{\left(\frac{i}{k_c}\right)^{m+n}\lim_{\glim}\left(
\frac{\gamma^{J}}{\Lambda'_{1}(z_0)}
\int^\infty_{-\infty}\,dv\,\frac{
D_m^\ast\,D_n\,\hat{I}_{k,l}(v)}{v-z_{k,l}}\right)=
-\sum^{l-1}_{j=0}\left[(k+j)b_{l-j}+jb_{l-j}^\ast\right]\,\bintlim
kj(m,n+1;\beta,\alpha')}\nonumber\\
&&+\sum_{i=1}^{m}\left\{\rule{0in}{0.2in}
\bintlim {k-1}{l}(m+1,n+1;\beta,\beta_i,\alpha')-
\bintlim {k+1}{l-1}(m+1,n+1;\beta,\beta_i,\alpha')\right.\label{eq:Ikleqn}\\
&&\hspace{0.75in}+\sum^{l-1}_{j=0}
c_{1,l-j-1}\bintlim {k-1}{j}(m+1,n+1;\beta,\beta_i,\alpha')-
\sum^{l-2}_{j=0}\left.
c_{1,l-j-2}^\ast\bintlim {k+1}{j}(m+1,n+1;\beta,\beta_i,\alpha')
\rule{0in}{0.2in}\right\}\nonumber\\
&&
+\sum_{i=1}^{n+1}\left\{\rule{0in}{0.2in}
\bintlim {k-1}{l}(m,n+2;\beta,\alpha',\alpha_i')-
\bintlim {k+1}{l-1}(m,n+2;\beta,\alpha',\alpha_i')\right\}\nonumber\\
&&\hspace{0.75in}+\sum^{l-1}_{j=0}
c_{1,l-j-1}\bintlim {k-1}{j}(m,n+2;\beta,\alpha',\alpha_i')-
\sum^{l-2}_{j=0}\left.
c_{1,l-j-2}^\ast\bintlim {k+1}{j}(m,n+2;\beta,\alpha',\alpha_i')
\rule{0in}{0.2in}\right\}.\nonumber
\eeq

\subsubsection{Evaluation of $c_{k,l}(\mu_c)$}

{}From (\ref{eq:ckldef}) and (\ref{eq:gmkj}), we have
\be
c_{k,l}(\mu_c)=\lim_{\glim}\left[\gamma^{(2k+4l-3+5\delta_{k,1})}\,
\left(\frac{-ik/k_c}{k^2-1+\Lambda_{1}(z_{k,l})}\right)
\int^\infty_{-\infty}\,dv\,\frac{I_{k,l}(v)}{v-z_{k,l}}\right]
\label{eq:ckleqn}
\ee
for $k\geq 1$ and $l\geq0$. Since $\Lambda_{1}(z_{k,l})=\Lambda_{1}(z_0+i\gamma
d_{k,l}/k_c)=i\gamma d_{k,l}
\Lambda'_{1}(z_{0})/k_c+\ord{\gamma^2}$, as $\glim$ the prefactor gives
\be
\left(\frac{-ik/k_c}{k^2-1+\Lambda_{1}(z_{k,l})}\right) =
\left\{\begin{array}{cc}
-\frac{1}{\gamma}\left[\frac{1}{ d_{1,l}\Lambda'_{1}(z_{0})}
+\ord{\gamma}\right]&k=1\\
&\\
\frac{-ik/k_c}{k^2-1}+\ord{\gamma}&k>1
\end{array}\right.
\ee
where $d_{1,l}=2(1+l)$. Thus for $k=1$
\be
c_{1,l}(\mu_c)=-\lim_{\glim}
\left(\frac{\gamma^{4l+3}}{d_{1,l}\,\Lambda'_{1}(z_0)}
\int^\infty_{-\infty}\,dv\frac{\hat{I}_{1,l}(v)}{v-z_{1,l}}\right),
\label{eq:clb}
\ee
and the right hand side is evaluated by setting $m=n=0$ in (\ref{eq:Ihat1}):
\beq
c_{1,l}(\mu_c)&=&-\left(\frac{1}{d_{1,l}}\right)\left\{
-\sum^{l-1}_{j=0}\left[(2+j)b_{l-j}+(1+j)b_{l-j}^\ast\right]
\bintlim 1j(0,1;z_{1,l})\right.\label{eq:clrecur}\\
&&\hspace{0.5in}
+\left[\rule{0in}{0.2in}
\bintlim 0l(0,2;z_{1,l},z_{1,l})-
\bintlim 2l(0,2;z_{1,l},z_{1,l})\right.\nonumber\\
&&\hspace{1.25in}
+\sum^{l-1}_{j=0}\left(\left.c_{1,l-j-1}
\bintlim 0j(0,2;z_{1,l},z_{1,l})
-c_{1,l-j-1}^\ast\bintlim 2j(0,2;z_{1,l},z_{1,l})\right)
\rule{0in}{0.2in}\right]\nonumber\\
&&\hspace{0.5in}
-\left[\rule{0in}{0.2in}
\bintlim 0l(0,2;z_0,z_0)-\bintlim 2l(0,2;z_0,z_0)\right.\nonumber\\
&&\hspace{1.75in}
+\left.\sum^{l-1}_{j=0}\left(\left.c_{1,l-j-1}\bintlim 0j(0,2;z_0,z_0)
-c_{1,l-j-1}^\ast\bintlim 2j(0,2;z_0,z_0)\right)
\rule{0in}{0.2in}\right]\nonumber\right\}.
\eeq
In deriving (\ref{eq:clrecur}), the relation $\sintlim(0,2;z_{1,l},z_0)=1$ from
(\ref{eq:mn2a}) in Appendix B has been used.

The corresponding expression for $c_{k,l}$ when $k>1$ is much simpler. Setting
$m=n=0$ in (\ref{eq:I2leqn}) and (\ref{eq:Ikleqn}), yields
\be
c_{k,l}(\mu_c)=\frac{i\Lambda'_{1}(\omega_c/k_c+i0^+)}{k_c}\bintlim
{k}{l}(0,0;\beta,\alpha)
\label{eq:cklrecur}
\ee
from (\ref{eq:ckleqn}) for $k\geq2$. Here $\omega_c/k_c$ is the phase velocity
at criticality; note also that when $m=n=0$ the arguments $(\beta,\alpha)$ in
$\bintlim {k}{l}(m,n;\beta,\alpha)$ are irrelevant.

\subsection{Evaluation of the lowest order singularities $b_1$ and $b_2$}

It is instructive to calculate the cubic singularity $b_1$ and the fifth order
singularity $b_2$ from these integrated recursion relations. In the cubic case
this merely recovers the previous result (\ref{eq:cubicb1}), but the
calculation illustrates the formalism. The procedure is to apply the recursion
relations until expressions involving $\sintlim(m,n;\beta,\alpha)$ are
obtained, then the results from Appendix B are used.

\subsubsection{Calculation of $b_1$}

For $j=1$, the expression for $b_j$ in (\ref{eq:bjfinal}) yields
\be
b_1(\mu_c)=\bintlim {2}{0}(0,2;z_0,z_0)-\bintlim
{0}{0}(0,2;z_0,z_0),\label{eq:b1recur}
\ee
and from (\ref{eq:B00}) - (\ref{eq:B20})  one has
\beq
\bintlim {0}{0}(0,2;z_0,z_0)&=&2\;\sintlim(1,4;z_0,z_0,z_{0},z_0,z_0)\\
\bintlim {2}{0}(0,2;z_0,z_0)&=&\sintlim(0,5;z_0,z_0,z_{2,0},z_0,z_0)
+\frac{1}{2}\sintlim(0,5;z_0,z_0,z_{2,0},z_{2,0},z_0).\label{eq:B20form}
\eeq
In  Appendix B the functions $\sintlim(m,n;\beta,\alpha)$ are evaluated;
applying these results yields
\beq
\bintlim
{0}{0}(0,2;z_0,z_0)&=&2\;\sintlim(1,4;z_0,z_0,z_{0},z_0,z_0)\nonumber\\
&=&2\left[d(1,4;0,0,0,0,0)-d(4,1;0,0,0,0,0)e^{i\xi}\right]\nonumber\\
&=&2\;d(1,4;0,0,0,0,0)\nonumber\\
&=&\frac{1}{4},\label{eq:b1calc}
\eeq
and it follows from (\ref{eq:0nform}) that $\bintlim {2}{0}(0,2;z_0,z_0)=0$.
Hence, as expected, $b_1=-1/4$ from (\ref{eq:b1recur}).

\subsubsection{Calculation of $b_2$}

The evaluation of $b_2$ proceeds similarly but is considerably more laborious;
I summarize the calculation below but omit the details. For $j=2$,
(\ref{eq:bjfinal}) yields
\be
b_2(\mu_c)=\bintlim {2}{1}(0,2;z_0,z_0)-\bintlim
{0}{1}(0,2;z_0,z_0)-c_{1,0}\bintlim {0}{0}(0,2;z_0,z_0)\label{eq:b2recura}
\ee
since $\bintlim {2}{0}(0,2;z_0,z_0)=0$ from (\ref{eq:B20form}) and
(\ref{eq:0nform}). Setting $l=0$ in (\ref{eq:clrecur}) gives
\be
c_{1,0}=\frac{1}{2}\left[\bintlim {0}{0}(0,2;z_0,z_0)-
\bintlim {0}{0}(0,2;z_{1,0},z_{1,0})\right].
\ee
With $\bintlim {0}{0}(0,2;z_0,z_0)=1/4$ from (\ref{eq:b1calc}) and a similar
evaluation yielding $\bintlim {0}{0}(0,2;z_{1,0},z_{1,0})=1/32$, this gives
$c_{1,0}={7}/{4^3}$
and (\ref{eq:b2recura}) becomes
\be
b_2(\mu_c)=\bintlim {2}{1}(0,2;z_0,z_0)-\bintlim
{0}{1}(0,2;z_0,z_0)-\left(\frac{7}{4^4}\right).
\label{eq:b1recurb}
\ee

Consider $\bintlim {0}{1}$ in (\ref{eq:b1recurb}) first; from the $k=0$
recursion relation (\ref{eq:B0lrecur})
\beq
\bintlim {0}{1}(0,2;z_0,z_0)&=&\frac{1}{4}\left[\frac{1}{2}\bintlim
{0}{0}(0,2;z_0,z_0)+2\left(e^{i\xi}\bintlim {1}{0}(3,0;z_0,z_0,z_0)^\ast
-\bintlim {1}{0}(0,3;z_0,z_0,z_0)\right) \right.\nonumber\\
&&\hspace{0.6in}\left.+
2\;c_{1,0}\,\sintlim(1,3;z_0,z_0,z_0,z_0)\rule{0in}{0.2in}\right],
\eeq
and since $c_{1,0}\,\sintlim(1,3;z_0,z_0,z_0,z_0)=(7/4^3)(-1/4)$,
this simplifies to
\be
\bintlim {0}{1}(0,2;z_0,z_0)=\frac{1}{2}\left[\frac{9}{4^4} +e^{i\xi}\bintlim
{1}{0}(3,0;z_0,z_0,z_0)^\ast
-\bintlim {1}{0}(0,3;z_0,z_0,z_0)\right].\label{eq:b01a}
\ee
For $\bintlim {2}{1}$ in (\ref{eq:b1recurb}), from (\ref{eq:B2leqn}) and
(\ref{eq:I2leqn}) and using the results $\bintlim
{2}{0}(0,3;z_0,z_0,z_{2,1})=0$ and $\sintlim(0,5;\beta,\alpha)=0$, one finds
\beq
\bintlim {2}{1}(0,2;z_0,z_0)&=&-\frac{1}{2}\left[\rule{0in}{0.25in} 2
\left(\rule{0in}{0.2in}\bintlim {1}{0}(0,4;z_0,z_0,z_{2,1},z_0)- \bintlim
{3}{0}(0,4;z_0,z_0,z_{2,1},z_0)\right)\right.\label{eq:b21a}\\
&&\hspace{0.4in}\left.+\bintlim {1}{0}(0,4;z_0,z_0,z_{2,1},z_{2,1})-\bintlim
{3}{0}(0,4;z_0,z_0,z_{2,1},z_{2,1})\rule{0in}{0.25in}\right].
\nonumber
\eeq
{}From the recursion for $k=3$ in (\ref{eq:Bkleqn}), two terms in
(\ref{eq:b21a}) vanish: $\bintlim {3}{0}(0,4;z_0,z_0,z_{2,1},z_0)=0$ and
$\bintlim {3}{0}(0,4;z_0,z_0,z_{2,1},z_{2,1})=0$, leaving
\be
\bintlim {2}{1}(0,2;z_0,z_0)=-\left[\bintlim {1}{0}(0,4;z_0,z_0,z_{2,1},z_0)
+\frac{1}{2}\bintlim {1}{0}(0,4;z_0,z_0,z_{2,1},z_{2,1})\right].\label{eq:b21b}
\ee

The remaining terms in (\ref{eq:b01a}) and (\ref{eq:b21b}) involving $\bintlim
{1}{0}$ are more tedious to evaluate from the $k=1$ recursion relation in
(\ref{b1lrecur}); they have the following values
\beq
\bintlim {1}{0}(3,0;z_0,z_0,z_0)&=& -\frac{15\;e^{i\xi(\mu_c)}}{2^6}\\
\bintlim {1}{0}(0,3;z_0,z_0,z_0)&=&\frac{31}{2^8}\\
\bintlim {1}{0}(0,4;z_0,z_0,z_{2,1},z_0)&=&-\frac{379}{(3^3)(2^8)}\\
\bintlim {1}{0}(0,4;z_0,z_0,z_{2,1},z_{2,1})&=&-\frac{107}{(3^3)(2^7)}.
\eeq
{}From these results (\ref{eq:b01a}) and (\ref{eq:b21b}) give
\beq
\bintlim {0}{1}(0,2;z_0,z_0)&=& -\frac{41}{2^{8}}\\
\bintlim {2}{1}(0,2;z_0,z_0)&=&\frac{9}{2^7},
\eeq
and from (\ref{eq:b1recurb})
\be
b_2(\mu_c)=\frac{13}{64}.\label{eq:b2final}
\ee
The most striking feature of this result is that, like the cubic coefficient
$b_1$, the fifth order coefficient does not depend on the critical equilibrium
$F_0(v,\mu_c)$.

\subsection{Dependence on $\mu_c$: the role of $e^{i\xi}$}

The recursion relations in (\ref{eq:bjfinal}), (\ref{eq:B00}) - (\ref{eq:B20}),
 (\ref{eq:B0lrecur}), (\ref{b1lrecur}) - (\ref{eq:Ihat1}), (\ref{eq:B2leqn}) -
(\ref{eq:I2leqn}), (\ref{eq:Bkleqn}) - (\ref{eq:Ikleqn}), and
(\ref{eq:clrecur})
determine a closed set of equations for the coefficients $\{b_j\}$ and
$\{c_{1,l}\}$ and the functions $\{\bintlim kl\}$ (also $\{\cintlim kl\}$
determined from (\ref{eq:cintid})). These relations are not independent of the
critical equilibrium $F_0(v,\mu_c)$, but the dependence on the parameters
$\mu_c$ is entirely through a functional dependence on the phase $\exp
i\xi(\mu_c)$. This implies that the higher order coefficients $b_3, b_4,
\ldots$ can only depend on $F_0(v,\mu_c)$ through their dependence on $\exp
i\xi(\mu_c)$; a more precise statement is given in Theorem V.1 below.

In order to describe the iterative procedure to be followed, it is helpful to
organize this collection into nested subsets ${\cal D}(N)$ as follows. For
$N=0$ let
\be
{\cal D}(0)\equiv\{\bintlim 00, \bintlim 20\}
\ee
and for $N\geq1$ define
\be
{\cal D}(N)\equiv\left\{\begin{array}{c}
\{b_j(\mu_c)\;|\;j=1,\ldots,N\}\\
\\
\{c_{1,l}(\mu_c)\;|\;l=0,\ldots,N-1\}\\
\\
\{\bintlim 0l\;|\;l=0,\ldots,N\}\\
\\
\{\bintlim 1l\;|\;l=0,\ldots,N-1\}\\
\\
\{\bintlim kl\;|\;k=2,\ldots k_{max}(N)\;\;\mbox{\rm and }
\,\,l=0,\ldots,(k_{max}(N)-k)\}
\end{array}\right.
\ee
where $k_{max}(N)\equiv N+2$. In this notation it is understood that $\bintlim
kl$ denotes the entire set of functions $\bintlim kl(m,n;\beta,\alpha)$ for all
non-negative integers $(m,n)$. Obviously ${\cal D}(N-1)$ is a subset of ${\cal
D}(N)$ and as $N\rightarrow\infty$ all coefficients and functions mentioned
above are included in ${\cal D}(N)$.

\begin{lemma}  For $N=1,2,\ldots$, the recursion relations determine all
elements of ${\cal D}(N)$ in terms of the elements of ${\cal D}(N-1)$, the
functions $\{\sintlim (m,n;\beta,\alpha)\}$, and the phase $e^{i\xi}$.
\end{lemma}
\noindent {\em {\bf Proof}.} \begin{quote}This is readily verified by
inspection of the recursion relations, and taking into account the identity
${\cintlim 1l(m,n;\beta,\alpha)=(-1)^{m+n-1}\,\bintlim
1l(n,m;\alpha,\beta)^\ast\,\exp{i\xi}}$
from (\ref{eq:cintid}).
{\bf $\Box$}\end{quote}

The recursion relations lead to our main result concerning the dependence of
the $\glim$ limit on the underlying critical equilibrium $F_0(v,\mu_c)$.

\begin{theorem} For $0\leq N<\infty$, the elements of ${\cal D}(N)$ depend on
the critical parameters $\mu_c$ only through a functional dependence on the
phase $e^{i\xi(\mu_c)}$. In particular, for $1\leq j<\infty$ there exist
functions $Q_j(z)$, satisfying
\be
Q_j(z)^\ast=Q_j(z^\ast),\label{eq:5.1b}
\ee
such that
\be
b_j(\mu_c)=Q_j(e^{i\xi(\mu_c)}).\label{eq:5.1a}
\ee
Each function $Q_j$ is universal in the sense that it is independent of $k_c$
and $F_0(v,\mu_c)$. The other elements of ${\cal D}(N)$ can
also be similarly expressed as functions of the phase $\exp (i\xi)$ which
satisfy \mbox{\rm (\ref{eq:5.1b})}.
\end{theorem}
\noindent {\em {\bf Proof}.} \begin{quote}The proof is by induction.
\begin{enumerate}
\item Lemma V.1 shows that $\sintlim (m,n;\beta,\alpha)$ depends on $\mu_c$
only through a functional dependence on $e^{i\xi}$ and the functions involved
are simple polynomials with real coefficients. Specifically
$\sintlim (m,n;\beta,\alpha)=s(e^{i\xi})$
where
\be
s(z)=d(m,n;\beta,\alpha)+(-1)^{m+n}\,d(n,m;\alpha,\beta)\,z.
\ee
It then follows immediately from (\ref{eq:B00}) - (\ref{eq:B00}) for $\bintlim
00$ and $\bintlim 20$ that these functions can also be expressed as polynomials
of $e^{i\xi}$. Since $d(m,n;\beta,\alpha)$ is real-valued, these polynomials
satisfy (\ref{eq:5.1b}). This proves the theorem for ${\cal D}(0)$.

\item For the induction step assume that the theorem holds for ${\cal D}(N-1)$
for some fixed $N\geq1$, and consider the recursion relations which determine
${\cal D}(N)$ from ${\cal D}(N-1)$ and $\{\sintlim (m,n;\beta,\alpha)\}$. Each
of these relations is a sum of terms and most of these terms fit the following
description: the term consists of either a single element of ${\cal D}(N-1)$, a
product of two elements of ${\cal D}(N-1)$, or a product of an element of
${\cal D}(N-1)$ with $\sintlim (m,n;\beta,\alpha)$ (for some $(m,n)$). In each
case the term is multiplied by a real coefficient which is independent of
$\mu_c$ and $k_c$. The exceptions to this description are discussed below. For
each term covered by this description if the individual elements of
${\cal D}(N-1)$ appearing in the term are functions of $\mu_c$ only through
$e^{i\xi}$, then the entire term has this property. In addition, if the
individual functions satisfy (\ref{eq:5.1b}), then their product will also.
Thus all terms, describable in this way, will depend on $\mu_c$ only through a
functional dependence on $e^{i\xi}$ and this function will satisfy
(\ref{eq:5.1b}). Hence these terms preserve for ${\cal D}(N)$ the functional
dependence on $\mu_c$ assumed for ${\cal D}(N-1)$.

\item The recursion relations also contain two types of terms that differ from
the above description. First, there are terms which are the complex conjugate
of an element of ${\cal D}(N-1)$, or which are the product of an element of
${\cal D}(N-1)$ with the complex conjugate of a second element of ${\cal
D}(N-1)$. In either case the coefficient of the term is real and independent of
$\mu_c$ and $k_c$. For example in (\ref{eq:bjfinal}) one finds the term
$c_{1,l}^\ast\;\bintlim 2{j-l-2}(0,2;z_0,z_0)$. Since by assumption there are
functions $f_1(z)$ and $f_2(z)$ such that $c_{1,l}=f_1(e^{i\xi})$ and $\bintlim
2{j-l-2}(0,2;z_0,z_0)=f_2(e^{i\xi})$, this term leads to a functional
dependence $c_{1,l}^\ast\;\bintlim 2{j-l-2}(0,2;z_0,z_0)=f(e^{i\xi})$ where
$f(z)\equiv f_1(1/z)\;f_2(z)$. Therefore this type of term also preserves for
for ${\cal D}(N)$ the functional dependence on $\mu_c$ assumed for ${\cal
D}(N-1)$.

\item The second type of exceptional term involves $\cintlim 1l$ and only
arises in the recursion relation for $\bintlim 0l$ in (\ref{eq:B0lrecur}).
Typical examples are $\cintlim 1{l-1}(m+1,n;\beta,\beta_i,\alpha)$ and
$c_{1,l-j-2}\,\cintlim 1{j}(m+1,n;\beta,\beta_i,\alpha)$. Using the identity
(\ref{eq:cintid}), these terms can be re-expressed in terms of $\bintlim 1l$
as:
\beq
\cintlim 1{l-1}(m+1,n;\beta,\beta_i,\alpha)&=&
(-1)^{m+n}\,e^{i\xi}\,\bintlim 1{l-1}(m+1,n;\beta,\beta_i,\alpha)^\ast
\nonumber\\
c_{1,l-j-2}\,\cintlim 1{j}(m+1,n;\beta,\beta_i,\alpha)&=&
(-1)^{m+n}\,e^{i\xi}\,\bintlim 1{j}(m+1,n;\beta,\beta_i,\alpha)^\ast.\nonumber
\eeq
In each case the essential difference from terms already discussed is that the
coefficient, $(-1)^{m+n}\,e^{i\xi}$, is complex. However since this complex
coefficient is obviously a function of the phase as well, it does not change
the conclusion: these terms also  preserve for for ${\cal D}(N)$ the functional
dependence on $\mu_c$ assumed for ${\cal D}(N-1)$. Hence if the theorem holds
for ${\cal D}(N-1)$, then it holds ${\cal D}(N)$. Since the conclusion has been
verified above for ${\cal D}(0)$, by induction the theorem  holds for all $N$.
\end{enumerate}
{\bf $\Box$}\end{quote}

When the equilibrium has reflection symmetry $F_0(v,\mu)=F_0(-v,\mu)$ then
$p(\sigma,\mu)$ is real and hence $b_j=Q_j(e^{i\xi})$ must be real also. One
can check that in the case of reflection symmetry $\Lambda'_1(z_0)$ is pure
imaginary and $\exp{i\xi}=-1$. Thus the reality condition (\ref{eq:5.1b})
ensures that in this case $b_j=Q_j(-1)$ is real as expected.

Note that the explicit calculation of $b_1$ and $b_2$ in the previous Section
reveals that the first two functions in (\ref{eq:5.1a}) are constant:
\beq
Q_1(z)&=&-\frac{1}{4}\\
Q_2(z)&=&\frac{13}{64};\label{eq:Q2}
\eeq
thus non-trivial dependence on $\exp{i\xi}$ can only arise in $b_j$ for
$j\geq3$. It is natural to wonder if the $Q_j$ are simply constants to all
orders. Although this cannot be categorically ruled out, an inspection of the
recursion relations does not seem encouraging. The evaluation of (\ref{eq:Q2})
involves a crucial and seemingly accidental cancellation that eliminates all
the terms depending on $\exp{i\xi}$. If such a cancellation persists at still
higher order, then detecting it will require a deeper understanding of the
integrated
recursion relations.

\subsection{Implications of Theorem V.1}

Theorem V.1 implies that as $\glim$ the dynamics represented by our amplitude
expansions can depend on the critical equilibrium $F_0(v,\mu_c)$
only through a functional dependence on derivative of the dielectric function
evaluated at the phase velocity of the critical linear mode $\omega_c/k_c$,
\be
\Lambda'_{1}(\omega_c/k_c)\equiv\lim_{\glim}\Lambda'_1(z_0)=\left[{\mbox{\rm
P.V.}} \int^\infty_{-\infty}\,dv\,
\frac{\partial_v\eta(v,\mu_c)}{(v-\omega_c/k_c)}\right]+i\pi
\frac{\partial\eta}{\partial v}(\omega_c/k_c,\mu_c).
\label{eq:eprime}
\ee
This may also be written in terms of the phase (\ref{eq:phase}),
\be
\Lambda'_{1}(\omega_c/k_c)=|\Lambda'_{1}(\omega_c/k_c)|\exp(-i\xi/2);
\ee
in fact for many features it is only the phase $\exp(i\xi)$ that matters and
not the magnitude.

For example, the asymptotic form of the amplitude equation (\ref{eq:g=0})
\be
\frac{dr}{d\tau}=r\,\left\{1+\sum^\infty_{j=1}\,
[\mbox{\rm Re}\;b_j(\mu_c)]\,r^{2j}\right\}
\ee
depends on $F_0(v,\mu_c)$ through $b_j$, and from (\ref{eq:5.1a}) this can be
re-expressed as
\be
\frac{dr}{d\tau}=r\,\left\{1+\sum^\infty_{j=1}\,
[\mbox{\rm Re}\;Q_j(e^{i\xi(\mu_c)})]\,r^{2j}\right\}
\ee
where $Q_j(z)$ itself is a universal function, i.e. independent of $F_0$.
For the electric field in (\ref{efield}), the Fourier component at the
wavelength of the unstable mode is
\be
|k_c\,E_1(t)|=\gamma^2\;r(\tau)\,\left|1+\Gamma^c_1r(\tau)^4\right|
\ee
where
\be
\Gamma^c_1=\sum_{l=0}^{\infty}\,c_{1,l}r^{2l}.
\ee
Since according to Theorem V.1, the coefficients $\{c_{1,l}\}$ also depend on
$F_0(v,\mu_c)$ only through a universal functional dependence on $\exp(i\xi)$,
this component of $E(x,t)$ is predicted to have an asymptotic dynamics that is
determined by $\exp(i\xi)$. For the other wavelengths, one finds that there is
an overall factor of $\Lambda'_{1}(\omega_c/k_c)$ in $E_k(t)$ (cf.
(\ref{eq:cklrecur})), so that $|k_c\,E_k(t)/\Lambda'_{1}(\omega_c/k_c)|$ is
determined by $\exp(i\xi)$.

\section{The distribution function: asymptotic behavior}

The evolving distribution function $F(x,v,t)=F_0(v,\mu) +f^u(x,v,t)$, rewritten
using (\ref{eq:fu}), (\ref{eq:Hexpand}) and (\ref{eq:hdef}), takes the form
\beq
F(x,v,t)&=&F_0(v,\mu) +\sigma h_0(v,\sigma)\label{eq:pdist}\\
&&\hspace{0.00in}+\left[\rho(t)\;(\psi_c(v)+\sigma
h_1(v,\sigma))\;e^{i(k_cx-\theta(t))}+ \sum_{k=2}^{\infty}\;\rho^k
h_k(v,\sigma)
e^{ik(k_cx-\theta(t))} +cc\right]\nonumber
\eeq
where the phase and amplitude variables $\rho(t)e^{-i\theta(t)}=A(t)$ have been
used and $\sigma=|A|^2$.
For fixed $\gamma>0$, our analysis of the mode amplitude dynamics leads to the
equations (\ref{eq:rhodot}) - (\ref{eq:thetadot}), but the long time evolution
of $\rho(t)$ is difficult to predict since the higher order nonlinear terms are
not negligible; this difficulty persists even when $\glim$ as shown in
(\ref{eq:g=0}). However {\em if} $\rho(t)$ asymptotically approaches a constant
value as $t\rightarrow\infty$,
\be
\rho(t)\rightarrow \rho_\infty,\label{eq:tlong}
\ee
then $\theta(t)$ must settle down to a fixed frequency
${d\theta}/{dt}\rightarrow\omega_\infty$
according to (\ref{eq:thetadot}). In this event, the form of $F(x,v,t)$ in
(\ref{eq:pdist}) must approach a travelling wave moving at a constant wave
velocity $\omega_\infty/k_c$. Hence the limiting behavior in (\ref{eq:tlong})
implies the flow on the unstable manifold asymptotically approaches a
travelling wave such as a Bernstein-Greene-Kruskal mode.\cite{bgk} This
asymptotic state is necessarily periodic in time due to the periodic boundary
conditions.

The asymptotic form of $F(x,v,t)$ as $\glim$ can be analyzed from
(\ref{eq:pdist}) without needing to know the long time evolution. In terms of
the rescaled amplitude $\rho(t)=\gamma^2r(\gamma t)$, (\ref{eq:pdist}) becomes
\beq
F(x,v,t)&=&F_0(v,\mu) +r^2\;\gamma^4 h_0(v,\sigma)\label{eq:rescaled1}\\
&& \hspace{0.00in}+\left[r\;\left(\gamma^2\psi_c(v)+\gamma^6\;r^2
h_1(v,\sigma)\right)\;e^{i(k_cx-\theta(t))}+
\sum_{k=2}^{\infty}\;\gamma^{2k}\;r^k h_k(v,\sigma) e^{ik(k_cx-\theta(t))}
+cc\right].\nonumber
\eeq
For {\em fixed} $v\neq v_p$, as $\glim$, this implies
\be
\frac{F(x,v,t)-F_0(v,\mu)}{\gamma^2}=
\left[r\;\psi_c(v)e^{i(k_cx-\theta(t))}+cc\right]+\ord{\gamma^2};
\label{eq:nonres}
\ee
thus away from the phase velocity the correction at the wavelength of the
unstable mode is dominant and is given by the critical eigenfunction.
At $v=v_p$, (\ref{eq:rescaled1}) must be examined more closely since the
functions $h_k(v,\sigma)$ are singular when $\gamma=0$.

\subsection{Asymptotic behavior near $v=v_p$}

Our main interest is in the structure of $F(x,v,t)$ near $v=v_p$ as $\glim$. It
is necessary to extract the singular behavior of $h_k(v_p,\sigma)$ and balance
it against the explicit factors of $\gamma$ in (\ref{eq:rescaled1}). The key
idea, pointed out by Larsen\cite{larsen}, is to appropriately magnify the
neighborhood of $v_p(\gamma)$ using a rescaled velocity coordinate for
$\gamma>0$
\be
u\equiv\frac{k_c}{\gamma}(v-v_p(\gamma)).\label{eq:larsen}
\ee
The motivation for this definition is the resulting factorization of a resonant
denominator:
\be
\frac{1}{v-z_{k,l}}=\frac{k_c}{\gamma}\left[\frac{1}{u-i(1+d_{k,l})}\right],
\ee
in which the $\gamma^{-1}$ singularity has been extracted and the remaining
function of $u$ is nonsingular as $\glim$.

Consider the effect of this coordinate change on the linear eigenfunction
(\ref{eq:lefcn}). Note that since $\partial_vF_0(\omega_c/k_c,\mu_c)=0$,
\beq
\eta(v_p+\gamma u/k_c,\mu)&=&-\left(\frac{1}{k_c^2}\right)\frac{\partial
F_0}{\partial v}(v_p(\gamma)+\gamma u/k_c,\mu(\gamma))\\
&=&-\gamma\left(\frac{1}{k_c^2}\right)
\left[\frac{\partial^2 F_0}{\partial v^2}(\omega_c/k_c,\mu_c)\left(\frac{d
v_p(0)}{d\gamma}+\frac{u}{k_c}\right)\right.\nonumber\\
&&\left.\hspace{1.0in}+\frac{\partial^2 F_0}{\partial v\partial
\mu}(\omega_c/k_c,\mu_c)\frac{d \mu(0)}{d\gamma}+\ord\gamma\right]\nonumber\\
&=&-\frac{\gamma}{k_c}\left[\frac{u}{k_c^2}+\mbox{\rm
Re}\,[\Lambda'_{1}(\omega_c/k_c)]+\ord\gamma\right]\label{eq:etalim}
\eeq
where in the last step the first order solutions for $v_p(\gamma)$ and
$\mu(\gamma)$ from (\ref{eq:roota}) in Appendix A have been used.
This motivates the definitions
\beq
\bar{\eta}(u,\mu)&\equiv&\left(\frac{k_c}{\gamma}\right)\eta(v_p+\gamma
u/k_c,\mu),\label{eq:bareta}\\
\bar{\psi}_c(u)&\equiv&\psi_c(v_p+\gamma u/k_c)\label{eq:resclefn}
\eeq
so that
\be
\bar{\psi}_c(u)=\frac{-\bar{\eta}(u,\mu)}{u-i}.
\ee
Thus the eigenfunction has a non-singular limit as $\gamma\rightarrow0$; from
(\ref{eq:etalim}) - (\ref{eq:bareta})
\be
\lim_{\glim}\bar{\psi}_c(u)=\frac{\left[\mbox{\rm
Re}\,[\Lambda'_{1}(\omega_c/k_c)]+{u}/{k_c^2}\right]}{(u-i)}.\label{eq:nonun}
\ee

The generalization of these definitions for the nonlinear theory is easily done
for any function characterized by an index as defined in Section IV. Let ${\cal
G}(v,\mu)$ have index $\ind [{\cal G}]$ then define $\bar{{\cal G}}(u,\mu)$
\be
\bar{{\cal G}}(u,\mu)\equiv{\gamma}^{1+\ind [{\cal G}]}\;
{\cal G}(v_p+\gamma u/k_c,\mu);
\ee
this is consistent with (\ref{eq:resclefn}) since the eigenfunction has index
$-1$. It is not hard to see that $\bar{{\cal G}}(u,\mu)$ will be non-singular
as
$\glim$. In particular this definition accomplishes the goal of extracting the
singularities in $h_{k,l}$:
\be
\bar{h}_{k,l}(u)\equiv{\gamma}^{1+\ind [h_{k,l}]}\;
h_{k,l}(v_p+\gamma u/k_c).\label{eq:reschkl}
\ee
Applying this to $h_k(v,\sigma)$ gives
\beq
h_k(v_p+\gamma u/k_c,\sigma)&=&
\sum_{l=0}^\infty\;h_{k,l}(v_p+\gamma u/k_c)\,\sigma^l\nonumber\\
&=&\sum_{l=0}^\infty\;\gamma^{4l}\,h_{k,l}(v_p+\gamma u/k_c)\,r^{2l}
\nonumber\\
&=&\sum_{l=0}^\infty\;\gamma^{4l}\left(\frac{1}{\gamma}\right)^{1+\ind
[h_{k,l}]}\,\bar{h}_{k,l}(u)\,r^{2l}
\nonumber\\
&=&\left(\frac{1}{\gamma}\right)^{2k-2+4(\delta_{k,0}+\delta_{k,1})}
\sum_{l=0}^\infty\;\bar{h}_{k,l}(u)\,r^{2l};
\eeq
so if $J=2k-2+4(\delta_{k,0}+\delta_{k,1})$ then $\gamma^{J}h_k(v_p+\gamma
u/k_c,\sigma)$ is non-singular as $\glim$. Formalize this observation in the
definition
\be
\bar{h}_k(u,r^2)\equiv
\left({\gamma}\right)^{2k-2+4(\delta_{k,0}+\delta_{k,1})}\;\left[h_k(v_p+\gamma
u/k_c,\gamma^4\,r^2)\right]=
\sum_{l=0}^\infty\;\bar{h}_{k,l}(u)\,r^{2l}.\label{eq:exseries}
\ee

The left hand side of (\ref{eq:nonres}) can now be evaluated at $v=v_p+\gamma
u/k_c$
in terms of $\bar{\psi}_c(u)$ and $\bar{h}_k(u,r^2)$,
\be
\frac{\left[{F(x,v_p+\gamma u/k_c,t)-F_0(v_p+\gamma u/k_c,\mu)}\right]}
{\gamma^{2}}=g(x,u,t,\mu)
\ee
where
\beq
g(x,u,t,\mu)&\equiv&r^2\bar{h}_0(u,r^2)
+\left[\rule{0in}{0.25in}r\;(\bar{\psi}_c(u)+r^2\;\bar{h}_1(u,r^2))\;
e^{i(k_cx-\theta(t))}\right.\nonumber\\
&&\hspace{0.3in}\left.+ \sum_{k=2}^{\infty}\;r^k\; \bar{h}_k(u,r^2)
e^{ik(k_cx-\theta(t))} +cc\right].
\eeq
As $\glim$, this yields a nonsingular expression $g(x,u,t,\mu_c)$ for the
distribution function in the neighborhood of the phase velocity. In contrast to
(\ref{eq:nonres}), here {\em all} wavelengths contribute to the leading
correction at $\ord{\gamma^2}$.

It is natural to consider whether $g(x,u,t,\mu_c)$ is in some sense universal,
or equivalently to ask how does $g(x,u,t,\mu_c)$ depend on $F_0(v,\mu_c)$?
Obviously there is a trivial dependence on the linear frequency $\omega$
through the factors
$\exp ik(k_cx-\theta(t))$ which can be suppressed by considering the Fourier
coefficients:
\beq
\left|g_0(u,t,\mu_c)\right|&=&
r^2\left|\bar{h}_0(u,r^2)\right|\label{eq:g0}\\
\left|g_1(u,t,\mu_c)\right|&=&
r\left|\bar{\psi}_c(u)+r^2\bar{h}_1(u,r^2)\right|\label{eq:g1}
\eeq
and for $k\geq2$
\be
\left|g_k(u,t,\mu_c)\right|=
r^k\left|\bar{h}_k(u,r^2)\right|;\label{eq:gk}
\ee
these depend on $F_0(v,\mu_c)$ through $\bar{\psi}_c(u)$ and the functions
$\bar{h}_k(u,r^2)$. At $\mu=\mu_c$, the eigenfunction (\ref{eq:nonun})
depends on the critical equilibrium only through the derivative of the
dielectric function $\Lambda'_{1}(\omega_c/k_c)$, and the dependence of
$\bar{h}_k(u,r^2)$
on $F_0(v,\mu_c)$ can be investigated by analyzing the series coefficients
$\bar{h}_{k,l}(u)$ in (\ref{eq:exseries}). The explicit forms for
$\bar{h}_{0,0}(u)$ and $\bar{h}_{2,0}(u)$ follow from (\ref{eq:h00}) -
(\ref{eq:h20})
\beq
\bar{h}_{0,0}(u)&=&-\frac{\partial}{\partial u}
\left[\frac{\bar{\eta}(u,\mu)}{(u-i)\,(u+i)}\right]=\frac{\partial}{\partial u}
\left[\frac{\bar{\psi}_c(u)}{(u+i)}\right]\label{eq:h00ex}\\
\bar{h}_{2,0}(u)&=&\frac{1}{2}\left[\frac{\partial_u\bar{\psi}_c}{(u-i)} +
\left(\frac{\gamma^2}{k_c^2}\right)
\frac{\Lambda^{(2)}_{1}(z_0)\,\bar{\eta}(u,\mu)}{6(u-i)}\right];
\label{eq:h20ex}
\eeq
as $\glim$, these expressions depend on $\mu_c$ only through $\bar{\psi}_c(u)$
and therefore depend on $F_0(v,\mu_c)$ only through
$\Lambda'_{1}(\omega_c/k_c)$.

The remaining coefficients are determined iteratively with recursion relations
that follow by making the change of variable $v=v_p+\gamma u/k_c$ in the
truncated relations in (\ref{eq:h0hat}) - (\ref{eq:Iklhat}) and allowing for
the asymptotic behaviors in (\ref{eq:ckldef}) and (\ref{eq:bjdef}). Let
\beq
\bar{I}_{k,l}(u)&\equiv&{\gamma}^{\ind [h_{k,l}]}\;
\left[I_{k,l}(v_p+\gamma u/k_c)\right]={\gamma}^{\ind [h_{k,l}]}\;
\left[\hat{I}_{k,l}(v_p+\gamma u/k_c)+\ord\gamma\right],
\eeq
then for $k=0$
\be
\bar{h}_{0,l}(u)=\frac{\bar{I}_{0,l}(u)}{2(1+l)}\label{eq:exh0hat}
\ee
with
\beq
\bar{I}_{0,l}(u)&=& \left\{\begin{array}{lc}
{i}\,\frac{\partial }{\partial u}
(\bar{\psi}_c^\ast(u)-\bar{\psi}_c(u))\hspace{1.0in}&
l=0\hspace{0.0in}\\
&\\
-\sum^{l-1}_{j=0}\,(1+j)(b_{l-j}+b_{l-j}^\ast)\,\bar{h}_{0,j}(u)&
l>0\\
+{i}\,\frac{\partial}{\partial u}
\left\{\left[\bar{h}_{1,l-1}^\ast(u)-c_{1,l-1}^\ast\bar{\psi}_c(u)
+\sum^{l-2}_{j=0}\,\bar{h}_{1,j}^\ast(u) c_{1,l-j-2}\right]-cc\right\}&\\
\hspace{0.5in}+\ord{\gamma}&
\end{array}\right.\label{eq:exI0hat}
\eeq

For $k\geq1$, the general relation
\be
\bar{h}_{k,l}(u)={\gamma}^{1+\ind [h_{k,l}]}\;
\left[\left(R_k(w_{k,l})\,{I}_{k,l}\right)(v_p+\gamma u/k_c)\right];
\ee
applies, although the evaluation of the resolvent depends on whether $k=1$ or
$k>1$. For $k=1$
\be
{\gamma}^{1+\ind [h_{1,l}]}
\left[\left(R_1(w_{1,l})\,{I}_{1,l}\right)(v_p+\gamma u/k_c)\right]=
\frac{\left[\bar{I}_{1,l}(u)+
\frac{i\bar{\eta}(u,\mu_c)}{d_{1,l}\,\Lambda'_{1}(\omega_c/k_c)}
\int_{-\infty}^{\infty}\,du'\;\frac{\bar{I}_{1,l}(u')}{u'-i(1+d_{1,l})}
\right]}{i[u-i(1+d_{1,l})]}+\ord\gamma
\ee
where
\beq
\bar{I}_{1,l}(u)&=&-\sum^{l-1}_{j=0}
\left[(2+j)b_{l-j}+(1+j)b_{l-j}^\ast\right]\bar{h}_{1,j}(u)\label{eq:exI1hat}\\
&&+i\frac{\partial}{\partial u}\left\{\bar{h}_{0,l}(u)-\bar{h}_{2,l}(u)
+\sum^{l-1}_{j=0}\left(\bar{h}_{0,j}(u)c_{1,l-j-1}-
\bar{h}_{2,j}(u)c^\ast_{1,l-j-1}\right)\right\}\nonumber\\
&&+\frac{i\bar{\psi}_c(u)}{\Lambda_1'(\omega_c/k_c)}
\int_{-\infty}^\infty\,\frac{du'}{(u'-i)^2}
\left[\bar{h}_{0,l}(u')-\bar{h}_{2,l}(u')
+\sum^{l-1}_{j=0}\left(\bar{h}_{0,j}(u')c_{1,l-j-1}-
\bar{h}_{2,j}(u')c^\ast_{1,l-j-1}\right)\right]\nonumber\\
&&\hspace{1.5in}+\ord\gamma,\nonumber
\eeq
and for $k\geq2$
\be
{\gamma}^{1+\ind
[h_{k,l}]}\left[\left(R_k(w_{k,l})\,{I}_{k,l}\right)(v_p+\gamma u/k_c)\right]=
\frac{\left[\bar{I}_{k,l}(u)\right]}{ik[u-i(1+d_{k,l})]}+\ord\gamma
\ee
where
\beq
\bar{I}_{2,l}(u)&=& \left\{\begin{array}{lc}{i}\,\frac{\partial}{\partial u}
\bar{\psi}_c(u)\hspace{1.0in}&l=0\hspace{0.0in}\\
&\\
-\sum^{l-1}_{j=0}\left[(2+j)b_{l-j}+jb_{l-j}^\ast\right]\,\bar{h}_{2,j}(u)&
l>0\hspace{0.0in}\\
\hspace{0.2in}+{i}\,\frac{\partial}{\partial u}
\left\{\bar{h}_{1,l-1}(u)-\bar{h}_{3,l-1}(u)+\bar{\psi}_c(u)c_{1,l-1}\right.&\\
\hspace{0.5in}+\sum^{l-2}_{j=0}\left.\left[\bar{h}_{1,j}(u)c_{1,l-j-2} -
\bar{h}_{3,j}(u)c^\ast_{1,l-j-2}\right]\right\}+\ord\gamma&
\end{array}\right.\label{eq:exI2hat}
\eeq
and
\beq
\bar{I}_{k,l}(u)&=&
-\sum^{l-1}_{j=0}\,[(k+j)b_{l-j}+j\,b_{l-j}^\ast]\,\bar{h}_{k,j}(u)
\label{eq:exIklhat}\\
&&+{i}\,\frac{\partial }{\partial u}
\left\{\bar{h}_{k-1,l}(u)-\bar{h}_{k+1,l-1}(u)
+\sum^{l-1}_{j=0}\,\bar{h}_{k-1,j}(u)c_{1,l-j-1}
-\sum^{l-2}_{j=0}\,\bar{h}_{k+1,j}(u)c^\ast_{1,l-j-2}\right\}\nonumber\\
&&\hspace{1.5in}+\ord\gamma\nonumber
\eeq
for $k\geq3$.

Inspection of these recursion relations leads to a simple characterization of
the $\mu_c$ dependence of $g(x,u,t,\mu_c)$.

\begin{theorem} The Fourier components $\left|g_k(u,t,\mu_c)\right|$ of
$g(x,u,t,\mu_c)$ in \mbox{\rm (\ref{eq:g0}) - (\ref{eq:gk})} depend on the
critical equilibrium $F_0(v,\mu_c)$ only through a functional dependence on
$\Lambda'_{1}(\omega_c/k_c)$ where
\be
\Lambda'_{1}(\omega_c/k_c)\equiv\lim_{\glim}\Lambda'_{1}(z_0).
\ee

\end{theorem}
\noindent {\em {\bf Proof}.} \begin{quote}
It suffices to examine the dependence of $\bar{\psi}_c(u)$ and
$\bar{h}_{k}(u,r^2)$ on $F_0(v,\mu_c)$. The $\mu_c$ dependence of
$\left.\bar{\psi}_c(u)\right|_{\mu=\mu_c}$ is through $\mbox{\rm
Re}\,[\Lambda'_{1}(\omega_c/k_c)]$ from (\ref{eq:nonun}), and the dependence of
$\bar{h}_{k}(u,r^2)=\sum \bar{h}_{k,l}(u)r^{2l}$ will be inferred from the
coefficients $\bar{h}_{k,l}(u)$.

\begin{enumerate}

\item By inspection from (\ref{eq:h00ex}) - (\ref{eq:h20ex}), the lowest order
coefficients in the organization of Table I also depend on $F_0(v,\mu_c)$
through $\mbox{\rm Re}\,[\Lambda'_{1}(\omega_c/k_c)]$:
\beq
\left.\bar{h}_{0,0}(u)\right|_{\mu=\mu_c}&=&\frac{\partial}{\partial u}
\left[\frac{\mbox{\rm
Re}\,[\Lambda'_{1}(\omega_c/k_c)]+u/k_c^2}{u^2+1}\right]\\
\left.\bar{h}_{2,0}(u)\right|_{\mu=\mu_c}&=&\frac{1}{2(u-i)}
\frac{\partial}{\partial u} \left[\frac{\mbox{\rm
Re}\,[\Lambda'_{1}(\omega_c/k_c)]+u/k_c^2}{u-i}\right].
\eeq

\item The remaining coefficients $\{\bar{h}_{k,l}(u)\}$ are generated from
$\{\bar{\psi}_c,\bar{h}_{0,0},\bar{h}_{2,0}\}$ using the recursion relations
(\ref{eq:exh0hat}) - (\ref{eq:exIklhat}) which depend explicitly on $\mu_c$
through $\Lambda'_{1}(\omega_c/k_c)$, $\bar{\eta}(u,\mu_c)$, and the
coefficients $\{b_j\}$ and $\{c_{1,l}\}$. By Theorem V.1, the coefficients
depend on $\mu_c$ only through the phase
\be
e^{i\xi}=\frac{\Lambda'_{1}(\omega_c/k_c)^\ast}{\Lambda'_{1}(\omega_c/k_c)},
\ee
and from (\ref{eq:etalim}) - (\ref{eq:bareta}) $\bar{\eta}(u,\mu_c)$ depends on
$\mu_c$ through $\mbox{\rm Re}\,[\Lambda'_{1}(\omega_c/k_c)]$. Thus by
induction, the property that $\{\bar{\psi}_c,\bar{h}_{0,0},\bar{h}_{2,0}\}$
depend on $\mu_c$ only through $\Lambda'_{1}(\omega_c/k_c)$ is passed on to all
the coefficients $\{\bar{h}_{k,l}(u)\}$.
\end{enumerate}
{\bf $\Box$}\end{quote}

\section{Discussion}

The approach followed in this paper is patterned on the well established
methods of center manifold reduction and normal form analysis which have proved
quite powerful in analyzing bifurcations in dissipative systems. However for
the Vlasov equation the methods must be adapted. For a dissipative system
undergoing a Hopf bifurcation, one can reduce the problem to a
finite-dimensional submanifold for all $\mu$ sufficiently close to $\mu_c$, and
 for $\mu>\mu_c$ this
corresponds to the finite-dimensional unstable manifold associated with the
equilibrium. By contrast, in the Vlasov instability there is no analogous
possiblility of reducing to a finite-dimensional submanifold in a full
neighborhood of $\mu=\mu_c$ since the critical eigenvalues first appear at
$\mu=\mu_c$ embedded in the continuum and then emerge for $\mu>\mu_c$. It is
only on one side of criticality ($\mu>\mu_c$) that the linear spectrum
indicates finite-dimensional submanifolds should exist and offer a useful
parallel to the dissipative case.

The center manifold and normal form analysis for Hopf bifurcation can be
rephrased in terms of the unstable manifold by deriving the dynamics on the
unstable manifold and then taking the $\glim$ limit  of the resulting vector
field  to obtain normal form equations for the amplitude $A(t)$. As pointed out
in the Introduction, the scaling $A(t)=\sqrt{\gamma}\,r(\gamma
t)e^{-i\theta(t)}$ then yields asymptotic equations\cite{endnote5}
\beq
\frac{dr}{d\tau}&=&rR_H(r,\mu)\label{eq:disc1}\\
\frac{d\theta}{d t}&=&\omega+\ord{\gamma^2}\label{eq:disc2}
\eeq
where
\be
R_H(r,\mu)=1+[a_1(\mu_c)+\ord\gamma]r^2+\ord{{\gamma^2}}.\label{eq:hopfexp}
\ee
This result reveals the correct scaling $\sqrt{\gamma}$ for the nonlinear
oscillation and shows that the expansion (\ref{eq:hopfexp}) for the amplitude
dynamics can be truncated after the lowest order nonlinear term. After
truncation, the remaining  dependence on the underlying critical equilibrium is
expressed entirely through the lowest order nonlinear coefficient $a_1(\mu_c)$
which is a calculable function of $\mu_c$. This truncation also allows the
time-asymptotic behavior to be determined since the solutions of $R_H=0$ are
then easily found.

Our application of this same procedure to the instability of an electrostatic
wave in a collsionless plasma leads to equations like (\ref{eq:disc1}) -
(\ref{eq:disc2}) with several qualitative differences. First, the scaling of
the mode amplitude required to obtain nonsingular asymptotic equations is quite
different: $A(t)={\gamma}^2\,r(\gamma t)e^{-i\theta(t)}$ where now
\beq
\frac{dr}{d\tau}&=&rR(r,\mu)\label{eq:disc3}\\
\frac{d\theta}{d t}&=&\omega+\ord\gamma\label{eq:disc4}
\eeq
with
\be
R(r,\mu)=1+\sum_{j=1}^{\infty}\,[\mbox{\rm Re}\; Q_j(e^{i\xi(\mu_c)})\,r^{2j}
 +\ord\gamma].\label{eq:vpexp}
\ee
Secondly, the expansion for the amplitude equation (\ref{eq:vpexp}) cannot be
truncated. This means that the solutions to $R(r,\mu_c)=0$ are not readily
found and consequently (\ref{eq:disc3}) makes no obvious prediction for the
long time behavior. However despite the greater complexity of $R(r,\mu_c)$, the
dependence on $\mu_c$ can be characterized as arising only through a functional
dependence on the phase $e^{i\xi(\mu_c)}$ which is in turn determined from the
derivative of the dielectric function.

This characterization of $R(r,\mu_c)$ (and similar statements for the electric
field and distribution function) makes testable predictions about instabilities
driven by physically very different distributions. If $e^{i\xi(\mu_c)}$ is
fixed, then any variations in densities or temperatures characterizing
$F_0(v,\mu_c)$ do not affect the evolution of
$r(\tau)$. For example, a beam-plasma instability (complex $\lambda$) and a
two-stream instability (real $\lambda$), compared at a common value of
$e^{i\xi}$, have identical amplitude equations up to ${\cal{O}}(\gamma)$
corrections. Examination  of this prediction through numerical solution of the
Vlasov equation will be undertaken in a future paper.

A final important difference with the dissipative case concerns the generality
of the analysis. In standard center manifold reduction, the local attractivity
of the submanifold ensures that the time-asymptotic behavior of any initial
condition near the equilibrium can be reliably predicted from the evolution on
the submanifold. For Vlasov, or more generally in a Hamiltonian bifurcation,
when there are neutral modes in addition to the critical modes, then there may
be little or no correlation between the evolution of an initial condition on
the unstable manifold and the evolution of an arbitrary initial condition.
Nevertheless numerical studies of the one mode instablility observe the
trapping scaling in the saturation amplitude of the electric field as a quite
robust phenomenon; there is no evidence that the initial condition must be
carefully chosen. This indicates that some features of the evolution on the
unstable
manifold have a wider validity, but it is not clear what selects these
features. One appealing conjecture is that for initial conditions near $F_0$,
but not on the unstable manifold, the electric field $E(x,t)$ evolves
asymptotically towards the electric field $E^u(x,t)$ associated with a solution
on the unstable manifold. This could occur without requiring a corresponding
asymptotic behavior in $F(x,v,t)$ and would suffice to explain the robustness
of the trapping scaling.  This picture seems difficult to investigate
analytically but can be tested through numerical experiments.

A basic feature of the dynamics expected from an autonomous one-dimensional
flow such as $dr/d\tau=R(r,\mu_c)$ is the absence of oscillations, provided $R$
is differentiable in $r$ (or at least Lipschitz continuous). This absence of
oscillations implies that if $r(\tau)$ approaches an asymptotic limit
$r(\tau)\rightarrow r_\infty$as $\tau\rightarrow\infty$, then the approach will
be monotonic from below. This is indeed found in Hopf bifurcation where
$R_H(r,\mu_c)$ is simply a quadratic polynomial in $r$. However in numerical
studies of the one mode instability such monotonic relaxation in the
time-asymptotic regime is not observed, rather one finds the familiar trapping
oscillations in the electric field.\cite{dru}

The explanation of trapping oscillations within the setting of unstable
manifold dynamics may be related to the survival of an infinite sum of terms in
(\ref{eq:vpexp}) as $\glim$. The fact that higher order terms in $r$ are not
higher order in $\gamma$ is a marked contrast with the much simpler limit found
in the Hopf normal form (\ref{eq:hopfexp}), and is directly related to the fact
that the critical eigenvalues merge with the continuous spectrum as $\glim$. In
exactly solvable models, where neutral modes also introduce singularities into
the amplitude equation, I have shown that the unstable manifold can develop a
spiral structure which persists as $\glim$ and this spiral allows the flow to
approach an asymptotic limit through a decaying oscillation.\cite{cra4} This
spiral structure is illustrated in Fig. 4.

When such a spiral is present, then describing the dynamics on
the manifold via a mapping $H$ from the unstable subspace (\ref{eq:graph})
yields a vector field on $E^u$ with branch point singularities at the points
where the flow moves from one branch of the spiral to the next. In the solvable
examples, the fact that $R(r,\mu_c)$ has a branch point within the domain of
flow implies that the higher order terms in the expansion of $R(r,\mu_c)$
remain essential to the dynamics even as $\glim$. Supposing that the amplitude
dynamics for the one mode problem has a similar structure, then $R(r,\mu_c)$
would have a branch point at $r_b$, the turning point of the spiral. As the
mode
grows, the increase of  $r(\tau)$ to $r_b$ would signal the onset of trapping
oscillations with the passage of the trajectory to the next branch of the
unstable manifold.

Note that a trajectory will reach such a branch point node in {\em finite}
time, unlike the more familiar situation of a node where the vector field is
differentiable and the approach time is infinite. In addition, the loss of
smoothness at $r=r_b$ introduces the lack of uniqueness needed by the
solution to pass through the branch point. Finally it should be clear that such
a spiral structure would present a significant obstacle to using the power
series (\ref{eq:vpexp}) to determine the time-asymptotic amplitude $r_\infty$.

Instabilities in other systems, including ideal shear
flows\cite{case2,bdl,cs}, solitary waves\cite{pegowein1,pegowein2,pegwein3},
bubble clouds\cite{russo}, and globally-coupled populations of
oscillators\cite{sm,smm}, also exhibit key features
of this problem, most notably that the unstable modes correspond to
eigenvalues emerging from a neutral continuum at onset. In the case of ideal
shear flows similar singularities arise in the amplitude equations for the
unstable modes ($\gamma^{-3}$ at cubic order)\cite{cs}; by contrast, in
globally coupled phase models for the onset of synchronized behavior in a
population of oscillators the critical eigenvalues emerge from the
continuum at the onset of instability but the amplitude equations are
nonsingular and $\sqrt\gamma$ scaling is found (at least in the best understood
case of a real eigenvalue).\cite{cra5,sm,daido} This difference in the
nonlinear
behavior seems noteworthy since the linear dynamics of the oscillator model is
qualitatively similar to Vlasov although apparently lacking a Hamiltonian
structure.\cite{smm} In the models of solitary waves and bubble clouds, the
scaling behavior in the weakly unstable regime has not been investigated.  It
would be interesting to determine if singularities arise in the amplitude
equations for the unstable modes in these problems with corresponding
implications for the scaling behavior of the nonlinear states.

The study of these novel bifurcations from a unified viewpoint is just
beginning, and it is not yet clear how to abstract the essential features
required to produce singular expansions and unusual scaling behavior.

\acknowledgments

This work has developed over a number of
years and in several stages. Allan Kaufman originally interested me in the one
mode problem as a graduate student at Berkeley. My appreciation of the
importance of nonlinear trapping and the significance of the trapping scaling
resulted from numerous conversations with Tom O'Neil during a post-doc at the
University of California, San Diego. Initial results on the $\gamma^{-3}$
singularity of the cubic coefficient were obtained during a visit to the
Institute for Fusion Studies at the University of Texas and were reported at a
1989 transport conference in Blacksburg. At this conference, I learned from Ed
Larsen of his unpublished work on the one mode problem using a multiple scale
expansion.  His calculations postulated a fast velocity scale equivalent to the
variable $u$ introduced in Section 6.

\appendix
\section{Parametrized roots of the dielectric function}

For $\gamma\geq0$, the roots of the dielectric function determine the
eigenvalues of the unstable modes. From (\ref{eq:paramrt}) this requires the
phase velocity $v_p(\gamma)$ and parameter $\mu(\gamma)$ to satisfy
\be
\Lambda_{1}(z_0(\gamma),\mu(\gamma))= 1-\frac{1}{k_c^2}I(\gamma)=0
\label{eq:app1}
\ee
where
\be
I(\gamma)\equiv\int_{-\infty}^{\infty}\,
\frac{dv\,\partial_v F_0(v,\mu(\gamma))}{v-v_p(\gamma)-i\gamma/k_c}.
\ee
For simplicity, $\mu$ is taken to be a single parameter.

{}From (\ref{eq:app1}) these functions can be calculated perturbatively for
small $\gamma$; let
\beq
v_p(\gamma)&=&\frac{\omega_c}{k_c}+v_1\,\gamma+\ord{\gamma^2}\\
\mu(\gamma)&=&\mu_c+\mu_1\,\gamma+\ord{\gamma^2},
\eeq
and also expand $I(\gamma)$:
\be
I(\gamma)=I(0)+\gamma\frac{dI}{d\gamma}(0)+\ord{\gamma^2}.
\ee
Then (\ref{eq:app1}) implies
\beq
I(0)&=&k_c^2\\
\frac{dI}{d\gamma}(0)&=&0\label{eq:first}
\eeq
and so forth; the coefficients $(v_1,\mu_1)$ are determined by
(\ref{eq:first}). Define  $v=v_p(\gamma)+u/k_c$  and take $u$ as the variable
of integration in $I$,
\be
I(\gamma)=\int_{-\infty}^{\infty}\,
\frac{du}{u-i\gamma}\,\frac{\partial F_0}{\partial
v}(v_p(\gamma)+u,\mu(\gamma)),
\ee
then
\beq
\frac{dI}{d\gamma}(0)&=&{\mbox{\rm
P.V.}}\int_{-\infty}^{\infty}\,\frac{du}{u}\,
\left[\left(\frac{i}{k_c}+v_1\right)\frac{\partial^2F_0}{\partial v^2}
(\omega_c/k_c+u/k_c,\mu_c)
+\mu_1\frac{\partial^2F_0}{\partial v\partial\mu}
(\omega_c/k_c+u/k_c,\mu_c)\right]\nonumber\\
&&+i\pi\left[\left(\frac{i}{k_c}+v_1\right)\frac{\partial^2F_0}{\partial v^2}
(\omega_c/k_c,\mu_c)
+\mu_1\frac{\partial^2F_0}{\partial v\partial\mu} (\omega_c/k_c,\mu_c)\right].
\eeq
Since the real and imaginary parts must vanish separately in (\ref{eq:first}),
there are two equations for $v_1$ and $\mu_1$:
\beq
\frac{1}{k_c}
\left[{\mbox{\rm P.V.}}\int_{-\infty}^{\infty}\,\frac{du}{u}\,
\frac{\partial^2F_0}{\partial v^2} (\omega_c/k_c+u/k_c,\mu_c)\right]
\hspace{1in}&&\nonumber\\
+\pi\left[v_1\frac{\partial^2F_0}{\partial v^2} (\omega_c/k_c,\mu_c)
+\mu_1\frac{\partial^2F_0}{\partial v\partial\mu}
(\omega_c/k_c,\mu_c)\right]&=&0\\
{\mbox{\rm P.V.}}\int_{-\infty}^{\infty}\,\frac{du}{u}\,
\left[v_1\frac{\partial^2F_0}{\partial v^2} (\omega_c/k_c+u/k_c,\mu_c)
+\mu_1\frac{\partial^2F_0}{\partial v\partial\mu} (\omega_c/k_c+u/k_c,\mu_c)
\right]\hspace{0in}&&\nonumber\\
-\frac{\pi}{k_c}
\left[\frac{\partial^2F_0}{\partial v^2} (\omega_c/k_c,\mu_c)\right]
&=&0
\eeq
With
\be
\Lambda'_{1}(\omega_c/k_c)=-\frac{1}{k_c^2}\left\{\left[{\mbox{\rm P.V.}}
\int^\infty_{-\infty}\,,\frac{du}{u}\,
\frac{\partial^2F_0}{\partial v^2} (\omega_c/k_c+u/k_c,\mu_c)\right]
+i\pi \frac{\partial^2F_0}{\partial v^2} (\omega_c/k_c,\mu_c)\right\}
\ee
from (\ref{eq:eprime}), these equations may be rewritten more simply as
\beq
v_1\left[\frac{\partial^2F_0}{\partial v^2} (\omega_c/k_c,\mu_c)\right]
+\mu_1\left[\frac{\partial^2F_0}{\partial v\partial\mu}
(\omega_c/k_c,\mu_c)\right]
&=&k_c\mbox{\rm Re}\,[\Lambda'_{1}(\omega_c/k_c)]\label{eq:roota}\\
v_1\left[{\mbox{\rm P.V.}}\int_{-\infty}^{\infty}\,\frac{du}{u}\,
\frac{\partial^2F_0}{\partial v^2} (\omega_c/k_c+u/k_c,\mu_c)\right]
\hspace{1in}&&\nonumber\\
+\mu_1\left[{\mbox{\rm P.V.}}
\int_{-\infty}^{\infty}\,\frac{du}{u}\,\frac{\partial^2F_0}{\partial
v\partial\mu} (\omega_c/k_c+u/k_c,\mu_c)\right]&=&-k_c\mbox{\rm
Im}\,[\Lambda'_{1}(\omega_c/k_c)].
\eeq

\section{Evaluation of singular integrals}

Let $D_0\equiv1$ and for $n>0$ define
\be
D_n(\alpha,v)\equiv\frac{1}{(v-\alpha_1)(v-\alpha_2)\cdots(v-\alpha_n)}
\ee
where  $\alpha\equiv(\alpha_1,\ldots,\alpha_n)$.
For non-negative integers $(m,n)$ such that $m+n\geq2$, we consider limits of
the following type (c.f. (\ref{eq:slimdef})):
\be
\sintlim(m,n;\beta,\alpha)\equiv
\left(\frac{i}{k_c}\right)^{m+n-2}
\lim_{\glim}\left(
\frac{\gamma^{m+n-2}}{\Lambda'_{1}(z_0)}
\int^\infty_{-\infty}\,dv\,D_m(\beta,v)^\ast\,D_n(\alpha,v)\,
\eta(v,\mu)\right),\label{eq:slimdefb}
\ee
where the poles of the integrand
\beq
\alpha_j&=&z_0+i\gamma\nu_j/k_c\hspace{0.5in}j=1,\ldots,n\\
\beta^\ast_j&=&z_0^\ast-i\gamma\zeta_j/k_c\hspace{0.5in}j=1,\ldots,m
\eeq
lie along the vertical line Re $v=v_p$. The non-negative constants $\nu_j\geq0$
and $\zeta_j\geq0$ are assumed to be independent of $F_0$ for all $j$; in
particular they are independent of $\gamma$. The limit
$\sintlim(m,n;\beta,\alpha)$ does depend on $F_0$ as determined below.

Several general relations can be noted immediately; first interchanging the
order of the arguments gives the simple identity
\be
\sintlim(n,m;\alpha,\beta)=
(-1)^{m+n}\,e^{i\xi(\mu_c)}\,\sintlim(m,n;\beta,\alpha)^\ast\label{eq:interS}
\ee
where
\be
e^{i\xi(\mu_c)}=
\lim_{\glim}\left(\frac{\Lambda'_{1}(z_0)^\ast}{\Lambda'_{1}(z_0)}\right)
\ee
is the phase defined previously in (\ref{eq:phase}). Thus it is sufficient to
evaluate (\ref{eq:slimdefb}) for $m\leq n$. Secondly, by expanding the
denominator with partial fractions,  the limit $\sintlim(m,n;\beta,\alpha)$ can
be expressed in terms of the limits for $m+n-1$. If $m>1$ then
\be
\sintlim(m,n;\beta,\alpha)=
\frac{[\sintlim(m-1,n;\beta'',\alpha)
-\sintlim(m-1,n;\beta',\alpha)]}{\zeta_{m-1}-\zeta_{m}}\label{eq:recur3}
\ee
can be used, if $n>1$ then
\be
\sintlim(m,n;\beta,\alpha)=
\frac{[\sintlim(m,n-1;\beta,\alpha'')
-\sintlim(m,n-1;\beta,\alpha')]}{\nu_{n}-\nu_{n-1}}\label{eq:recur2}
\ee
applies, and if $m>1$ and $n>1$ then
\be
\sintlim(m,n;\beta,\alpha)=
\frac{[\sintlim(m-1,n;\beta',\alpha)-\sintlim(m,n-1;\beta,\alpha')
]}{2+\nu_n+\zeta_m}\label{eq:recur1}
\ee
can be used. In these recursion relations, the primed arguments are defined by
$\alpha'=(\alpha_1,\ldots,\alpha_{n-1})$,
$\alpha''=(\alpha_1,\ldots,\alpha_{n-2},\alpha_{n})$,
$\beta'=(\beta_1,\ldots,\beta_{m-1})$, and
$\beta''=(\beta_1,\ldots,\beta_{m-2},\beta_{m})$.

\subsection{Proof of Lemma V.1}

These general properties allow $\sintlim(m,n;\beta,\alpha)$ to be calculated
recursively from results for $m+n=2,3$. For $m+n=2$ the integral is
straightforward to evaluate and the limits are given by
\beq
\sintlim(0,2;\alpha_1,\alpha_2)&=&1\label{eq:mn2a}\\
\sintlim(2,0;\beta_1,\beta_2)&=&e^{i\xi}\label{eq:mn2b}\\
\sintlim(1,1;\beta_1,\alpha_1)&=&\frac{\nu_1+\zeta_1\,e^{i\xi}}
{2+\nu_1+\zeta_1}.\label{eq:mn2c}
\eeq
For $m+n\geq2$, when either $m=0$ or $n=0$, the integral is nonsingular as
$\glim$, so the results in (\ref{eq:mn2a}) and (\ref{eq:mn2b}) generalize
easily
\beq
\sintlim(0,n;\alpha)&=&\delta_{n,2}.\label{eq:0nform}\\
\sintlim(m,0;\beta)&=&\delta_{m,2}\,e^{i\xi}.\label{eq:m0form}
\eeq
A similar evaluation for $m+n=3$ gives two more limits
\beq
\sintlim(1,2;\beta_1,\alpha_1,\alpha_2)&=&\frac{(2+\zeta_1)-\zeta_1\,e^{i\xi}}
{(2+\nu_1+\zeta_1)(2+\nu_2+\zeta_1)}\label{eq:mn3a}\\
\sintlim(2,1;\beta_1,\beta_2,\alpha_1)&=&\frac{\nu_1-(2+\nu_1)\,e^{i\xi}}
{(2+\nu_1+\zeta_1)(2+\nu_1+\zeta_2)}.\label{eq:mn3b}
\eeq

By inspection of (\ref{eq:mn2a}) - (\ref{eq:mn3b}), the results for $m+n=2,3$
have the following form:
\be
\sintlim(m,n;\beta,\alpha)=d(m,n;\zeta,\nu)+
(-1)^{m+n}d(n,m;\nu,\zeta)\,e^{i\xi}
\label{eq:genform}
\ee
where $\zeta=(\zeta_1,\ldots,\zeta_m)$, $\nu=(\nu_1,\ldots,\nu_n)$, and the
functions $d(m,n;\zeta,\nu)$ are given by
\beq
d(0,2;\nu_1,\nu_2)&=&1\\
d(2,0;\zeta_1,\zeta_2)&=&0\\
d(1,1;\zeta_1,\nu_1)&=&\frac{\nu_1}{2+\nu_1+\zeta_1}
\eeq
for $m+n=2$ and
\beq
d(1,2;\zeta_1,\nu_1,\nu_2)&=&\frac{2+\zeta_1}
{(2+\nu_1+\zeta_1)(2+\nu_2+\zeta_1)}\label{eq:d12}\\
d(2,1;\zeta_1,\zeta_2,\nu_1)&=&\frac{\nu_1}
{(2+\nu_1+\zeta_1)(2+\nu_1+\zeta_2)}\label{eq:d21}
\eeq
for $m+n=3$.

The representation in (\ref{eq:genform}) in fact holds for all $m+n\geq2$. From
(\ref{eq:0nform}) and (\ref{eq:m0form})
\beq
d(0,n;\nu)&=&\delta_{n,2}\\
d(m,0;\zeta)&=&\delta_{m,2}.
\eeq
More generally, when $m$ and $n$ are both non-zero, inserting
(\ref{eq:genform}) into (\ref{eq:recur3}) - (\ref{eq:recur1}) yields the
corresponding recursion relations for $d(m,n;\zeta,\nu)$:
\beq
d(m,n;\zeta,\nu)&=&\frac{d(m-1,n;\zeta',\nu)-d(m,n-1;\zeta,\nu')}
{2+\nu_n+\zeta_m}\label{eq:recurd1}\\
&&\nonumber\\
d(m,n;\zeta,\nu)&=&\frac{d(m,n-1;\zeta,\nu'')-d(m,n-1;\zeta,\nu')}
{\nu_{n}-\nu_{n-1}}\label{eq:recurd2}\\
&&\nonumber\\
d(m,n;\zeta,\nu)&=&\frac{d(m-1,n;\zeta'',\nu)-d(m-1,n;\zeta',\nu)}
{\zeta_{m-1}-\zeta_{m}}\label{eq:recurd3}
\eeq
where $\zeta'=(\zeta_1,\ldots,\zeta_{m-1})$, $\nu'=(\nu_1,\ldots,\nu_{n-1})$,
$\zeta''=(\zeta_1,\ldots,\zeta_{m-2},\zeta_m)$, and
$\nu''=(\nu_1,\ldots,\nu_{n-2},\nu_n)$. In (\ref{eq:recurd2}) and
(\ref{eq:recurd3}) there is no singularity when the denominator vanishes
because the numerator also vanishes; more convenient forms for these recursions
are given below.

Thus the recursive evaluation of $\sintlim(m,n;\beta,\alpha)$ reduces to the
recursive evaluation of $d(m,n;\zeta,\nu)$. Note that the functions
$d(m,n;\zeta,\nu)$ are real-valued and universal in the sense of being
independent of the equilibrium $F_0(v,\mu_c)$. The limit
$\sintlim(m,n;\beta,\alpha)$ depends on $F_0(v,\mu_c)$
only through the phase $e^{i\xi}$. This completes the proof of Lemma V.1.

\subsection{Evaluation of $b_2$}

In the remainder of this Appendix, some results for the functions
$d(m,n;\zeta,\nu)$ that are useful in the calculation of $b_2$ are briefly
summarized. The case when $m=0$ or $n=0$ is trivial, and from (\ref{eq:d12})
and the recursion relation in (\ref{eq:recurd1}), it is easy to show that for
$n\geq2$
\be
d(1,n;\zeta_1,\nu)=\frac{(-1)^n(2+\zeta_1)}{\prod_{i=1}^{n}(2+\nu_i+\zeta_1)}.
\ee
Thus the functions are explicitly known unless both $m$ and $n$ are greater
than $1$ so that $m+n\geq4$.

Henceforth assume that both $m$ and $n$ are non-zero and that $m+n\geq3$.
In (\ref{eq:d12}) - (\ref{eq:d21}), note that $d(2,1;\zeta_1,\zeta_2,\nu_1)=
d(1,2;\nu_1-2,2+\zeta_1,2+\zeta_2)$; more generally such a relation also holds
for $m+n>3$:
\be
d(m,n;\zeta_1,\ldots,\zeta_m,\nu_1,\ldots,\nu_n)=(-1)^{m+n-1}
d(n,m;\nu_1-2,\ldots,\nu_n-2,2+\zeta_1,\ldots,2+\zeta_m).\label{eq:dd1}
\ee
This identity is readily verified by induction from the recursion relations for
$d(m,n;\zeta,\nu)$ in (\ref{eq:recurd1}) - (\ref{eq:recurd3}).
Hence it is sufficient to calculate $d(m,n;\zeta,\nu)$ for $m\leq n$.

{}From the definitions it is clear that $d(m,n;\zeta,\nu)$ must be symmetric
under interchange of the $m$ arguments $(\zeta_1,\ldots,\zeta_m)$ and also
under interchange of the $n$ arguments $(\nu_1,\ldots,\nu_n)$. It is convenient
to have a notation that makes this more explicit and which is manifestly
nonsingular when some of these arguments coincide and the denominators in
(\ref{eq:recurd2}) and (\ref{eq:recurd3}) vanish. For $m\geq1$ and $n\geq1$,
define $N(m,n;\zeta,\nu)$ by the formula
\be
d(m,n;\zeta,\nu)=\frac{N(m,n;\zeta,\nu)}
{\prod_{i=1}^{n}\prod_{j=1}^{m}(2+\nu_i+\zeta_j)};
\ee
so that
\beq
N(1,1;\zeta_1,\nu_1)&=&\nu_1\\
N(1,2;\zeta_1,\nu_1,\nu_2)&=&2+\zeta_1\\
N(2,1;\zeta_1,\zeta_2,\nu_1)&=&\nu_1.
\eeq
{}From (\ref{eq:recurd1}) the corresponding recursion relation for
$N(m,n;\zeta,\nu)$ when $m$ and $n$ are greater than $1$ and $m+n\geq4$ is
\be
N(m,n;\zeta,\nu)=
N(m-1,n;\zeta',\nu)\left[\prod_{i=1}^{n-1}(2+\nu_i+\zeta_m)\right]
-N(m,n-1;\zeta,\nu')\left[\prod_{j=1}^{m-1}(2+\nu_n+\zeta_j)\right],
\label{eq:Nrecur}
\ee
and from (\ref{eq:dd1}) we have the identity
\be
N(m,n;\zeta_1,\ldots,\zeta_m,\nu_1,\ldots,\nu_n)=(-1)^{m+n-1}
N(n,m;\nu_1-2,\ldots,\nu_n-2,2+\zeta_1,\ldots,2+\zeta_m)\label{eq:Nexch}
\ee
when $m+n\geq3$.

Clearly the functions $N(m,n;\zeta,\nu)$ are also symmetric under interchange
of the $m$ arguments $(\zeta_1,\ldots,\zeta_m)$ and also under interchange of
the $n$ arguments $(\nu_1,\ldots,\nu_n)$ but this is not manifest in
(\ref{eq:Nrecur}). Some functions $N(m,n;\zeta,\nu)$, useful in the evaluation
of $b_2$, are listed below showing explicitly the interchange symmetry; in
these formulas the notation $P_n$ corresponds to the symmetric polynomials
defined by
$P_1(x,y)\equiv x+y$, $P_2(x,y)\equiv x^2+xy+y^2$, $P_3(x,y)\equiv
x^3+x^2y+xy^2+y^3$, and $P_4(x,y)\equiv x^4+x^3y+x^2y^2+xy^3+y^4$.
\beq
N(2,2;\zeta,\nu)&=&-\nu_1\nu_2+(2+\zeta_1)(2+\zeta_2)\\
N(2,3;\zeta,\nu)&=&\nu_1\nu_2\nu_3-(2+\zeta_1)(2+\zeta_2)
\left\{\rule{0in}{0.2in}P_1(2+\zeta_1,2+\zeta_2)+\nu_1+\nu_2+\nu_3\right\}\\
N(2,4;\zeta,\nu)&=&-\nu_1\nu_2\nu_3\nu_4+(2+\zeta_1)(2+\zeta_2)
\left\{\rule{0in}{0.2in} P_2(2+\zeta_1,2+\zeta_2)
+P_1(2+\zeta_1,2+\zeta_2)
\left[\rule{0in}{0.17in}\nu_1+\nu_2+\nu_3+\nu_4\right]
\right.\nonumber\\
&&\hspace{2.0in}\left.
+\nu_1(\nu_2+\nu_3+\nu_4)+\nu_2(\nu_3+\nu_4)+\nu_3\nu_4
\rule{0in}{0.2in}\right\}
\eeq
\beq
N(2,5;\zeta,\nu)&=&\nu_1\nu_2\nu_3\nu_4\nu_5-(2+\zeta_1)(2+\zeta_2)
\left\{\rule{0in}{0.2in}P_3(2+\zeta_1,2+\zeta_2)\right.\\
&&\hspace{1.0in}+P_2(2+\zeta_1,2+\zeta_2)
\left[\rule{0in}{0.17in}\nu_1+\nu_2+\nu_3+\nu_4+\nu_5\right]\nonumber\\
&&\hspace{1.0in}+P_1(2+\zeta_1,2+\zeta_2)
\left[\rule{0in}{0.17in}
\nu_1(\nu_2+\nu_3+\nu_4+\nu_5)+\nu_2(\nu_3+\nu_4+\nu_5)\right.\nonumber\\
&&\hspace{2.5in}\left.+\nu_3(\nu_4+\nu_5)+
\nu_4\nu_5\rule{0in}{0.17in}\right]\nonumber\\
&&\hspace{1.0in}
+\nu_1\left[\rule{0in}{0.17in}\nu_2(\nu_3+\nu_4+\nu_5)
+\nu_3(\nu_4+\nu_5)+\nu_4\nu_5)\right]\nonumber\\
&&\hspace{2.0in}\left.
+\nu_2\left[\rule{0in}{0.17in}\nu_3(\nu_4+\nu_5)+\nu_4\nu_5\right]
+\nu_3\nu_4\nu_5\rule{0in}{0.2in}\right\}\nonumber
\eeq
and
\beq
N(3,3;\zeta,\nu)&=&-(2+\zeta_1)(2+\zeta_2)(2+\zeta_3)
\left\{\rule{0in}{0.2in}
(2+\zeta_1)(2+\zeta_2+2+\zeta_3)+(2+\zeta_2)(2+\zeta_3)\right.\\
&&\hspace{2.0in}+(6+\zeta_1+\zeta_2+\zeta_3)(\nu_1+\nu_2+\nu_3)\nonumber\\
&&\left.\hspace{2.0in}
+\nu_1^2+\nu_2^2+\nu_3^2+\nu_1(\nu_2+\nu_3)+\nu_2\nu_3
\rule{0in}{0.2in}\right\}\nonumber\\
&&+\nu_1\nu_2\nu_3\left\{\rule{0in}{0.2in}
\nu_1(\nu_2+\nu_3)+\nu_2\nu_3+(6+\zeta_1+\zeta_2+\zeta_3)(\nu_1+\nu_2+\nu_3)
+(2+\zeta_1)^2\right.\nonumber\\
&&\hspace{0.75in}\left.+(2+\zeta_2)^2+(2+\zeta_3)^2
+(2+\zeta_1)(2+\zeta_2+2+\zeta_3)+(2+\zeta_2)(2+\zeta_3)
\rule{0in}{0.2in}\right\}\nonumber
\eeq
\beq
N(3,4;\zeta,\nu)&=&(2+\zeta_1)(2+\zeta_2)(2+\zeta_3)
\left\{\rule{0in}{0.2in}
(2+\zeta_1)^2\left[\rule{0in}{0.17in}(2+\zeta_2)^2+(2+\zeta_3)^2\right]
\right.\\
&&\hspace{0.5in}+ (2+\zeta_2)^2(2+\zeta_3)^2+(2+\zeta_1)(2+\zeta_2)(2+\zeta_3)
[6+\zeta_1+\zeta_2+\zeta_3]\nonumber\\
&&\hspace{0.5in}+(\nu_1+\nu_2+\nu_3+\nu_4)
\left[\rule{0in}{0.17in}(2+\zeta_1)^2(4+\zeta_2+\zeta_3)
+(2+\zeta_2)^2(4+\zeta_1+\zeta_3)
\right.\nonumber\\
&&\hspace{2.0in}\left.
+(2+\zeta_3)^2(4+\zeta_1+\zeta_2)+2(2+\zeta_1)(2+\zeta_2)(2+\zeta_3)
\rule{0in}{0.17in}\right]
\nonumber\\
&&\hspace{0.5in}+(\nu_1^2+\nu_2^2+\nu_3^2+\nu_4^2)
\left[\rule{0in}{0.17in}(2+\zeta_1)(2+\zeta_2+2+\zeta_3)
+(2+\zeta_2)(2+\zeta_3)\right]\nonumber\\
&&\hspace{0.5in}+[\nu_1(\nu_2+\nu_3+\nu_4)+\nu_2(\nu_3+\nu_4)+\nu_3\nu_4]
(6+\zeta_1+\zeta_2+\zeta_3)^2\nonumber\\
&&\hspace{0.5in}+(6+\zeta_1+\zeta_2+\zeta_3)
\left[\rule{0in}{0.17in}
\nu_1^2(\nu_2+\nu_3+\nu_4)+\nu_2^2(\nu_1+\nu_3+\nu_4)
+\nu_3^2(\nu_1+\nu_2+\nu_4)\right.\nonumber\\
&&\hspace{1.5in}\left.
+\nu_4^2(\nu_1+\nu_2+\nu_3)
+2(\nu_1\nu_2\nu_3+\nu_1\nu_2\nu_4+\nu_1\nu_3\nu_4+\nu_2\nu_3\nu_4)
\rule{0in}{0.17in}\right]
\nonumber\\
&&\hspace{0.5in}+\nu_1^2(\nu_2^2+\nu_3^2+\nu_4^2)+\nu_2^2(\nu_3^2+\nu_4^2)
+\nu_3^2\nu_4^2+\nu_1^2(\nu_2\nu_3+\nu_2\nu_4+\nu_3\nu_4)\nonumber\\
&&\hspace{0.5in}
+\nu_2^2(\nu_1\nu_3+\nu_1\nu_4+\nu_3\nu_4)
+\nu_3^2(\nu_1\nu_2+\nu_1\nu_4+\nu_2\nu_4)\nonumber\\
&&\hspace{0.5in}\left.
+\nu_4^2(\nu_1\nu_2+\nu_1\nu_3+\nu_2\nu_3)+3\nu_1\nu_2\nu_3\nu_4
\rule{0in}{0.2in}\right\}
\nonumber\\
&&-\nu_1\nu_2\nu_3\nu_4
\left\{\rule{0in}{0.2in}
(2+\zeta_1)^3+(2+\zeta_2)^3+(2+\zeta_3)^3
+(2+\zeta_1)^2(4+\zeta_2+\zeta_3)\right.
\nonumber\\
&&\hspace{0.8in}
+(2+\zeta_2)^2(4+\zeta_1+\zeta_3)+(2+\zeta_3)^2(4+\zeta_1+\zeta_2)
+2(2+\zeta_1)(2+\zeta_2)(2+\zeta_3)
\nonumber\\
&&\hspace{0.8in}+(\nu_1+\nu_2+\nu_3+\nu_4)\,
\left[\rule{0in}{0.17in}(2+\zeta_1)^2+(2+\zeta_2)^2+(2+\zeta_3)^2\right.
\nonumber\\
&&\hspace{2.5in}\left.
+(2+\zeta_1)(4+\zeta_2+\zeta_3)+(2+\zeta_2)(2+\zeta_3)\rule{0in}{0.17in}\right]
\nonumber\\
&&\hspace{1.25in}+(6+\zeta_1+\zeta_2+\zeta_3)
\left[\rule{0in}{0.17in}\nu_1(\nu_2+\nu_3+\nu_4)+\nu_2(\nu_3+\nu_4)+\nu_3\nu_4
\right]\nonumber\\
&&\hspace{1.25in}+\left.
\nu_1(\nu_2\nu_3+\nu_2\nu_4+\nu_3\nu_4)+\nu_2\nu_3\nu_4
\rule{0in}{0.2in}\right\}\nonumber
\eeq

\begin{figure}
\caption{Linear stability boundary for the beam-plasma instability described by
a Lorentzian plasma and a Lorentzian beam as in (\protect\ref{eq:family}) with
$L=2\pi$, $n=0.8$, $\Delta=0.3$, and $u_p=0.0$. The instability occurs at the
longest wavelength corresponding to $k=1$; modes with $k=2,3,\ldots$ are always
stable.
The linear spectra for (a) stable equilibria, (b) the critical equilibrium and
(c) unstable equilibria are illustrated in Fig. 2.}
\label{fig1}
\end{figure}

\begin{figure}
\caption{Spectrum of $\lop$ near criticality for the beam-plasma instability of
Fig. 1. (a) The subcritical spectrum contains only the continuous spectrum and
a (degenerate) eigenvalue at zero which is related to the degenerate
Hamiltonian structure of the Vlasov
equation.\protect\cite{cra3} The continuous spectrum coincides with the
imaginary axis but is slightly thickened for ease of visualization. (b) At
criticality, the conjugate pair of eigenvalues, $(\lambda,\lambda^\ast)$ with
$\lambda=\gamma-i\omega$, appears for the first time embedded in the continuous
spectrum. (c) The supercritical spectrum shows a quadruplet of eigenvalues in
addition to
the continuum and zero eigenvalue.}
\label{fig2}
\end{figure}

\begin{figure}
\caption{Local geometry of the unstable manifold; the equilibrium $F_0$ is at
the origin.}
\label{fig3}
\end{figure}

\begin{figure}
\caption{Conjectured spiral structure in the global unstable manifold shown in
cross section with the $\theta$ coordinate suppressed. The turning points of
the spiral correspond to branch points, e.g. $r=r_b$, in the mapping functions
describing the manifold. The time-asymptotic state at $r=r_\infty$ is not on
the branch of the manifold connected to the equilibrium at $r=0$.}
\label{fig4}
\end{figure}

\widetext
\begin{table}
\caption{Order of calculation of $h_{k,l}(v)$ and $p_j$ from $\psi_c(v)$. The
flow of calculation of the $h_{k,l}(v)$ is indicated
by moving downward. From $\psi_c(v)$, $h_{0,0}$ and $h_{2,0}$ can be calculated
 and then $p_1$ determined; $h_{1,0}$ and $h_{3,0}$ are calculated next from
$\{p_1, h_{0,0}, h_{2,0}\}$ and then $h_{0,1}$ and $h_{2,1}$ can be evaluated.
This then determines $p_2$, and so forth. For $N\geq2$, $p_N$ requires prior
calculation of $h_{k,l}$ for $0\leq k\leq N+1$ and $0\leq l\leq N-k+1-2
(\delta_{k,0}+\delta_{k,1}).$}
\begin{tabular}{l|cccccccc}
   &$ k=0$&$k=1$&$k=2$&$k=3$&$k=4$&$k=5$&$k=6$&$\cdots$\\ \hline
\\
$p_0$&  & $\psi_c(v)$   &   & & & & & \\
\hline
\\
$p_1$&$h_{0,0}$&  -         &$h_{2,0}$& & & & &\\
\hline
     &            &$h_{1,0}$&            &$h_{3,0}$& & & &\\
$p_2$&$h_{0,1}$&            &$h_{2,1}$&            & & & & \\
\hline

     &            &$h_{1,1}$&            &            &$h_{4,0}$& & & \\
$p_3$&$h_{0,2}$&            &$h_{2,2}$&$h_{3,1}$&            & & &\\
\hline
     &            &$h_{1,2}$&            &         &      &$h_{5,0}$& &\\
$p_4$&$h_{0,3}$&            &$h_{2,3}$& $h_{3,2}$&$h_{4,1}$ && &\\
\hline
     &            &$h_{1,3}$&            &         &      &&$h_{6,0}$ &\\
$p_5$&$h_{0,4}$&            &$h_{2,4}$& $h_{3,3}$&$h_{4,2}$ &$h_{5,1}$& &\\
\hline
     &            &&            &         &      && &\\
$\vdots$&&            && & && &\\
\end{tabular}
\label{table1}
\end{table}


\begin{references}

\bibitem{frieman} E. Frieman, S. Bodner, and P. Rutherford, Some new results on
the quasi-linear theory of plasma instabilities, { Phys. Fl.} {\bf 6} 1298
(1963).

\bibitem{bald} D.E. Baldwin, Perturbation method for waves in a slowly varying
plasma, { Phys. Fl.} {\bf 7} 782 (1964).

\bibitem{dru} W.E. Drummond, J.H. Malmberg, T.M. O'Neil and J.R.
Thompson, Nonlinear development of the beam-plasma instability, { Phys. Fl.}
{\bf 13} 2422 (1970).

\bibitem{oni}  I.N. Onischenko, A.R. Linetskii, N.G. Matsiborko, V.D.
Shapiro and V.I. Shevchenko, Contribution to the nonlinear theory of excitation
of a monochromatic plasma wave by an electron beam, { JETP Lett.} {\bf 12} 281
(1970).

\bibitem{owm}  T.M. O'Neil, J.H. Winfrey and J.H. Malmberg, Nonlinear
interaction of a small cold beam and a plasma, { Phys. Fl.} {\bf 14}
1204 (1971).


\bibitem{dewar} R.L. Dewar, Saturation of kinetic plasma instabilities by
particle trapping, { Phys. Fl.} {\bf 16} 431 (1973).

\bibitem{sim1} A. Simon and M. Rosenbluth, Single-mode saturation of the
bump-on-tail instability: immobile ions, { Phys. Fl.} {\bf
19} 1567 (1976).

\bibitem{den} J. Denavit, Simulations of the single-mode, bump-on-tail
instability, { Phys. Fl.} {\bf 28} 2773 (1985).

\bibitem{sim2} A. Simon, S. Radin, and R.W. Short, Long-time simulation of
the single-mode bump-on-tail instability, { Phys. Fl.} {\bf
31} 3649 (1988).

\bibitem{janssen}  P. Janssen and J. Rasmussen, Limit cycle behavior
of the bump-on-tail instability, { Phys. Fl.}
{\bf 24} 268 (1981).

\bibitem{burnap}   C. Burnap, M. Miklavcic, B. Willis and P.
Zweifel, Single-mode saturation of a linearly unstable plasma, { Phys. Fl.}
{\bf 28} 110 (1985).

\bibitem{guc} J. Guckenheimer and P. Holmes, {\em Nonlinear
oscillations, dynamical systems and bifurcations of vector
fields}, (Springer-Verlag, New York, 1983).

\bibitem{jdc2} J.D. Crawford, Introduction to bifurcation theory, { Rev. Mod.
Phys.} {\bf 63} 991-1037 (1991).

\bibitem{morrison} P.J. Morrison, The Maxwell-Vlasov equations as a continuous
Hamiltonian system, { Phys. Lett. A} {\bf 80} 383-386 (1980).

\bibitem{marwein} J.E. Marsden and A. Weinstein, The Hamiltonian structure of
the Maxwell-Vlasov equations, { Physica D} {\bf 4} 394-406 (1982).

\bibitem{holm} D.D. Holm, J.E. Marsden, T. Ratiu, and A. Weinstein, Nonlinear
stability of fluid and plasma equilibria, { Phys. Reports} {\bf 123} 1-116
(1985).

\bibitem{cra1} J.D. Crawford and P. Hislop, Application of the method
of spectral deformation to the Vlasov-Poisson model, { Ann. Phys.}
{\bf 189} 265-317 (1989).

\bibitem{jdc} J.D. Crawford, Universal trapping scaling on the unstable
manifold
for an unstable electrostatic mode, { Phys. Rev. Lett.} {\bf 73} 656-659
(1994).

\bibitem{tsu} S.I. Tsunoda, F. Doveil and J.H. Malmberg, Nonlinear interaction
between a warm electron beam and a single wave, {Phys. Rev. Lett.} {\bf 59}
2752 (1987).

\bibitem{endnote} In addition to the lack of consensus on the correct scaling,
these authors are often unaware of previous work. From the viewpoint of this
paper, the most important theoretical discussion is Baldwin's
analysis\cite{bald}, and he is not referenced in any of the subsequent
publications.\cite{dru,oni,owm,dewar,sim1,den,sim2,janssen,burnap} Baldwin
finds the pinching singularity at lowest nonlinear order and correctly notes
the effect this singularity will have on the saturation level of the mode. He
does not appear to recognize the complete generality of this result, and argues
that it can be reconciled with the results of Frieman {\em et
al.}\cite{frieman} if the second velocity derivative of $F_0(v,\mu_c)$
vanishes. But this is not the
essential discrepancy between Baldwin's calculation and the analysis by Frieman
{\em et al.}, rather the latter authors apply the Plemelj formula incorrectly
and thereby miss the pinching singularity that Baldwin finds a year later.

\bibitem{marsmac} J.E. Marsden and M. McCracken, {\em The Hopf Bifurcation and
Its Applications}, (Springer-Verlag, New York, 1976). pp. 1-135, 250-304.

\bibitem{henry} D. Henry, {\em Geometric Theory of Semilinear Parabolic
Equations}, Lecture Notes in Mathematics {\bf 840} (Springer-Verlag, New York,
1981).

\bibitem{carr} J. Carr, {\em The Centre Manifold Theorem and its Applications},
(Springer-Verlag, New York, 1983).

\bibitem{chow} S-N. Chow and K. Lu, Invariant manifolds for flows in Banach
spaces, { J. Diff. Eqns.} {\bf 74} 285-317 (1988).

\bibitem{mielke} A. Mielke, Locally invariant manifolds for quasilinear
parabolic equations, { Rocky Mountain J. Math.} {\bf 21} 707-714 (1991).

\bibitem{renardy} M. Renardy, A centre manifold theorem for hyperbolic PDE's,
{ Proc. Roy. Soc. Edin} {\bf 122A} 363-377 (1992).

\bibitem{vi}  A. Vanderbauwhede and G. Iooss, 1992, Centre manifold theory in
infinite dimensions, {\em Dynamics Reported}, New Series, Vol. 1,
Springer-Verlag, New York, 125-163.

\bibitem{vkamp} N.G. van Kampen, On the theory of stationary waves in plasmas,
{ Physica} {\bf 21} (1955) 949; see also N.G. van Kampen and B.U. Felderhof,
{\em Theoretical Methods in Plasma Physics}, (North-Holland, Amsterdam, 1967).

\bibitem{case} K. Case, Plasma oscillations, { Ann. Phys.}
{\bf 7} (1959) 349-364; also { Phys. Fl.} {\bf 21} 249-257 (1978).

\bibitem{cowley} S.W.H. Cowley, Growing plasma oscillations for symmetrical
douple-humped velocity distributions, { J. Plasma Phys.} {\bf 4} 297 (1970).

\bibitem{musk} N.I. Muskhelishvili, {\em Singular Integral Equations}, (Dover,
New York, 1992). pp 42-43.

\bibitem{endnote1} I do not reserve this
notation for the specific choice of a Maxwellian $F_0$; $\epsilon_{{k}}(z)$
denotes the analytic continuation for an arbitrary equilibrium.

\bibitem{text} S. Ichimaru, {\em Basic Principles of Plasma Physics},
(Benjamin-Cummings, Reading, MA, 1973). pp. 43-46.

\bibitem{shad} B.A. Shadwick and P.J. Morrison, On neutral plasma oscillations,
{ Phys. Lett. A} {\bf 184} 277-282 (1994).

\bibitem{endnote2} By {\em invariant} we mean that if the initial condition
$f(x,v,0)$ belongs to the subspace (manifold), then the solution $f(x,v,t)$
remains in the subspace (manifold).

\bibitem{endnote3} This is a standard result symmetric bifurcation theory; see
reference \cite{cra5} for a more detailed discussion.

\bibitem{cra5} J.D. Crawford, Amplitude expansions for instabilities in
populations of globally-coupled oscillators, { J. Stat. Phys} {\bf 74}
1047-1084 (1994).

\bibitem{cra2} P. Hislop and J.D. Crawford, Application of spectral
deformation to the Vlasov-Poisson system II:  mathematical
results, J. Math. Phys. {\bf 30} 2819-2837 (1989).

\bibitem{endnote4} The occurrence of such a pinching singularity appears to
have been first discovered by D. Baldwin thirty years ago.\cite{bald} Unaware
of Baldwin's paper, I first found the $\gamma^{-3}$ singularity of the cubic
coefficient several years ago and reported it in the proceedings of a
Blacksburg Transport conference.\cite{cra4} This original calculation used the
``leaf'' representation of the Vlasov equation developed with
Hislop.\cite{cra3} In this representation the real coefficient of $\gamma^{-3}$
is manifestly negative but has a nontrivial dependence on the critical
distribution function $F_0(v,\mu_c)$.

\bibitem{bgk} I. Bernstein, J.M. Greene, and M.D. Kruskal, Exact nonlinear
plasma oscillations, { Phys. Rev.} {\bf 108} 546-550 (1957).

\bibitem{larsen} E. Larsen, private communication, (1989).

\bibitem{endnote5} This may require a near-identity transformation on the
$(r,\theta)$ variables to remove terms quadratic in $A$ as described elsewhere
\cite{jdc2}.

\bibitem{cra4} J.D. Crawford, Amplitude equations on unstable
manifolds: singular behavior from neutral modes, in {\em Modern
Mathematical Methods in Transport Theory} (Operator Theory:
Advances and Applications, vol. 51), W. Greenberg and J.
Polewczak, eds., Birkhauser Verlag, Basel, 1991, pp. 97-108.

\bibitem{case2} K. Case, Stability of inviscid plane Couette flow, { Phys. Fl.}
{\bf 3} 143 (1960).

\bibitem{bdl} R.J. Briggs, J.D. Daugherty, and R.H. Levy, Role of Landau
damping in crossed-field electron beams and inviscid shear flow, { Phys. Fl.}
{\bf 13}  421-432 (1970).

\bibitem{cs} S.M. Churilov and I.G. Shukhman, Nonlinear stability of a
stratified shear flow in the regime with an unsteady critical layer, { J. Fluid
Mech.} {\bf 194} 187-216 (1988).


\bibitem{pegowein1} R. Pego and M.I. Weinstein, Eigenvalues and instabilities
of solitary waves, { Phil. Trans. R. Soc. Lond. A} {\bf 340} 47-94 (1992).

\bibitem{pegowein2} R. Pego and M.I. Weinstein, Evans function, Melnikov's
integral, and solitary wave instabilities,  in {\em Differential Equations with
Applications to Mathematical Physics}, W.F. Ames, E.M. Harrell II, and J.V.
Herod, eds., Academic Press, Orlando, 1993. pp. 273-286.

\bibitem{pegwein3} R. Pego, P. Smereka, and M.I. Weinstein, Oscillatory
instability of solitary waves in a continuum model of lattice vibrations, {
Nonlinearity}, submitted, (1994).

\bibitem{russo} G. Russo and P. Smereka, Kinetic theory for bubbly flow I:
collisionless case, { SIAM J. Appl. Math.}, submitted, (1994).

\bibitem{sm} S. Strogatz and R. Mirollo, Stability of incoherence in a
population of coupled oscillators, { J. Stat. Phys.} {\bf 63} 613-635 (1991).

\bibitem{smm} S. Strogatz, R. Mirollo and P.C. Matthews, Coupled nonlinear
oscillators below the synchronization threshold: relaxation by generalized
Landau damping, { Phys. Rev. Lett.} {\bf 68} 2730-2733 (1992).

\bibitem{daido} Very recent calculations by Daido indicate that this conclusion
depends on the specific form of the coupling between oscillators. See H. Daido,
Generic scaling at the onset of macroscopic mutual entrainment in limit-cycle
oscillators with uniform all-to-all coupling, {Phys. Rev. Lett.} {\bf 73} 760
(1994).

\bibitem{cra3} J.D. Crawford and P. Hislop, Vlasov equation on a
symplectic leaf, { Phys. Lett. A}
{\bf 134} 19-24 (1988).

\end{references}
\end{document}